\def\tr{{\rm tr}\,}
\def\Tr{{\rm Tr}\,}

\def\sgn{{\rm sgn\,}}

\def\be{\begin{equation}}
\def\ee{\end{equation}}
\def\bea{\begin{eqnarray}}
\def\eea{\end{eqnarray}}
\def\bml{\begin{mathletters}}
\def\eml{\end{mathletters}}
\def\bse{\begin{subequations}}
\def\ese{\end{subequations}}
\documentclass[rmp,twocolumn,showpacs,amsmath,amssymb,eqsecnum]
              {revtex4}
\usepackage{graphicx}
\usepackage{dcolumn}
\usepackage{bm}
\begin{document}
\title{Influence of Generic Scale Invariance on Classical and Quantum Phase
       Transitions}
\author{D.Belitz}\thanks{belitz@physics.uoregon.edu}
\affiliation{Department of Physics and Materials Science Institute,
         University of Oregon,
         Eugene, OR 97403}
\author{T.R.Kirkpatrick}\thanks{tedkirkp@umd.edu}
\affiliation{Institute for Physical Science and Technology, and Department of
         Physics,
         University of Maryland, College Park, MD 20742}
\author{Thomas Vojta}\thanks{vojtat@umr.edu}
\affiliation{Dept. of Physics, University of Missouri-Rolla, Rolla, MO 65409}
\date{\today}

\begin{abstract}
This review discusses a paradigm that has become of increasing importance in
the theory of quantum phase transitions; namely, the coupling of the
order-parameter fluctuations to other soft modes, and the resulting
impossibility of constructing a simple Landau-Ginzburg-Wilson theory in terms
of the order parameter only. The soft modes in question are manifestations of
generic scale invariance, i.e., the appearance of long-range order in whole
regions in the phase diagram. The concept of generic scale invariance, and its
influence on critical behavior, is explained using various examples, both
classical and quantum mechanical. The peculiarities of quantum phase
transitions are discussed, with emphasis on the fact that they are more
susceptible to the effects of generic scale invariance than their classical
counterparts. Explicit examples include: the quantum ferromagnetic transition
in metals, with or without quenched disorder; the metal-superconductor
transition at zero temperature; and the quantum antiferromagnetic transition.
Analogies with classical phase transitions in liquid crystals and classical
fluids are pointed out, and a unifying conceptual framework is developed for
all transitions that are influenced by generic scale invariance.
\end{abstract}
\maketitle
\tableofcontents

\section*{Preamble}
\label{sec:0}

The theoretical understanding of classical, or thermal, phase transitions,
which occur at a nonzero temperature, is very well developed. A characteristic
feature of continuous phase transitions, or critical points, is that the free
energy, as well as time correlation functions, obey generalized homogeneity
laws. This leads to scale invariance, i.e., power-law behavior of various
thermodynamic derivatives and correlation functions as functions of the
temperature, external fields, wave number, time, etc. Quantum phase
transitions, which occur at zero temperature as a function of some nonthermal
control parameter such as composition, or pressure, are not yet as well
understood. One feature that has slowed down progress has turned out to be a
connection between quantum phase transitions and the phenomenon known as
generic scale invariance, which refers to power-law decay of correlation
functions in entire phases, not just at an isolated critical point. It turns
out that the critical behavior at some classical phase transitions is also
heavily influenced by a coupling between generic and critical scale invariance,
but these phenomena have usually not been cast in this language. The goal of
this review article is to give a unifying discussion of the coupling between
critical behavior and generic scale invariance, for both classical and quantum
phase transitions, and for both static and dynamic critical behavior.
Accordingly, we limit our discussion to phase transitions where this coupling
is known or suspected to be important.

The structure of this paper is as follows. In Sec. \ref{sec:I} we give a brief
introduction to phase transitions, both classical and quantum. In Sec.
\ref{sec:II} we discuss the concept of generic scale invariance, and illustrate
it by means of various examples. These concepts are less well known than those
pertaining to phase transitions, so our exposition is more elaborate. In Secs.
\ref{sec:III} and \ref{sec:IV} we return to phase transition physics and
discuss the influence of generic scale invariance on classical and quantum
transitions, respectively. We conclude in Sec. \ref{sec:V} with a summary and a
discussion of open problems.

\section{Phase Transitions}
\label{sec:I}

Phase transitions are among the most fascinating phenomena in nature. They also
have far-reaching implications. The liquid-gas and liquid-solid transitions in
water, for instance, are common occurrences of obvious importance. The
transition from a paramagnetic phase to a ferromagnetic one in the elements
iron, nickel, and cobalt made possible the invention of the compass. The
structural transition in tin from the $\beta$-phase (white tin) to the
$\alpha$-phase (gray tin) is responsible for the degradation of tin artifacts,
below a temperature of about 286~K, known as the tin pest. One could continue
with a long list, but these three examples may suffice to demonstrate that
phase transitions come in a wide variety of phenomenologies with no entirely
obvious unifying features. Accordingly, early attempts at theoretically
understanding phase transitions focused on particular examples.
\textcite{van_der_Waals_1873} developed the first example of what was later to
become known as a ``mean-field theory'', in this case, to describe the
liquid-gas transition. \textcite{Weiss_1907} gave a mean-field theory of
ferromagnetism which was based on the concept of the ``mean field'' seen by
each `elementary magnet' (spin had yet to be discovered at that time) that is
produced by all other elementary magnets. A unification of all mean-field
theories was achieved by \textcite{Landau_1937a, Landau_1937b, Landau_1937c,
Landau_1937d} (all of these papers, or their translations, are reprinted in
\onlinecite{Landau_1965}). He introduced the general concept of the ``order
parameter'', a thermodynamic observable that vanishes in one of the phases
separated by the transition (the ``disordered phase''\footnote{\label{fn:1} We
will use the term ``disordered phase'' in the sense of ``phase without
long-range order''. This is not to be confused with the presence of quenched
disorder, which we will also discuss.}), and is nonzero in the other (the
``ordered phase''). In the case of the ferromagnet, the order parameter is the
magnetization; in the case of the liquid-gas transition, the density difference
between the two phases. Landau theory underlies all later theories of phase
transitions, and we therefore discuss it first.

\subsection{Landau Theory}
\label{subsec:I.A}

Landau theory is based on one crucial assumption, namely, that the free energy
$F$ is an analytic function of the order parameter $m$,\footnote{\label{fn:2}
In this section we consider a scalar order parameter for simplicity, but later
we will encounter more general cases.} and hence can be expanded in a power
series,
\be
F \approx F_{\text{L}}(m) = r\,m^2 + v\,m^3 + u\,m^4 + O(m^5)\quad.
\label{eq:1.1}
\ee
Here $r$, $v$, $u$, etc., are parameters of the Landau theory that depend on
all of the degrees of freedom other than $m$. $F_{\text{L}}$ is sometimes
referred to as the Landau functional, although it actually is just a function
of the variable $m$. The physical value of $m$ is the one that minimizes
$F_{\text L}$.

\begin{figure}[t]
\includegraphics[width=8cm]{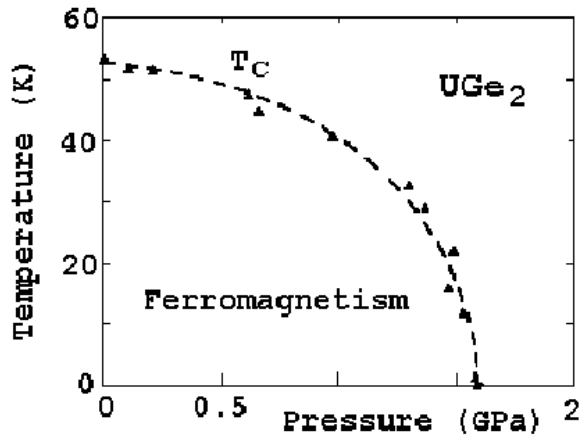}
\caption{\label{fig:1} Phase diagram of UGe$_2$. After
   \protect{\textcite{Saxena_et_al_2000}}.}
\end{figure}

Landau theory is remarkably versatile. For sufficiently large $r$, the minimum
of $F$ is always located at $m=0$, while for sufficiently small $r$ it is
located at some $m\neq 0$. If $v\neq 0$, the transition from $m=0$ to $m\neq 0$
occurs discontinuously, and the theory describes a ``first-order transition'',
with the liquid-gas transition, except at the liquid-gas critical point, being
the prime example. If $v = 0$, either accidentally or for symmetry reasons, it
describes a second-order transition, or critical point, at $r=0$, provided
$u>0$. Prime examples are the ferromagnetic transition in zero magnetic field,
and the liquid-gas transition at the critical point. Furthermore, Landau theory
applies to both classical and quantum systems, including systems at zero
temperature ($T=0$). In the latter case, the free energy $F = U - TS$ reduces
to the internal energy $U$.\footnote{\label{fn:3} At $T>0$, a stable phase can
have a higher internal energy $U$ than an unstable one, as long as its entropy
$S$ is also higher. At $T=0$, the stable phase must represent a minimum of the
internal energy.} This is not an academic case. Consider a ferromagnet with a
low Curie temperature $T_{\text{c}}$, e.g., UGe$_2$. By varying a non-thermal
control parameter, e.g., hydrostatic pressure, one can suppress the Curie
temperature to zero, see Fig. \ref{fig:1}.
The phase transition can then be triggered in particular by varying the
pressure at $T=0$. Obviously, Landau theory predicts the same critical behavior
in either case: The magnetization $m$ vanishes as $m(t\to 0) = \sqrt{-r/2u}$,
irrespective of how $r$ is driven to zero. This is an example of the
`super-universality' inherent in Landau theory: It predicts the critical
exponents to be the same for {\em all} critical points. For instance, the
critical exponent $\beta$, defined by $m\propto \vert r\vert^{\beta}$, is
predicted to have the value $\beta = 1/2$.

A complete description of a statistical mechanics system requires, in addition
to the thermodynamic properties encoded in the free energy, knowledge of time
correlation functions. Of particular interest is the order parameter
susceptibility $\chi_m$. As a critical point is approached from the disordered
phase, the harbingers of the latter's instability are diverging order-parameter
fluctuations, which lead to a divergence of $\chi_m$ at zero frequency and wave
number. Within Landau theory, these fluctuations are described in a Gaussian
approximation \cite{Landau_Lifshitz_V_1980}. For the wave number dependent
static order parameter susceptibility $\chi_m({\bm k})$ this yields the
familiar Ornstein-Zernike form
\bse
\label{eqs:1.2}
\be
\label{eq:1.2a}
\chi_m({\bm k}) \propto \frac{1}{r + c\,{\bm k}^2}\ ,
\ee
with $c$ a constant. In real space, this corresponds to exponential decay,
\be
\label{eq:1.2b}
\chi_m({\bm x}) \propto \vert{\bm x}\vert^{-1}\ e^{-\vert\bm{x}\vert/\xi},
\ee
\ese
where the correlation length $\xi$ diverges, for $r\to 0$, according to $\xi
\propto r^{-1/2}$.

The order-parameter fluctuations at criticality, $r=0$, are an example of a
``soft mode'',\footnote{\label{fn:4} The terms ``soft'', ``gapless'', and
``massless'' are used interchangeably in this context, with the first and
second most popular in classical and quantum statistical mechanics,
respectively, and the third one borrowed from particle physics.} i.e.,
fluctuations that diverge in the limit of small frequencies and wave numbers as
illustrated by Eqs.\ (\ref{eqs:1.2}). The soft-mode concept will play a crucial
role in the remainder of this article. In addition to modes that are soft only
at the critical point, we will discuss, in Sec.\ \ref{sec:II}, modes that are
soft in an entire phase. The coupling between such ``generic soft modes'' and
the critical order-parameter fluctuations can have profound consequences for
the critical behavior, as we will see in Secs.\ \ref{sec:III} and \ref{sec:IV}.
These consequences are the main topic of this review, but before we can discuss
them we need to recall some additional aspects of critical behavior.

Equations (\ref{eqs:1.2}) define three additional critical exponents, namely,
the correlation length exponent $\nu$, defined by $\xi\propto \vert
r\vert^{-\nu}$, and the susceptibility exponents $\gamma$ and $\eta$, defined
by $\chi_m({\bm k}=0) \propto r^{-\gamma}$, and $\chi_m(r=0) \propto \vert{\bm
k}\vert^{-2+\eta}$. Landau theory universally predicts $\nu = 1/2$, $\gamma =
1$, and $\eta = 0$.

Universality is actually observed in experiments, but it is weaker than the
superuniversality predicted by Landau theory, and the observed values of the
exponents are in general different from what Landau theory predicts. For
instance, all bulk Heisenberg ferromagnets have a $\beta\approx 0.35$ (e.g.,
\onlinecite{Zinn-Justin_1996}), which is significantly smaller than the Landau
value. Similarly, all bulk Ising ferromagnets have a common value of $\beta$,
but that value, $\beta\approx 0.32$, is different from the one in Heisenberg
systems. Also, the critical exponents turn out to be different for systems of
different dimensionality, again in contrast to the prediction of Landau theory.
For instance, in two-dimensional Ising ferromagnets, $\beta = 1/8$
\cite{Onsager_1948, Yang_1952}.

\subsection{Landau-Ginzburg-Wilson Theory, the Renormalization Group, and Scaling}
\label{subsec:I.B}

The reason for the failure of Landau theory to correctly describe the details
of the critical behavior is that it does not adequately treat the fluctuations
of the order parameter about its mean value. The deviations of these
fluctuations from a Gaussian character in general are stronger for lower
dimensionalities, and for order parameters with fewer components. This explains
why the critical behavior of Ising magnets deviates more strongly from the
Landau value than that of Heisenberg magnets, and why the exponents of bulk
systems are closer to the mean-field values than those of thin films. This
observation suggests that Landau theory might actually yield the correct
critical behavior in systems with a sufficiently high dimensionality $d$.
Indeed it turns out that in general there is an upper critical dimensionality,
$d_{\text{c}}^{\,+}$, such that for $d>d_{\text{c}}^{\,+}$ fluctuations are
unimportant for the leading critical behavior and Landau theory gives the
correct answer (this follows from the Ginzburg criterion; see, e.g.,
\onlinecite{Cardy_1996}, Sec. 2.4). For the ferromagnetic transition at $T>0$,
$d_{\text{c}}^{\,+} = 4$. For $d<d_{\text{c}}^{\,+}$, fluctuations need to be
taken into account beyond the Gaussian approximation in order to obtain the
correct critical behavior. This problem was solved by Wilson
(\onlinecite{Wilson_Kogut_1974}; see also, e.g.,
\onlinecite{Ma_1976,Fisher_1983}). He generalized the Landau functional, Eq.\
(\ref{eq:1.1}), by writing the partition function $Z = e^{-F/T}$ as a
functional integral,\footnote{\label{fn:5} We use units such that Boltzmann's
constant, Planck's constant, and the Bohr magneton are equal to unity.}
\bse
\label{eqs:1.3}
\be
Z = e^{-F/T} = \int D[\phi]\ e^{-S[\phi]}\,,
\label{eq:1.3a}
\ee
where
\be
S[\phi] = \frac{1}{TV}\int d{\bm x}\ \left[F_{\text{L}}(\phi(\bm{x})) +
c\,(\bm{\nabla}\phi(\bm{x}))^2\right]
\label{eq:1.3b}
\ee
\ese
is usually referred to as the ``action''. Here $V$ is the system volume, and
$\phi$ is a fluctuating field whose average value with respect to the
statistical weight $\exp(-S)$ is equal to $m$. This Landau-Ginzburg-Wilson
(LGW) functional is then analyzed by means of the renormalization group
(RG);\footnote{\label{fn:6} Pedagogical expositions of the Wilsonian RG have
been given by, e.g., \textcite{Wilson_Kogut_1974, Ma_1976,Fisher_1983,
Cardy_1996, Goldenfeld_1992}.} the lowest order, i.e., saddle-point/Gaussian,
approximation recovers Landau theory. Wilson's RG takes advantage of the fact
that at a critical point there is a diverging length scale, namely, the
correlation length $\xi$, which dominates the long-wavelength physics. By
integrating out all fluctuations on smaller length scales, one can derive a
succession of effective theories that describe the behavior near criticality.
The RG made possible the derivation and proof of the behavior near criticality
known as ``scaling'', which previously had been observed empirically and
summarized in the ``scaling hypothesis''.\footnote{\label{fn:7} The state of
affairs just before the invention of the RG was reviewed by
\textcite{Kadanoff_et_al_1967} and \textcite{Stanley_1971}.} Due to scaling,
all static critical phenomena are characterized by two independent critical
exponents, one for the reduced temperature $r \propto \vert
T-T_{\text{c}}\vert$,\footnote{\label{fn:8} One needs to distinguish between
the exact or fully renormalized value of $r$, which appears in formal scaling
arguments, the bare value of $r$, which appears in Landau theory or in the LGW
functional, and any partially renormalized values. We will explicitly make this
distinction when doing so is essential, but will suppress it otherwise.} and
one for the field $h$ that is conjugate to the order parameter. Let $\mu =
(r,h)$ define the space spanned by these two parameters. Under RG iterations,
the system moves away from criticality according to
\bse
\label{eqs:1.4}
\be
\mu \to \mu(b) = \left(r\,b^{1/\nu},h\,b^{\,y_h}\right),
\label{eq:1.4a}
\ee
where $b>1$ is the RG rescaling factor. $\nu$ is the correlation length
exponent defined above, and $y_h$ is related to the susceptibility exponent
$\eta$ by
\be
y_h = (d+2-\eta)/2.
\label{eq:1.4b}
\ee
\ese
The exponents $1/\nu$ and $y_h$ illustrate the more general concept of scale
dimensions, which determine how parameters change under RG transformations. If
$p$ is some parameter with scale dimension $[p\,]$, then $p\,(b) =
p_{\,0}\,b^{\,[p\,]}$, with $p_{\,0} = p\,(b=1)$. Accordingly, $[r] = 1/\nu$,
and $[h] = y_h$. Parameter values $\mu^*$ with the property $\mu^*(b) = \mu^*$
constitute a RG fixed point. Parameters with positive, zero, and negative scale
dimensions with respect to a particular fixed point are called relevant,
marginal, and irrelevant, respectively. A critical fixed point is characterized
by the presence of two and only two relevant parameters, $r$ and $h$. The
critical behavior of all thermodynamic quantities follows from a generalized
homogeneity law satisfied by the free energy density $f=-(T/V)\ln Z$,
\be
f(r,h) = b^{-d}f(r\,b^{1/\nu},h\,b^{\,y_h}).
\label{eq:1.5}
\ee
The derivation of this relation was one of the major triumphs of the RG. In
conjunction with the Wilson-Fisher $\epsilon$-expansion
\cite{Wilson_Fisher_1972}, it allows for the computation of the critical
exponents in an asymptotic series about
$d_{\text{c}}^{\,+}$.\footnote{\label{fn:9} It is a popular misconception that
the RG is useful {\em only} for dealing with critical phenomena. In fact, the
technique is much more powerful and versatile, and can describe entire phases
as well as the transitions between them, see Sec.\ \ref{sec:II} below. Although
this was realized by the founding fathers of the RG (see,
\onlinecite{Anderson_1984} ch. 5, \onlinecite{Fisher_1998}), it has been
exploited only recently (e.g., \onlinecite{Shankar_1994}). It is interesting to
note that the systematic application of the Wilsonian RG to condensed matter
systems implements a program that in a well-defined way is the opposite of that
in high-energy physics. In condensed matter physics, the microscopic theory is
known for all practical purposes, viz., the many-body Schr{\"o}dinger equation.
By applying the RG one derives effective theories valid at lower and lower
energy scales. In high-energy physics, some effective theory valid at
relatively low energies is known (say, the standard model), and the goal is to
deduce a `more microscopic' theory valid at higher energies. }

In addition to the diverging length scale set by $\xi$, there is a diverging
time scale $\tau_{\xi}$. Its divergence is governed by the dynamical critical
exponent $z$, according to $\tau_{\xi}\propto\xi^z$, and it results in the
phenomenon of ``critical slowing down'' (\onlinecite{van_Hove_1954};
\onlinecite{Landau_Khalatnikov_1954}; for a translation of the latter paper,
see, \onlinecite{Landau_1965} p.626), that is, the very slow relaxation towards
equilibrium of systems near a critical point. The critical behavior of time
correlation functions $C(\bm{k},\Omega;r,h)$ is given by
\be
C(\bm{k},\Omega;r,h) =
b^{\,x_C}C(\bm{k}b,\Omega\,b^z;r\,b^{1/\nu},h\,b^{\,y_h}).
\label{eq:1.6}
\ee
Here $\Omega$ denotes the frequency, which derives from a Fourier transform of
the time dependence. $x_C$ is an exponent that is characteristic of the
correlation function $C$. This relation was first postulated as the ``dynamical
scaling hypothesis'' \cite{Ferrell_et_al_1967, Ferrell_et_al_1968,
Halperin_Hohenberg_1967}, and later also derived by dynamical RG techniques
\cite{Hohenberg_Halperin_1977}.

\subsection{Classical versus Quantum Phase Transitions}
\label{subsec:I.C}

In classical statistical mechanics, the dynamical critical exponent $z$ is
independent of the static critical behavior \cite{Ma_1976,
Hohenberg_Halperin_1977}. The reason is as follows. Classically, the canonical
partition function for a system consisting of $N$ particles with $f$ degrees of
freedom is given as an integral over phase space of the Boltzmann factor,
\bea
Z &=& \frac{1}{N!}\int dp\,dq\ e^{-\beta H(p,q)}
\nonumber\\
  &=& \frac{1}{N!}\int dp\ e^{-\beta H_{\text{kin}}(p)}\int dq\ e^{-\beta
       H_{\text{pot}}(q)}
\nonumber\\
       &=& {\text{const.}}\int dq\ e^{-\beta H_{\text{pot}}(q)}.
\label{eq:1.7}
\eea
Here $\beta=1/T$ is the inverse temperature, and $p$ and $q$ represent the
generalized momenta and coordinates, respectively, $H$ is the Hamiltonian, and
$H_{\text{kin}}$ and $H_{\text{pot}}$ are the kinetic and potential energy,
respectively.\footnote{\label{fn:10} We assume that there are no velocity
dependent potentials.} Due to the factorization of the phase-space integral
indicated in Eq.\ (\ref{eq:1.7}), one can integrate over the momenta and solve
for the thermodynamic critical behavior without reference to the dynamics.

In quantum statistical mechanics, the situation is different. The Hamiltonian
${\hat H}$, and its constituents ${\hat H_{\text{kin}}}$ and ${\hat
H_{\text{pot}}}$, are now operators, and ${\hat H_{\text{kin}}}$ and ${\hat
H_{\text{pot}}}$ do not commute. Consequently, the grand canonical partition
function
\be
Z = \Tr e^{-\beta({\hat H}-\mu{\hat N})}
  = \Tr e^{-\beta({\hat H_{\text{kin}}} + {\hat H_{\text{pot}}}-\mu{\hat N})},
\label{eq:1.8}
\ee
does not factorize, and one must solve for the dynamical critical behavior
together with the thermodynamics. This becomes even more obvious if one
rewrites the partition function as a functional integral
\cite{Casher_Lurie_Revzen_1968,Negele_Orland_1988}. We will consider fermionic
systems, in which case the latter is taken with respect to Grassmann-valued
(i.e., anti-commuting) fields ${\bar\psi}$ and $\psi$,\footnote{\label{fn:11}
For a thorough treatment of Grassmannian algebra and calculus, see,
\textcite{Berezin_1966}.}
\bse
\label{eqs:1.9}
\be Z = \int D[{\bar\psi},\psi]\ e^{S[{\bar\psi},\psi]}.
\label{eq:1.9a}
\ee
The action $S$ is determined by the Hamiltonian,
\bea
S[{\bar\psi},\psi]&=&\int_0^{\beta}d\tau \int d{\bm x}\sum_{\sigma}
    {\bar\psi}_{\sigma}({\bm x},\tau)\,
      \left(-\partial_{\tau} + \mu\right)\,\psi_{\sigma}({\bm x},\tau)
\nonumber\\
&&\hskip -20pt - \int_0^{\beta} d\tau\int d{\bm x}\
     H\left({\bar\psi}_{\sigma}({\bm x},\tau),\psi_{\sigma}({\bm x},\tau)\right),
\label{eq:1.9b}
\eea
\ese
Here ${\bm x}$ denotes the position in real space, $\sigma$ is the spin
index,\footnote{\label{fn:12} $\sigma$ may also comprise other quantum numbers,
e.g., a band index, depending on the model considered.} and the fields
${\bar\psi}_{\sigma}({\bm x},\tau)$ and $\psi_{\sigma}({\bm x},\tau)$ are, for
each value of $\tau$, in one-to-one correspondence with the creation and
annihilation operators of second quantization,
${\hat\psi}^{\dagger}_{\sigma}({\bm x})$ and ${\hat\psi}_{\sigma}({\bm x})$,
respectively. They obey antiperiodic boundary conditions, $\psi_{\sigma}({\bm
x},\tau=0) = -\psi_{\sigma}({\bm x},\tau=\beta)$. The function $H$ is defined
such that ${\hat H} = \int d{\bm x}\ H({\hat\psi}^{\dagger}_{\sigma}({\bm
x}),{\hat\psi}_{\sigma}({\bm x}))$. $\mu$ is the chemical potential, and ${\hat
N}$ in Eq.\ (\ref{eq:1.8}) is the particle number operator. A quantum
mechanical generalization of the LGW functional can be derived from Eq.\
(\ref{eq:1.9b}) by constraining appropriate linear combinations of products
${\bar\psi}\psi$ of fermion fields to a bosonic order parameter field $\phi$,
and integrating out all other degrees of freedom. Often one can also just write
down a LGW functional based on symmetries and other general considerations.

Due to the coupling of statics and dynamics, the scaling relation
(\ref{eq:1.5}) for the free energy must be generalized to (e.g.,
\onlinecite{Sachdev_1999})
\be
f(r,h,T) = b^{-(d+z)}f(r\,b^{1/\nu},h\,b^{\,y_h},T\,b^z).
\label{eq:1.10}
\ee
This relation reflects the fact that the temperature is necessarily a relevant
operator at a $T=0$ critical point, and that temperature and frequency are
expected to scale the same way.\footnote{\label{fn:13} As we will see later,
the latter expectation can be violated, due to, (1) the existence of multiple
temperature and/or frequency scales, and (2) the existence of dangerous
irrelevant variables \cite{Fisher_1983}.}

\begin{figure}[t,b]
\vskip 0 cm
\includegraphics[width=6cm,angle=-90]{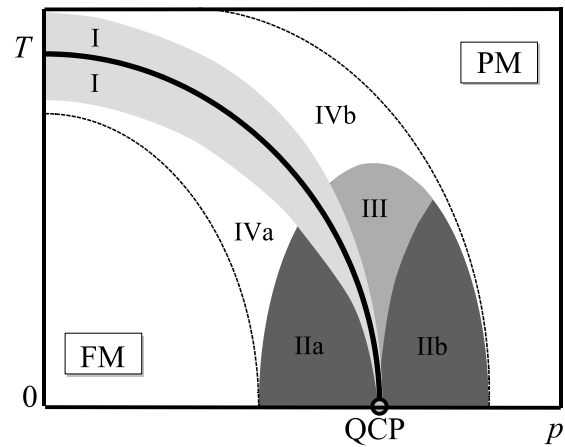}
\vskip 0.5 cm \caption{ \label{fig:2} Schematic phase diagram in the $T$-$p$
   plane, as in Fig.\ \ref{fig:1}. The solid line is the phase separation line,
   separating the ferromagnetic (FM) phase from the paramagnetic (PM) one,
   which ends in the quantum critical point (QCP). The critical region in the
   vicinity of the phase transition line is bounded by the dashed lines, which
   are defined as the loci of points where the correlation length is a certain
   multiple of the microscopic length scale. The critical region is separated
   into four regimes that show different
   scaling behaviors: Region I (light shaded) displays classical scaling governed
   by the classical fixed point. Regions IIa and IIb (dark shaded) display static
   or temperature-independent quantum scaling governed by the zero-temperature
   fixed point, and region III (medium shaded) displays dynamic or temperature
   quantum scaling. In regions IVa and IVb one has crossover scaling governed by
   both fixed points. The edges of the critical region, as well as the boundaries
   between the various scaling regimes, are not sharp. Notice that the dashed lines
   are not parallel to the phase separation line, since both the exponents $\nu$
   and the critical amplitudes associated with the classical and quantum
   fixed points, respectively, in general have different values.}
\end{figure}

Equation (\ref{eq:1.9b}) displays another remarkable property of quantum
statistical mechanics. The auxiliary variable $\tau$, usually referred to as
imaginary time, acts effectively as an extra dimension. For nonzero
temperature, $\beta = 1/T < \infty$, this extra dimension extends only over a
finite interval. If one is sufficiently close to criticality such that the
condition $\tau_{\xi} > 1/T_{\text{c}}$ is fulfilled, the extra dimension will
not affect the leading critical behavior. Rather, it will only lead to
corrections to scaling that are determined by finite-size scaling effects
\cite{Barber_1983,Cardy_1996}. We thus conclude that the asymptotic critical
behavior at any transition with a nonzero critical temperature is purely
classical. However, a transition at $T=0$ is described by a theory in an
effectively different dimension, and will therefore in general be in a
different universality class. This raises the question how continuity is
ensured as one moves to $T=0$ along the phase separation line. The resolution
is that, at low temperatures, the critical region, i.e. the region around the
phase separation line where critical behavior can be observed, is divided into
several regimes. Asymptotically close to the transition the critical behavior
is classical at any nonzero temperature,\footnote{\label{fn:14} This includes
the transitions at nonzero temperature in, say, superconductors or superfluids.
Although the occurrence of the transition in these cases, and indeed the very
existence of the order parameter, depend on quantum mechanics, the critical
properties are described by classical physics.} but since this asymptotic
classical regime is bounded by a crossover at $1/\tau_{\xi}\approx T$, it
shrinks to zero as $T\to 0$. In the vicinity of the quantum critical point,
quantum critical behavior is observed except in the immediate vicinity of the
phase boundary. This quantum critical regime is in turn divided into a region
characterized by $T\alt r^{\nu z}$, where one sees static quantum critical
behavior approximately independent of the temperature, and a region
characterized by $T\agt r^{\nu z}$, where one observes dynamic or temperature
scaling. The latter is often called ``quantum critical regime'' in the
literature. Finally, there is a regime inside the critical region, but outside
both the asymptotic classical and quantum scaling regimes. This is
characterized by crossover scaling governed by both the classical and quantum
fixed points. These various regimes are shown schematically in Fig.\
\ref{fig:2}.

Equation (\ref{eq:1.9b}) suggests that imaginary time or inverse temperature
always scale like a length, i.e., that the dynamical critical exponent $z=1$ at
any quantum phase transition. Indeed, early work generalizing the Wilsonian RG
to quantum systems either dealt with systems where $z=1$ \cite{Suzuki_1976}, or
assumed that $z=1$ universally \cite{Beal-Monod_1974}. This is misleading,
however. The point is that Eq.\ (\ref{eq:1.9b}) represents the bare microscopic
action. Under renormalization, or even when integrating out degrees of freedom
to derive a bare effective action, the relation between space and imaginary
time can change. As a result, $z$ can assume any positive value, as was
realized by \textcite{Hertz_1976}.\footnote{\label{fn:15} One might object that
the underlying microscopic theory must be Lorentz invariant, so $z=1$ after
all. Again, this argument ignores the fact that the RG derives an effective
low-energy theory valid at small wave numbers and frequencies. In the critical
theory, the relevant frequency, $1/\tau_{\xi}$, is zero. Another way to say
this is that the speed of light has been renormalized to $c=\infty$. The
critical theory is therefore {\em rigorously} nonrelativistic. As we have seen
above, at any nonzero temperature $\hbar$ renormalizes to zero, and the
critical theory is also rigorously non-quantum mechanical.} Hertz also noted
the following interesting consequence of the space-time nature of quantum
statistical mechanics, which holds independently of the exact value of $z$.
Since the quantum mechanical theory is effectively higher dimensional than the
corresponding classical one, fluctuation effects are weaker, and mean-field
theory will be more stable. Indeed, since Landau theory is valid classically
for $d > d_{\text{c}}^{\,+}$, quantum mechanically it should be valid for $d >
d_{\text{c}}^{\,+} - z$. Since $d_{\text{c}}^{\,+} = 4$ for many phase
transitions, and usually $z\geq 1$, it seems to follow that most quantum phase
transitions in the most interesting dimension, $d=3$, show mean-field critical
behavior. \textcite{Hertz_1976} demonstrated this conclusion by means of a
detailed LGW theory of the quantum phase transition in itinerant Heisenberg
ferromagnets, where $d_{\text{c}}^{\,+} = 4$ and $z=3$ in Hertz's theory.

\subsection{The Soft-Mode Paradigm}
\label{subsec:I.D}

The above chain of arguments implies that a quantum phase transition is closely
related to the corresponding classical phase transition in a higher dimension.
Since this higher dimension will usually be above the upper critical dimension,
the conclusion seems to be that quantum phase transitions generically show
mean-field critical behavior, and are thus uninteresting from a critical
phenomena point of view. In this review we discuss one of the reasons why this
is in general not correct. More generally, we will discuss how the entire
concept of a LGW theory in terms of a single order-parameter field can break
down for either classical or quantum transitions, and why this breakdown is
more common for the latter than for the former. The salient point is that a
correct description of a phase transition, i.e., of phenomena at long length
and time scales, must take into account {\em all} soft modes. If there are soft
modes in addition to the order-parameter fluctuations, and if the former couple
sufficiently strongly to the latter, then the behavior cannot be described in
terms of a local field theory for the order-parameter fluctuations only. Away
from the quantum phase transition, these extra soft modes, via interactions or
nonlinear couplings, lead to power-law correlations in various physical
correlation functions. This phenomenon is called generic scale invariance and
we will discuss it in Sec.\ \ref{sec:II}.

In a derivation of a LGW functional from a microscopic theory this manifests
itself as follows. As mentioned above, such a derivation involves the
integrating-out of all degrees of freedom other than the order-parameter field.
The parameters in the LGW functional are given in terms of integrals over the
correlation functions of these degrees of freedom. If there are soft modes
other than the order-parameter fluctuations, then these integrals may not
exist, which can lead to diverging coefficients in the LGW theory. At the level
of Landau or mean-field theory this manifests itself as a nonanalytic
dependence of the free energy on the mean-field order parameter. We will see
later, in Secs.\ \ref{subsubsec:III.A.3} and \ref{subsubsec:IV.A.3},
respectively, that both in liquid crystals and in quantum ferromagnets, extra
soft modes lead to an effective Landau free energy
\be
F = r\,m^2 + v\,m^4\ln m^2 + u\,m^4 + \ldots
\label{eq:1.11}
\ee
as opposed to Eq.\ (\ref{eq:1.1}). The situation gets worse if order parameter
fluctuations are taken into account. Writing down a LGW theory based on
symmetry considerations, or even deriving one in some crude approximation, will
in general lead to erroneous conclusions if additional soft modes are present.
A correct derivation, on the other hand, will lead to coefficients that are
singular functions of space and imaginary time, which renders the theory
useless. A more general approach, which keeps all of the soft modes on equal
footing, is thus called for. Getting in trouble by integrating out, explicitly
or implicitly, additional excitations is not unprecedented in physics. For
instance, the Fermi theory of weak interactions can be considered as resulting
from integrating out the $W$ gauge bosons in the standard model. This results
in a theory that is not renormalizable (see, e.g., \onlinecite{LeBellac_1991}).

Before we enter into details, let us elaborate somewhat on the general aspects
of this breakdown of LGW theory. Imagine a transition from a phase with an
already-broken continuous symmetry to one with an additional broken symmetry.
Then the Goldstone modes of the former will in general influence the
transition. A good classical example for this mechanism is the
nematic--smectic-A transition in liquid crystals \cite{DeGennes_Prost_1993}
which we will discuss in Sec.\ \ref{subsec:III.A}. However, for the most
obvious classical phase transitions, e.g., for the liquid-gas critical point,
the effects of generic soft modes affect only the dynamics, i.e., the critical
behavior of transport coefficients \cite{Hohenberg_Halperin_1977}. Quantum
mechanically, the concept is more important, for two reasons: (1) There are
more soft modes at $T=0$ than at $T>0$. An example that will be important for
our discussion are the particle-hole excitations in an electron fluid that are
soft at $T=0$ and acquire a mass proportional to $T$ at $T>0$. They couple
strongly to the magnetization, with consequences for the quantum ferromagnetic
transitions that will be discussed in Sec.\ \ref{sec:IV} below. They also
provide a good example of the phenomenon of generic scale invariance, which we
will discuss in detail in Sec.\ \ref{sec:II}. (2) Quantum mechanically, the
statics and the dynamics are coupled, as we have discussed above, and therefore
effects that classically would affect the dynamics only influence the static
critical behavior as well. As a result of these two points, the original
concept of mapping a quantum phase transition onto the corresponding classical
transition in a higher dimensionality turns out to be mistaken. Rather, the
mapping is in general onto a classical transition with additional soft modes.

In systems with generic scale invariance the correlation length, which diverges
at the critical point, must be defined differently from Eq.\ (\ref{eq:1.2b}).
The reason is that the order parameter correlations decay as power laws both at
and away from the critical point, albeit with different powers, the decay being
slower at criticality. In this case, $\xi$ is the length scale that separates
the generic power-law correlations from the critical ones. Specifically, in the
scaling region, and for $\vert{\bm x}\vert < \xi$, the correlations obey the
critical power law, while for $\vert{\bm x}\vert > \xi$ the correlations show
the generic power law.

In addition to this mechanism, there are other reasons why quantum phase
transitions can be more complicated than one might naively expect. In
low-dimensional systems, topological order can lead to very nontrivial effects.
One example is the intricate behavior of spin chains \cite{Haldane_1982,
Haldane_1983}, another, the recent proposals of exotic quantum critical
behavior in two-dimensional quantum antiferromagnets \cite{Senthil_et_al_2003a,
Senthil_et_al_2003b}. Also, the quantum phase transition may not have any
classical counterpart, and hence no classical upper critical dimension.
Examples include, metal-insulator transitions \cite{Mott_1990,
Belitz_Kirkpatrick_1994, Kramer_MacKinnon_1993}, and transitions between
quantum Hall states \cite{Sondhi_et_al_1997}. These phenomena are outside the
scope of the present review.

Finally, we mention that there are also examples of quantum phase transitions
where none of the above complications arise, and Hertz theory \cite{Hertz_1976}
is valid. This class of transitions include, quantum metamagnetic transitions
(\onlinecite{Millis_et_al_2002}, see Fig.\ \ref{fig:15} for an example), and
the liquid-crystal--like transitions in high-$T_{\text{c}}$ superconductors and
quantum Hall systems \cite{Emery_Kivelson_Tranquada_1999, Lilly_et_al_1999,
Du_et_al_1999, Oganesyan_Kivelson_Fradkin_2001}. One would also expect it to
include clean antiferromagnetic systems without local moments, although no
convincing experimental examples have been found so far. A general
classification of which quantum phase transition should be affected by generic
soft modes has been attempted by \textcite{Belitz_Kirkpatrick_Vojta_2002}.
Although they are very interesting, we will not further consider these `simple'
quantum phase transitions in this review.

\section{Generic Scale Invariance}
\label{sec:II}

Correlation functions characterize how fluctuations at one space-time point are
correlated with fluctuations at another point (see, e.g.,
\onlinecite{Forster_1975}). The default expectation is that they decay
exponentially for large distances in space or time; fluctuations that are far
apart we expect to be weakly correlated. If they do, then there is a
characteristic length or time scale associated with the decay, and the
correlations are said to be of short range. It is well known that at special
points in the phase diagram, at critical points in particular, certain
correlations may decay only as power laws in space or/and time (e.g.,
\onlinecite{Ma_1976}). In this case, the correlations are called long-ranged
(in space) or long-lived (in time); we will call them long-ranged irrespective
of whether the reference is to space or to time. Such correlations functions
exhibit scale invariance: A scale change in space and/or time can be
compensated by multiplying the correlation function by a simple scale factor.
This property is conveniently expressed in terms of generalized homogeneous
functions (e.g., \onlinecite{Chaikin_Lubensky_1995}), see, e.g., Eqs.\
(\ref{eq:1.5}, \ref{eq:1.6}). In addition, systems can exhibit scale invariance
in whole regions of the phase diagram, rather than only at special points. In
this case one speaks of ``generic scale invariance''
(GSI)\cite{Dorfman_Kirkpatrick_Sengers_1994, Law_Nieuwoudt_1989, Nagel_1992}.
In this section we discuss various mechanisms for GSI in both classical and
quantum systems, highlighting their similarities and differences. Before we go
into details, we list four mechanisms that can lead to this phenomenon.
\begin{description}
\item[\rm{(1)}] Spontaneous breaking of a global continuous symmetry leads to
Goldstone modes that are massless everywhere in the broken-symmetry phase
(e.g., \onlinecite{Forster_1975}; \onlinecite{Zinn-Justin_1996}). This
mechanism is operative in both classical and quantum systems.
\item[\rm{(2)}] Gauge symmetries lead to soft modes since a mass would be
incompatible with gauge invariance \cite{Weinberg_I_1996, Weinberg_II_1996,
Ryder_1985, Zinn-Justin_1996}. The only case we will be interested in is the
U(1) gauge symmetry that underlies the masslessness of the photon.
\end{description}
Both of these mechanisms, if operative, lead to soft modes that directly result
in obvious ways from the symmetry in question, and hence to GSI that we will
refer to as `direct' GSI effects. What is usually less obvious is that these
soft modes, via mode-mode coupling effects, can lead to other modes becoming
soft as well. We will see various examples for these `indirect' GSI effects.
Another source of direct GSI are
\begin{description}
\item[\rm{(3)}] conservation laws, which lead to power-law temporal decay of
local time correlation functions \cite{Forster_1975}. There are two ways in
which conservation laws can lead to indirect GSI, namely:
\begin{description}
\item[\rm{(3a)}] The conservation laws can appear in conjunction with mode-mode
coupling effects, or nonlinearities in the equations of motion
\cite{Pomeau_Resibois_1975, Boon_Yip_1991}. In classical systems, this leads to
long-ranged time correlation functions; in quantum systems, it leads to
long-ranged time correlation functions {\em and} thermodynamic quantities.
\item[\rm{(3b)}] Conservation laws in conjunction with a nonequilibrium situation
\cite{Schmittmann_Zia_1995, Kirkpatrick_Cohen_Dorfman_1982a,
Kirkpatrick_Cohen_Dorfman_1982b} can leads to long-ranged time correlation
functions and thermodynamic quantities even in classical systems.
\end{description}
\end{description}

In what follows we discuss specific examples, starting with classical systems.

\subsection{Classical Systems}
\label{subsec:II.A}

We now illustrate the four mechanisms listed above by means of four examples
that show how they lead to long-ranged correlations in classical systems. For
each case we give a general discussion, followed by the specification of a
suitable model, and a demonstration of how explicit calculations and, where
applicable, RG arguments, yield GSI.

\subsubsection{Goldstone modes}
\label{subsubsec:II.A.1}

To illustrate the Goldstone mechanism for long-ranged spatial correlations in
equilibrium, we choose as an example a generalized Heisenberg ferromagnet with
an $N$-dimensional order parameter. $N=3$ represents the usual Heisenberg
magnet, which will be relevant later in this review. Let ${\bm \phi}({\bm x})$
be the fluctuating magnetization, with ${\bm m} = \langle{\bm\phi}({\bm
x})\rangle$ its average value, and let ${\bm h}$ be an external magnetic field
conjugate to the order parameter. Goldstone's theorem \cite{Goldstone_1961,
Goldstone_Salam_Weinberg_1962} states that whenever there is a spontaneously
broken global continuous symmetry, there will be massless modes. (This is not
true for local, or gauge, symmetries, see Sec.\ \ref{subsubsec:II.A.4} below.)
To understand this concept, suppose that the action is invariant under
rotations of the vector field ${\bm\phi}$ provided that ${\bm h}=0$. This
implies that the order parameter in zero field vanishes, ${\bm m}({\bm
h}=0)=0$, due to rotational invariance. However, in the limit ${\bm h}\to 0$
there are two possible behaviors, depending on the temperature, which
determines whether the system is in the disordered phase, or in the ordered
phase. Namely,
\bse
\label{eqs:2.1}
\bea
{\bm m} ({\bm h}\to 0) &=& 0 \hspace{0.4in}(\text{disordered phase}),
\label{eq:2.1a}\\
{\bm m} ({\bm h}\to 0) &\neq& 0 \hspace{0.4in}(\text{ordered phase}).
\label{eq:2.1b}
\eea
\ese
The situation in the ordered phase is sometimes also characterized by saying
that the action obeys the symmetry while the state does not, and it is referred
to as a spontaneously broken continuous symmetry.\footnote{\label{fn:16} In
this case, the continuous symmetry is characterized by the group $O(N)$, but
the concept is valid for arbitrary Lie groups. A useful reference for Lie
groups, or continuous groups, is \textcite{Gilmore_1974}.} In the ordered
phase, one still has invariance with respect to rotations that leave the vector
${\bm m}$ fixed, i.e., with respect to the subgroup $O(N-1)$ that is the little
group of ${\bm m}$. Goldstone's theorem says that this results in as many soft
modes as the quotient space $O(N)/O(N-1)$ has dimensions, i.e., there are
$\text{dim}\,\left(O(N)/O(N-1)\right) = N-1$ Goldstone modes. These soft modes
are `perpendicular' or `transverse' to the direction ${\bm m}$ of the
spontaneous ordering. For example, if ${\bm\phi} = (\phi_1, \ldots,\phi_N)$ and
there is spontaneous ordering in the $N$-direction, then one has, in the limit
${\bm h}\to 0$ and for asymptotically small wave numbers,
\bse
\label{eqs:2.2}
\be
\langle\phi_i({\bm k})\,\phi_j(-{\bm k})\rangle = \delta_{ij}/\zeta\,
                   {\bm k}^2\quad,\quad (i,j = 1,\ldots,N-1).
\label{eq:2.2a}
\ee
$\zeta$ is called the stiffness coefficient of the Goldstone modes. In real
space in, say, three-dimensions, this implies that for large distances,
$\vert\bm{x}_1 - \bm{x}_2\vert\to\infty$,
\be
\langle\phi_i({\bm x}_1)\,\phi_j({\bm x}_2)\rangle \propto 1/\vert\bm{x}_1 -
\bm{x}_2\vert.
\label{eq:2.2b}
\ee
\ese
We see that Goldstone's theorem gives rise to power-law correlations, or GSI,
in the entire ordered phase. In the terminology explained above, this is an
example of direct GSI.

\paragraph{Nonlinear $\sigma$ model}
\label{par:II.A.1.a}

To illustrate the Goldstone mechanism more explicitly, and for later reference,
let us consider the derivation of a nonlinear $\sigma$ model
\cite{Gell-Mann_Levy_1960, Brezin_Zinn-Justin_1976, Polyakov_1975,
Nelson_Pelcovits_1977, Zinn-Justin_1996} for our $O(N)$-symmetric Heisenberg
ferromagnet. We specify the action by
\bea
S[{\bm\phi}] &=& \int d{\bm x}\ \bigl[r\,{\bm\phi}^2({\bm x})
     + c\,\left({\bm\nabla}{\bm\phi}({\bm x})\right)^2
     + u\,{\bm\phi}^4({\bm x})
\nonumber\\
&& \hskip 30pt - h\,\phi_N({\bm x})\bigr].
\label{eq:2.3}
\eea
Here ${\bm\phi}^2 = \phi_i\phi^i$, and $({\bm\nabla}{\bm\phi})^2 =
\partial_{\alpha}\phi_i\partial^{\alpha}\phi^i$, with $i=1,\ldots N$,
and $\alpha=1,\ldots d$ in $d$-dimensions. Summation over repeated indices is
implied, and we have assumed an external field in $N$-direction. $S$ determines
the partition function via Eq.\ (\ref{eq:1.3a}). For $h=0$, it is obviously
invariant under $O(N)$-rotations of the vector field $\bm{\phi}$. In the
low-temperature phase, where the $O(N)$-symmetry is spontaneously
broken,\footnote{\label{fn:17} Provided that $d>2$; no spontaneous symmetry
breaking occurs in $d\leq 2$ \cite{Mermin_Wagner_1966}.} it is convenient to
decompose $\bm{\phi}$ into its modulus $\rho$ and a unit vector field
$\hat{\bm{\phi}}$,
\be
\bm{\phi}(\bm{x}) = \rho(\bm{x})\,{\hat{\bm{\phi}}}(\bm{x})\quad,\quad
            {\hat{\bm{\phi}}}^2(\bm{x}) = 1.
\label{eq:2.4}
\ee
$\hat{\bm{\phi}}$ parameterizes the $(N-1)$-sphere which is isomorphic to the
coset space $O(N)/O(N-1)$. In terms of $\rho$ and $\hat{\bm{\phi}}$, the action
is\footnote{\label{fn:18} The change of variables from ${\bm\phi}$ to
$(\rho,{\hat\phi})$ also changes the integration measure in Eq.\
(\ref{eq:1.3a}). If one eliminates $\sigma$ in terms of $\pi$, Eqs.\
(\ref{eqs:2.6}), the measure changes again, see \textcite{Zinn-Justin_1996}.}
\bea
S[\rho,\hat{\bm{\phi}}] &=& c^{(1)}\int d\bm{x}\ \rho^2(\bm{x})\,
   [\bm{\nabla}\hat{\bm{\phi}}(\bm{x})]^{2}
\nonumber\\
&&\hskip -0pt + \int d{\bm x}\ \left[r\rho^2(\bm{x})
               + c^{(2)}[\bm{\nabla}\rho(\bm{x})]^2
               + u\rho^4(\bm{x})\right]
\nonumber\\
&&\hskip -0pt -h\int d{\bm x}\ \rho({\bm x})\,{\hat\phi}_N({\bm x}).
\label{eq:2.5}
\eea
The bare values of $c^{(1)}$ and $c^{(2)}$ are equal to $c$. Notice that the
field $\hat{\bm{\phi}}$ appears only in conjunction with two gradient
operators. This implies that the $\hat{\bm{\phi}}$-fluctuations represent the
$N-1$ Goldstone modes of the problem, while the $\rho$-fluctuations represent
the massive mode. Assuming the system is in the ordered phase, and taking the
order to be in the $N$-direction, we parameterize $\hat{\bm{\phi}}$ as follows,
\bse
\label{eqs:2.6}
\be
\hat{\bm{\phi}}(\bm{x}) = \left(\bm{\pi}(\bm{x}),\sigma(\bm{x})\right),
\label{eq:2.6a}
\ee
where the vector $\bm{\pi}$ represents the $N-1$ transverse directions, and
\be
\sigma(\bm{x}) = \left[1 - \bm{\pi}^2(\bm{x})\right]^{1/2}.
\label{eq:2.6b}
\ee
\ese
Next we split off the expectation value of the massive field by writing
$\rho({\bm x}) = M + \delta\rho({\bm x})$, with $M = \langle\rho({\bm
x})\rangle$. Absorbing appropriate powers of $M$ into the coupling constants,
we can write the action
\bse
\label{eqs:2.7}
\be
S[\rho,\bm{\pi}] = S_{\text{NL}\sigma\text{M}}[\bm{\pi}]
                   + \delta S[\rho,\bm{\pi}].
\label{eq:2.7a}
\ee
Here
\bea
S_{\text{NL}\sigma\text{M}}[\bm{\pi}] &=& \frac{\zeta}{2}\int d\bm{x}\
\left[({\bm\nabla\pi}({\bm x}))^2 + (\bm{\nabla}\sigma(\bm{x}))^2\right]
\nonumber\\ && - h\int d{\bm x}\ \sigma({\bm x}),
\label{eq:2.7b}
\eea
is the action of the $O(N)$-nonlinear $\sigma$ model, with the bare value of
$\zeta$ equal to $2cM^2$. $\delta S$ is the part of the action that contains
the $\delta\rho\,$-fluctuations, as well as the coupling between $\delta\rho$
and $\hat{\phi}$,
\be
\delta S = r\int d{\bm x}\ (\delta\rho({\bm x}))^2 + O\left(\delta\rho^3,
({\bm\nabla}\delta\rho)^2, \sigma\delta\rho,
\delta\rho({\bm\nabla}{\hat{\bm\phi}})^2\right).
\label{eq:2.7c}
\ee
\ese
This parameterization of the model accomplishes an explicit separation into
modes that are soft in the broken-symmetry phase (i.e., the ${\bm\pi}$-fields),
modes that are massive (i.e., the $\delta\rho\,$-field), and couplings between
the two.

\paragraph{GSI from explicit calculations}
\label{par:II.A.1.b}

At the Gaussian level, the action reads
\bea
S_{\text{G}} &=& \frac{\zeta}{2}\int d{\bm x}\ ({\bm\nabla\pi}({\bm x}))^2
                + \frac{h}{2}\int d{\bm x}\ ({\bm\pi}({\bm x}))^2
\nonumber\\
&& \hskip -30pt   +\ r\int d{\bm x}\ (\delta\rho({\bm x}))^2
                  + c^{(2)}\int d{\bm x}\ ({\bm\nabla}\delta\rho({\bm x}))^2.
\label{eq:2.8}
\eea
Here we see explicitly that the Gaussian ${\bm\pi}$-propagators are soft as
$h\to 0$,
\be
\langle\pi_i(\bm{k})\pi_j(-\bm{k})\rangle = \delta_{ij}/(\zeta{\bm k}^2 + h).
\label{eq:2.9}
\ee
That is, they have the form given by Eq.\ (\ref{eq:2.2a}), while the
$\delta\rho\,$-propagator is massive. Explicit perturbative calculations
\cite{Nelson_Pelcovits_1977} show that this does not change within the
framework of a loop expansion. It is also interesting to consider the explicit
perturbation theory for the magnetization $m = M\langle\sigma({\bm x})\rangle$.
To one-loop order, Eq.\ (\ref{eq:2.6b}) yields
\bea
m/M &=& 1 - \frac{1}{2}\int_{\bm p}\ \langle\pi_i({\bm p})\pi_j(-\bm{p})\rangle
\nonumber\\
    &=& m(h=0)/M + \text{const.}\times h^{(d-2)/2}.
\label{eq:2.10}
\eea
Here $\int_{\bm p} = \int d{\bm p}/(2\pi)^d$, and we show only the leading
nonanalytic dependence of $m$ on $h$.

\paragraph{Generic scale invariance from RG arguments}
\label{par:II.A.1.c}

The next question is whether these perturbative results are generic, and valid
independent of perturbation theory. Within a perturbative RG
\cite{Brezin_Zinn-Justin_1976, Nelson_Pelcovits_1977} this can be checked order
by order in a loop expansion. The fact that Eqs.\ (\ref{eq:2.9}, \ref{eq:2.10})
are valid to all orders in perturbation theory, and are indeed exact properties
independent of any perturbative scheme, hinges on the proof of the
renormalizability of the nonlinear $\sigma$ model
\cite{Brezin_Zinn-Justin_LeGuillou_1976}.

Here we present a much simpler, if less complete, argument based on power
counting \cite{Belitz_Kirkpatrick_1997}. We employ the concept of
\textcite{Ma_1976}, whereby one postulates a fixed point, and then
self-consistently checks its stability. Accordingly, we assign a scale
dimension $[L]=-1$ to lengths $L$, and look for a fixed point where the fields
have scale dimensions
\bse
\label{eqs:2.11}
\bea
\left[\pi_i({\bm x})\right] &=& (d-2)/2,
\label{eq:2.11a}\\
\left[\delta\rho({\bm x})\right] &=& d/2,
\label{eq:2.11b}
\eea
\ese
in $d$-dimensions. This {\it ansatz} is motivated by the expectation that the
Gaussian approximation for the $\pi$-correlation, Eq.\ (\ref{eq:2.9}), is
indeed exact, and that $\delta\rho$ is massive. Power counting shows that
$\zeta$ and $r$ are marginal, while all other coupling constants are irrelevant
except for $h$, which is relevant with $[h]=2$. In particular, all terms in
$\delta S$ that do not depend on $h$ are irrelevant with respect to the
putative fixed point, and so is the term $({\bm\nabla}\sigma)^2$ in
$S_{\text{NL}\sigma{\text{M}}}$. The fixed point is thus stable and describes
the ordered phase. The fixed-point action plus the most relevant external-field
term is Gaussian,
\bea
S_{\text{FP}} &=& \frac{\zeta}{2}\int d{\bm x}\ ({\bm\nabla\pi}({\bm x}))^2
                + \frac{h}{2}\int d{\bm x}\ ({\bm\pi}({\bm x}))^2
\nonumber\\
&&                +\ r\int d{\bm x}\ (\delta\rho({\bm x}))^2.
\label{eq:2.12}
\eea
We see that the $\pi$-$\pi$ correlation function for asymptotically small wave
numbers is indeed given by Eq.\ (\ref{eq:2.9}), i.e., one has soft Goldstone
modes, or GSI, everywhere in the ordered phase. Notice that the above RG
arguments {\em prove} this statement, albeit the proof is not rigorous in a
mathematical sense. Furthermore, the fixed point value of the magnetization, $m
= M\langle\sigma({\bm x})\rangle$, is given by $M$, and the leading correction
is given by the $\pi$-$\pi$ correlation function.\footnote{\label{fn:19} A
detailed analysis shows that $\delta S$ does not contribute to the leading
corrections to scaling at the stable fixed point.} The scale dimensions of
${\bm\pi}$ and $h$ then yield
\bse
\label{eqs:2.13}
\be
m(h) = m(h=0) + \text{const.}\times h^{(d-2)/2},
\label{eq:2.13a}
\ee
in agreement with the perturbative result, Eq.\ (\ref{eq:2.10}). This in turn
implies that the longitudinal susceptibility
\be
\chi_{\text{L}} = \partial m/\partial h \propto h^{(d-4)/2}
\label{eq:2.13b}
\ee
diverges in the limit $h\to 0$ for all $d<4$ \cite{Brezin_Wallace_1973}.
Alternatively, the zero-field susceptibility diverges in the homogeneous limit
as
\be
\chi_{\text{L}}({\bm k}\to 0) \propto \vert{\bm k}\vert^{-(4-d)}.
\label{eq:2.13c}
\ee
\ese
In real space this corresponds to a decay as $1/\vert{\bm x}\vert^{2(d-2)}$.

The above arguments assume that no structurally new terms are generated under
renormalization which might turn out to be marginal or relevant; this is one of
the reasons why the proof is not rigorous. With this caveat, they show that
several properties of isotropic Heisenberg ferromagnets that are readily
obtained by means of perturbation theory are indeed exact. They also illustrate
how much more information can be extracted from simple power counting after
performing a symmetry analysis, and separating the soft and massive modes, as
opposed to doing power counting on the original action, Eq.\ (\ref{eq:2.4})
\cite{Ma_1976}. More importantly for our present purposes, they illustrate how
soft modes in the presence of nonlinearities lead to slow decay of {\em
generic} correlation functions. In the present case, the transverse Goldstone
modes couple to the longitudinal fluctuations, which leads to the long-range
correlations expressed in Eq.\ (\ref{eq:2.13c}). This is an example of indirect
GSI due to mode-mode coupling effects. In Sec.\ \ref{subsubsec:II.A.2} we will
see that a very similar mechanism leads to long-ranged time correlation
functions in classical fluids.

\subsubsection{$U(1)$ gauge symmetry}
\label{subsubsec:II.A.4}

The second mechanism on our list is GSI caused by a gauge symmetry. Let us
consider $\phi^4$-theory again, Eq.\ (\ref{eq:2.3}), with $N=2$ or,
equivalently, with a complex scalar field $\phi$. Let us further demand that
the theory is invariant under {\em local} $U(1)$ gauge transformations,
\bse
\label{eqs:2.14}
\be
\phi({\bm x}) \to \phi'({\bm x}) = e^{i\Lambda({\bm x})}\,\phi({\bm x}),
\label{eq:2.14a}
\ee
with an arbitrary real field $\Lambda$. It is well known (e.g.,
\onlinecite{Ryder_1985}) that this demand forces the introduction of a gauge
field ${\bm A}$ with components $A_{\alpha}$ ($\alpha=1,2,3$ in
three-dimensions) that transforms as
\be
{\bm A}({\bm x}) \to {\bm A}'({\bm x}) = {\bm A}({\bm x}) + \frac{1}{q}\,
   {\bm\nabla}\Lambda({\bm x}),
\label{eq:2.14b}
\ee
\ese
and a new action
\bse
\label{eqs:2.15}
\bea
S[\phi,{\bm A}] &=& \int d{\bm x}\ \Bigl[r\,\vert\phi({\bm x})\vert^2
   + c\,\vert[{\bm\nabla} - iq{\bm A}({\bm x})]\,\phi({\bm x})\vert^2
\nonumber\\
&&  \hskip -20pt +\,u\,\vert\phi({\bm x})\vert^4
          + \frac{1}{16\pi\mu}\,F_{\alpha\beta}({\bm x})\,F^{\alpha\beta}({\bm x})
          \Bigr],
\label{eq:2.15a}
\eea
where
\be
F_{\alpha\beta}({\bm x}) = \partial_{\alpha}\,A_{\beta}({\bm x})
   - \partial_{\beta}\,A_{\alpha}({\bm x}).
\label{eq:2.15b}
\ee
\ese
Here $\mu$ and $q$ are coupling constants that characterize the gauge field
${\bm A}$ and its coupling to $\phi$, respectively. The usual interpretation of
this action is that of a charged particle, or excitation, with charge $q$,
described by $\phi$, that couples to electromagnetic fluctuations, or photons,
described by the vector potential ${\bm A}$. We will see in Sec.\
\ref{subsubsec:III.A.1}, however, that it can describe other systems as well,
at least in certain limits.

\paragraph{Symmetric phase}
\label{par:II.A.4.a}

Let us first discuss the symmetric phase, where both fields have zero
expectation values. In this phase, we expect the Gaussian $\bm{A}$-propagator
to be massless, as the ${\bm A}$-field appears only in conjunction with
derivatives. This is a consequence of the local gauge invariance; a term of the
form $m^2{\bm A}^2({\bm x})$ would violate the latter. However, the action $S$,
Eq.\ (\ref{eq:2.15a}), comes with the usual problems related to gauge fields.
That is, the ${\bm A}$-propagator does not exist since the ${\bm A}$-vertex has
a zero eigenvalue. We deal with this problem by ``gauge fixing'', i.e., we work
in Coulomb gauge, ${\bm\nabla}~\cdot~{\bm A}({\bm x}) = 0$, which we enforce by
adding to the action a gauge fixing term (e.g., \onlinecite{Ryder_1985}, ch.
7.1)
\be
S_{\text{GF}} = \frac{1}{\eta}\int d{\bm x}\ \left({\bm\nabla}\cdot{\bm A}({\bm
x})\right)^2,
\label{eq:2.16}
\ee
with $\eta\to 0$. One finds from Eqs.\ (\ref{eqs:2.15}, \ref{eq:2.16})
\bse
\label{eqs:2.17}
\be
\left\langle A_{\alpha}(\bm{k})\,A_{\beta}(-\bm{k})\right\rangle =
   4\pi\mu\,\frac{\delta_{\alpha\beta} - {\hat k}_{\alpha}{\hat k}_{\beta}}
                 {\bm{k}^2},
\label{eq:2.17a}
\ee
so the photon is indeed soft. Local gauge invariance thus leads to GSI in an
obvious way; this is another example of direct GSI. Notice that this mechanism
is distinct from the one related to conservation laws to be discussed in Sec.\
\ref{subsubsec:II.A.2} below: While local gauge invariance is sufficient for
the conservation of the charge $q$, it is not necessary; a {\em global} $U(1)$
suffices to make $q$ a conserved quantity.

The $\phi$-field, on the other hand, has two massive components with equal
masses. Writing $\phi = (\phi_1 + i\phi_2)/\sqrt{2}$, with $\phi_1$ and
$\phi_2$ real, we have
\be
\langle\phi_i({\bm k})\,\phi_j(-{\bm k})\rangle = \delta_{ij}/(r + c\,{\bm
k}^2)\quad,\quad (i,j=1,2).
\label{eq:2.17b}
\ee
\ese

We thus have two massive scalar fields, and one massless vector field with two
degrees of freedom.\footnote{\label{fn:20} The vector field, or photon, has
only two degrees of freedom, rather than three. In the gauge we have chosen
this is obvious, since the propagator, Eq.\ (\ref{eq:2.17a}), is purely
transverse.}

\paragraph{The broken-symmetry phase}
\label{par:II.A.4.b}

We now turn to the phase where the local $U(1)$-symmetry is spontaneously
broken. Suppose the spontaneous expectation value of $\phi$ is in the
$\phi_1$-direction, so we have
\begin{equation*}
\phi({\bm x}) = v + \left(\phi_1({\bm x}) + i\phi_2({\bm x})\right)/\sqrt{2},
\end{equation*}
with $v$ real and equal to $\sqrt{-r/2u}$ at tree level. If we expand about the
saddle-point solution $\phi=v$, ${\bm A}=0$, we find to Gaussian order
\bea
S &=& \int d{\bm x}\ \biggl[ 2u\,v^2(\phi_1({\bm x}))^2 +
\frac{c}{2}\,({\bm\nabla}\phi_1({\bm x}))^2 +
\nonumber\\
&& \hskip -20pt +\, c\,q^2v^2({\bm A}({\bm x}))^2 +
\frac{1}{16\pi\mu}\,F_{\alpha\beta}({\bm x})\,F^{\alpha\beta}({\bm x})
\nonumber\\
&& \hskip -20pt + \frac{c}{2}\,({\bm\nabla}\phi_2({\bm x}))^2 -
\sqrt{2}c\,q\,v{\bm A}({\bm x})\cdot{\bm\nabla}\phi_2({\bm x})
 \biggr].
\label{eq:2.18}
\eea

There are two interesting aspects of this action. First, the vector field has
acquired a mass that is proportional to $v^2$. Second, the field $\phi_2$,
which is often called the Higgs field in this context, and which we would
expect to form the Goldstone mode associated with the spontaneously broken
$U(1)$ or $O(2)$-symmetry, couples to the now-massive vector field. Indeed,
$\phi_2$ can be eliminated from the Gaussian action by shifting ${\bm A}$,
${\bm A}({\bm x})\to {\bm A}({\bm x})-{\bm\nabla}\phi_2({\bm x})/\sqrt{2}qv$.
Notice that this shift just amounts to a change of gauge, and hence does not
change the physical nature of the vector field.\footnote{\label{fn:21} By
writing $\phi$ in terms of an amplitude and a phase, one can choose a gauge
such that the entire action, not just the Gaussian part, is independent of
$\phi_2$. This is known as the ``physical'' or ``unitary'' gauge (e.g.,
\onlinecite{Ryder_1985} ch. 8.3). Other choices, which retain $\phi_2$ and lead
to a different ${\bm A}$-propagator, are possible. It has been shown that all
these formulations are indeed physically equivalent; the contributions from any
nonzero $\phi_2$-propagator cancel against pieces of the ${\bm A}$-propagator
and thus do not contribute to any observable properties. See,
\textcite{Zinn-Justin_1996,Weinberg_II_1996,Ryder_1985} for detailed
discussions of this point.} It does, however, give ${\bm A}$ a longitudinal
component and thus increases the number of photon degrees of freedom from two
to three. With this shift, the Gaussian action consists of only the first four
terms in Eq.\ (\ref{eq:2.18}), which leads to the following propagators,
\bse
\label{eqs:2.19}
\bea
\langle A_{\alpha}({\bm k})\,A_{\beta}(-{\bm k})\rangle &=& 4\pi\mu\,
\frac{\delta_{\alpha\beta} + k_{\alpha}k_{\beta}/m^2}{m^2 + {\bm k}^2},
\label{eq:2.19a}\\
\langle\phi_1({\bm k})\,\phi_1(-{\bm k})\rangle &=& \frac{1}{4u\,v^2 + c\,{\bm
k}^2},
\label{eq:2.19b}
\eea
\ese
with $m^2 = 4\pi\mu cq^2v^2$.

We see that now there are two massive fields, namely, a massive scalar field
with one degree of freedom, and a massive vector field with three degrees of
freedom. In particular, there is {\em no} Goldstone mode, despite the
spontaneously broken $O(2)$-symmetry. Pictorially speaking, the gauge field has
eaten the Goldstone mode and has become massive in the process. This phenomenon
is commonly referred to as the Higgs mechanism \cite{Anderson_1963,
Higgs_1964a, Higgs_1964b}. Notice that the situation is complementary, in a
well-defined sense, to the case without a gauge field in Sec.\
\ref{subsubsec:II.A.1}: Without local gauge invariance, one has two massive
modes in the disordered phase, and one massive mode and one soft Goldstone mode
in the ordered phase. In the gauge theory, there is a massless mode (the
photon) in the disordered phase, and no massless modes in the ordered phase.
While there are no obvious examples of indirect GSI involving photons, we will
see in Sec.\ \ref{subsubsec:III.A.1} that the nature of the mass acquired by
$\phi_1$ in the ordered phase can have a drastic influence on the nature of the
phase transition.

We close this subsection with one additional remark. In particle physics, local
gauge invariance is an indispensable requirement, since it is necessary for
Lorentz invariance. In the theory of phase transitions there is no such
requirement, since the critical theory is rigorously nonrelativistic, see
footnote \ref{fn:15}. Nevertheless, it is important if an order parameter field
couples to photons, and also in some other cases. We will elaborate on this in
Sec.\ \ref{sec:III}.

\subsubsection{Long-time tails}
\label{subsubsec:II.A.2}

We now turn to the third mechanism on our list; namely, long-ranged
correlations in time correlation functions due to conservation laws. Let us
consider time correlation functions involving the local currents of conserved
quantities, e.g., mass, momentum, or energy. Time integrals over the spatial
averages of these correlation functions, referred to as Green-Kubo expressions
\cite{Green_1954, Kubo_1957, Kubo_1959}, determine the transport coefficients
of fluids. A simple example is the velocity autocorrelation function of a
tagged particle. It is defined as
\bse
\label{eqs:2.20}
\be
C_{D}(t) = \langle{\bm v}(t)\cdot{\bm v}(0)\rangle_{\text{eq}},
\label{eq:2.20a}
\ee
where ${\bm v}(0)$ is the initial velocity of the tagged particle, ${\bm v}(t)$
its velocity at a later time $t$, and $\langle\ldots\rangle_{\text{eq}}$
denotes an equilibrium ensemble average. The coefficient of self diffusion in a
$d$-dimensional system is then given by (e.g., \onlinecite{Boon_Yip_1991})
\be
D = \frac{1}{d}\int_{0}^{\infty }dt\ C_{D}(t).
\label{eq:2.20b}
\ee
\ese
Analogous expressions determine the coefficients of shear viscosity, $\eta$,
bulk viscosity, $\zeta$, and heat conductivity, $\lambda$, in a classical fluid
(and, when appropriate, in a classical solid). In general we denote these
current correlation functions by $C_{\mu}(t)$, where $\mu$ can stand for $D$,
$\eta$, $\zeta$, or $\lambda$.

\begin{figure}[t]
\includegraphics[width=6cm]{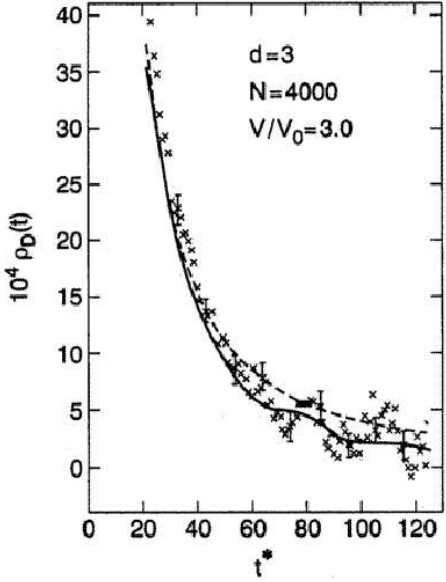}
\caption{\label{fig:3} Normalized velocity autocorrelation function
   $\rho_{\text{D}}(t) = C_{D}(t)/\langle{\bm v}^2(0)\rangle$ as
   a function of the dimensionless time $t^* = t/t_0$, where $t_0$ is the
   mean-free time. The crosses indicate computer results obtained by
   \protect\textcite{Wood_Erpenbeck_1975} for a
   system of $4000$ hard spheres at a reduced density corresponding to
   $V/V_0 = 3$, where $V$ is the actual volume and $V_0$ is the close-packing
   volume. The dashed curve represents the theoretical curve
   $\rho_{\text{D}}(t) = \alpha_D\,(t^*)^{-3/2}$. The solid curve represents a
   more complete evaluation of the mode-coupling formula with contributions from
   all possible hydrodynamic modes and with finite-size corrections
   included \protect\cite{Dorfman_1981}.
   From \protect\textcite{Dorfman_Kirkpatrick_Sengers_1994}.
 }
\end{figure}

In traditional many-body theories (e.g., the Boltzmann equation) the $C_{\mu}$
were always found to decay exponentially in time \cite{Chapman_Cowling_1952,
Dorfman_van_Beijeren_1977, Cercignani_1988}, with a characteristic decay time
on the order of the mean-free time between collisions, and until the mid-1960s
this was believed to be generally true. It thus came as a great surprise when
both numerical molecular-dynamics studies \cite{Alder_Wainwright_1967,
Alder_Wainwright_1968, Alder_Wainwright_1970}, and, shortly thereafter, more
sophisticated theories \cite{Dorfman_Cohen_1970, Dorfman_Cohen_1972,
Dorfman_Cohen_1975, Ernst_Hauge_van_Leeuwen_1970, Ernst_Hauge_van_Leeuwen_1971,
Ernst_Hauge_van_Leeuwen_1976a, Ernst_Hauge_van_Leeuwen_1976b}, showed that all
of these correlations decay for asymptotically long times as $1/t^{d/2}$. This
power-law decay in time is a type of GSI that is referred to as long-time tails
(LTT). In Fig.\ \ref{fig:3} we show results of both computer simulations and
theoretical calculations for $C_{D}(t)$ for a hard-sphere fluid in
three-dimensions. The LTT is clearly visible.

\paragraph{Fluctuating hydrodynamics}
\label{par:II.2.a}

The simplest and most general way to understand the LTT mechanism in a
classical fluid is via the equations of fluctuating hydrodynamics
\cite{Ernst_Hauge_van_Leeuwen_1971}, i.e., the Navier-Stokes equations with
appropriate Langevin forces added, although explicit many-body calculations,
within the framework of a generalized kinetic theory, are also possible and
give identical results for the LTTs \cite{Dorfman_Cohen_1972,
Dorfman_Cohen_1975}.\footnote{\label{fn:22} Notice that the term ``kinetic
theory'', which is often used synonymously with ``Boltmann theory'', is used
much more generally by theses authors.} The hydrodynamic approach gives the
exact long-time behavior because the slowest-decaying fluctuations in a
classical fluid are the fluctuations of the conserved variables, which are
described by the hydrodynamic equations. The conserved variables are, the mass
density $\rho$, the momentum density ${\bm g} = \rho\,\bm{u}$, with $\bm{u}$
the fluid velocity, and the energy density $\epsilon$ or, alternatively, the
entropy density $s$. In the long-wavelength and low-frequency limit the
fluctuations of these variables are described exactly by the equations
\cite{Landau_Lifshitz_VI_1987}
\bse
\label{eqs:2.21}
\bea
\partial_t\rho + \partial_{\alpha} g^{\alpha} &=& 0,
\label{eq:2.21a}\\
\partial_t g_{\alpha} + \partial_{\beta} g_{\alpha } u^{\beta} &=&
   -\partial_{\alpha}p + \partial_{\beta}\,
   \Bigl[\,\eta\, (\partial_{\alpha}u^{\beta} +
\partial^{\beta}u_{\alpha})
\nonumber\\
&&\hskip -60pt +\,\biggl(\zeta
-\frac{(d-1)}{d}\,\eta\biggr)\,\delta_{\alpha\beta}\,\partial_{\gamma}
u^{\gamma} + P_{\alpha\beta}\Bigr],
\label{eq:2.21b}\\
\rho\, T(\partial_t + u_{\alpha}\partial^{\alpha})\,s &=&
\partial_{\alpha}(\lambda\,\partial^{\alpha}T + q^{\alpha}).
\label{eq:2.21c}
\eea
\ese
Here $p$ denotes the  pressure, and summation over repeated indices is implied.
In Eq.\ (\ref{eq:2.21c}) we have neglected a viscous dissipation term that
represents entropy production since it is irrelevant to the leading LTTs. The
Langevin forces $P_{\alpha\beta}$ and $q_{\alpha}$ are uncorrelated with the
initial hydrodynamic variables, and satisfy
\bse
\label{eqs:2.22}
\bea
\left\langle P_{\alpha\beta}(\bm{x},t)\,P_{\mu\nu}(\bm{x}',t')\right\rangle &=&
2T\biggl[\eta (\delta_{\alpha\mu}\delta_{\beta\nu}
                                    + \delta_{\alpha\nu}\delta_{\beta\mu})
\nonumber\\
&&\hskip -120pt + \left(\zeta
-\frac{(d-1)}{d}\,\eta\right)\,\delta_{\alpha\beta}\,\delta_{\mu\nu}\biggr]\,
               \delta(\bm{x}-\bm{x}')\,\delta(t-t'),
\label{eq:2.22a}
\eea
\bea
\left\langle q_{\alpha}(\bm{x},t)\,q_{\beta}(\bm{x}',t')\right\rangle &=&
2\lambda\, T^{2}\,\delta_{\alpha\beta}\,\delta(\bm{x}-\bm{x}')\,\delta (t-t'),
\nonumber\\
\label{eq:2.22b}\\
\left\langle P_{\alpha\beta}(\bm{x},t)\,q_{\mu}(\bm{x}',t')\right\rangle &=& 0.
\label{eq:2.22c}
\eea
\ese
The above equations can be derived in a number of ways
\cite{Landau_Lifshitz_VI_1987, Fox_Uhlenbeck_1970} and are known to exactly
describe the long-wavelength and low-frequency fluctuations in a fluid.

\paragraph{GSI from explicit calculations}
\label{par:II.A.2.b}

To illustrate how the LTTs arise we choose a slightly different example than
the self-diffusion coefficient discussed above, namely, the shear viscosity
$\eta$. The appropriate time correlation function that enters the Green-Kubo
formula in this case is the autocorrelation of the transverse velocity current,
or the stress tensor \cite{Ernst_Hauge_van_Leeuwen_1971}. The basic idea is
that $\eta$ in the above equations is really a bare transport coefficient that
gets renormalized by the nonlinearities and fluctuations. We will present a
simplified calculation of $\eta$ \cite{Kirkpatrick_Belitz_Sengers_2002} and
then discuss more general results.

For simplicity, we assume an incompressible fluid. The mass conservation law
then reduces to the condition
\be
{\bm\nabla} \cdot {\bm u}({\bm x},t) = 0,
\label{eq:2.23}
\ee
and the momentum conservation law is a closed partial differential equation for
$\bm{u}$,
\be
\partial_t u_{\alpha} + u_{\beta}\partial^{\beta}u_{\alpha} = -\partial_{\alpha}
p/\rho + \gamma\,\nu\,\partial_{\beta}\partial^{\beta} u_{\alpha} +
\partial_{\beta}P_{\alpha\beta}/\rho .
\label{eq:2.24}
\ee
Here $\nu =\eta /\rho$ is the kinematic viscosity, which we assume to be
constant. The pressure term serves only the enforce the condition of
incompressibility. In fact, it can be eliminated by taking the curl of Eq.\
(\ref{eq:2.24}), which turns it into an equation for the transverse velocity.
The cause of the LTTs is the coupling of slow hydrodynamic modes due to the
nonlinear term in Eq.\ (\ref{eq:2.24}); this is another example of the
mode-mode coupling effects mentioned in the introduction to the current
section. Since we will treat this term as a perturbation, we have formally
multiplied it by a coupling constant $\gamma$ whose physical value is unity. We
will take the nonlinearity into account to lowest nontrivial order in $\gamma$.

Consider the velocity autocorrelation function tensor,
\be
C_{\alpha\beta}(\bm{k},t) = \langle
                        u_{\alpha}(\bm{k},t)\,u_{\beta}(-\bm{k},0)\rangle.
\label{eq:2.25}
\ee
An equation for $C$ can be obtained by Fourier transforming Eq.\
(\ref{eq:2.24}), multiplying by $u_{\beta}(-\bm{k},0)$, and averaging over the
noise while keeping in mind that the noise is uncorrelated with the initial
fluid velocity. In the case of an incompressible fluid we need to consider only
the transverse-velocity correlation function, $C_{\perp}$. This is easily done
by multiplying with unit vectors, $\hat{\bm{k}}_{\perp}^{(i)}$
($i=1,2,...,d-1$), that are perpendicular to $\bm{k}$, which eliminates the
pressure term. We obtain
\bea
(\partial_t + \nu \bm{k}^2)C_{\perp}(\bm{k},t) &=& -i\gamma\, k_{\mu}
        \hat{k}_{\perp\alpha}^{(i)}\hat{k}_{\perp\beta}^{(i)}\sum_{\bm{q}}
\nonumber\\
&&\hskip -60pt \times \langle u^{\mu}(\bm{k}-\bm{q},t)\,u^{\alpha}(\bm{q},t)\,
   u^{\beta}(-\bm{k},0)\rangle.
\label{eq:2.26}
\eea
Here we have used the incompressibility condition, Eq.\ (\ref{eq:2.23}), to
write all gradients as external ones. To zeroth order in $\gamma$ we find, with
the help of the f-sum rule \cite{Forster_1975} $C_{\perp}(\bm{k},0) = T/\rho$,
\bse
\label{eqs:2.27}
\be
C_{\perp}(\bm{k},t) = (T/\rho)\,e^{-\nu \bm{k}^2 t} + O(\gamma).
\label{eq:2.27a}
\ee
This is the standard result obtained from linearized hydrodynamics
\cite{Chapman_Cowling_1952}, which predicts exponential decay for $\bm{k}\neq
0$.\footnote{\label{fn:23} The reader might find it curious that this
derivation did not make explicit use of the Langevin force correlations, Eqs.\
(\ref{eqs:2.22}). However, it needs various equal-time correlation functions as
input, which contain the same information as Eqs.\ (\ref{eqs:2.22}).} Notice
that it amounts to GSI in time space for the local time correlation function,
\be
C_{\perp}({\bm x}=0,t) \propto 1/t^{d/2}.
\label{eq:2.27b}
\ee
\ese
This is an immediate consequence of the conservation law for the transverse
momentum, and hence an example of direct GSI.

To calculate corrections due to the nonlinearity, we need an equation for the
three-point correlation function in Eq.\ (\ref{eq:2.26}). The simplest way to
obtain this is to use the time translational invariance properties of this
correlation function to put the time dependence in the last velocity, and then
use Eq.\ (\ref{eq:2.24}) again. The result is an equation for the three-point
function in terms of a four-point one that is analogous to Eq.\
(\ref{eq:2.26}). To solve this equation, we note that, due to the velocity
being odd under time reversal, the equal-time three-point correlation vanishes.
By means of a Laplace transform, one can therefore express the three-point
function as a product (in frequency space) of the zeroth-order result for
$C_{\bot }$, Eq.\ (\ref{eq:2.27a}), and the four-point function. To leading
(i.e., zeroth) order in $\gamma $ the latter factorizes into products of
velocity autocorrelation functions. Upon transforming back into time space, and
to quadratic order in the coupling constant $\gamma$, we obtain
\bse
\label{eqs:2.28}
\be
(\partial_t + \nu \bm{k}^2)\,C_{\perp}(\bm{k},t) + \int_0^t d\tau\ \Sigma
(\bm{k},t-\tau)\,C_{\perp}(\bm{k},\tau) = 0,
\label{eq:2.28a}
\ee
with
\bea
\Sigma (\bm{k},t) &=&\gamma^2\frac{\rho}{T}\,k_{\mu}k_{\nu}
   \sum_{\bm{q}}
    \left[C^{\alpha\beta}(\bm{q},t)\,C^{\mu\nu}(\bm{k}-\bm{q},t)\right.
\nonumber\\
&&\hskip -50pt \left.
+\,C^{\alpha\nu}(\bm{q},t)\,C^{\mu\beta}(\bm{k}-\bm{q},t)\right]
  \hat{k}_{\perp,\alpha}^{(i)}\hat{k}_{\perp,\beta}^{(i)}+O(\gamma^3).
\label{eq:2.28b}
\eea
\ese
The self-energy $\Sigma\,$ is proportional to $\bm{k}^2$, and thus provides a
renormalization of the bare viscosity $\nu $, or the time correlation function
that determines the shear viscosity, $C_{\eta }(t)$. This correction to $\nu$
is time and wave number dependent.

The source of the LTTs is now evident. In our model of an incompressible fluid,
only the transverse component of $C_{\alpha\beta}$ is nonzero,
\be
C_{\alpha \beta }(\bm{q,}t)=(\delta _{\alpha \beta }-\hat{q}_{\alpha
}\hat{q}_{\beta })C_{\bot }(\bm{q,}t).
\label{eq:2.29}
\ee
Putting $\gamma =1$, and defining $\delta C_{\eta}(t) = \rho
\lim_{\bm{k}\rightarrow 0}\Sigma (\bm{k},t)/\bm{k}^2$, one obtains for
asymptotically long times
\bse
\label{eqs:2.30}
\be
\delta C_{\eta}(t) = T\rho \left(\frac{ d^2-2}{d(d+2)}\right)
                 \frac{1}{(8\pi\nu t)^{d/2}}\,.
\label{eq:2.30a}
\ee
The correction to the static, or zero-frequency, shear viscosity is given by
$\delta\eta \propto \int_0^{\infty}dt\,\delta C_{\eta}(t)$, cf. Eq.\
(\ref{eq:2.20b}). We thus have
\be
C_{\eta}(t\to\infty) = \delta C_{\eta}(t\to\infty)\propto t^{-d/2}.
\label{eq:2.30b}
\ee
\ese
This is the well-known contribution of the transverse velocity modes to the LTT
of the viscosity \cite{Ernst_Hauge_van_Leeuwen_1976a,
Ernst_Hauge_van_Leeuwen_1976b}. Notice that $C_{\eta}$ describes correlations
of shear stress, which is {\em not} conserved. The LTT therefore is a result of
mode-mode coupling effects, and hence an example of indirect GSI. In a
compressible fluid, a similar process coupling two longitudinal modes also
contributes to the leading LTT. The other transport coefficients, e.g.,
$\lambda$ and $\zeta$, also have LTTs proportional to $1/t^{d/2}$, and all of
them have less-leading LTTs proportional to $1/t^{(d+1)/2}$ or weaker
\cite{Dorfman_1981, Ernst_Hauge_van_Leeuwen_1976a,
Ernst_Hauge_van_Leeuwen_1976b}.

For the frequency-dependent kinematic viscosity, the nonexponential decay of
$\delta\nu(t)$ implies a nonanalyticity at zero frequency. More generally, the
algebraic $1/t^{d/2}$ LTTs in the long-time limit imply, for the frequency or
wave number-dependent transport coefficients $\mu$, a nonanalyticity at zero
frequency $\Omega$, or wave number $\vert{\bm k}\vert$,\footnote{\label{fn:24}
To avoid misunderstandings, we note that a static transport coefficient,
$\mu({\bm k},\Omega=0)$, is a time integral over a time correlation function,
see the Green-Kubo formula, Eq.\ (\ref{eq:2.20b}). As such, it is long-ranged
in space, just as $\mu({\bm k}=0,\Omega)$ is in time. A static susceptibility,
on the other hand, is related to an {\em equal-time} correlation function via
the fluctuation-dissipation theorem \cite{Forster_1975} and is {\em not}
long-ranged in space in classical equilibrium systems.}
\bse
\label{eqs:2.31}
\bea
\mu (\Omega)/\mu (0) &=& 1 - c_{d}^{\mu}\,\Omega^{(d-2)/2},
\label{eq:2.31a}\\
\mu (\bm{k})/\mu(0) &=& 1 - b_{d}^{\mu}\,\vert\bm{k}\vert^{d-2},
\label{eq:2.31b}
\eea
\ese
where the prefactors $c_{d}^{\mu}$ and $b_{d}^{\mu}$ are positive, and only the
leading nonanalyticities are shown. For the implications of these results in
$d\leq 2$, see footnote \ref{fn:49} below.

\subsubsection{Spatial correlations in nonequilibrium steady states}
\label{subsubsec:II.A.3}

We lastly consider a fluid in a nonequilibrium steady state; the equations of
fluctuating hydrodynamics can be extended to this case
\cite{Ronis_Procaccia_Machta_1980}. It has been known for some time that these
systems in general exhibit GSI in both time correlation functions {\em and}
thermodynamic quantities (see, e.g.,
\onlinecite{Dorfman_Kirkpatrick_Sengers_1994}). The spatial correlations
responsible for the GSI in thermodynamic susceptibilities are closely related
to LTTs of the equilibrium time correlation functions. We will consider a fluid
in a steady, spatially uniform, temperature gradient ${\bm\nabla} T$, but far
from any convective instability. Further, we use a number of approximations
that enable us to focus on the most interesting effects of such a gradient. For
a justification of this procedure, as well as the underlying details, we refer
the reader to the original literature \cite{Kirkpatrick_Cohen_Dorfman_1982a,
Kirkpatrick_Cohen_Dorfman_1982b}.

We write the temperature $T = T_0 + \delta T$ as fluctuations $\delta T$ about
an average value $T_0$, and focus on the coupling between fluctuations of the
transverse fluid velocity $\bm{u}_{\perp }$ and $\delta T$. If we neglect the
nonlinearity in Eq.\ (\ref{eq:2.21b}), Eqs.\ (\ref{eqs:2.21}) can be written
\bse
\label{eqs:2.32}
\be
\partial_t\, u_{\perp,\alpha} = \nu\,\partial_{\beta}\partial^{\beta}
                                           u_{\perp,\alpha}
   +\frac{1}{\rho_0}\,(\partial^{\beta} P_{\alpha\beta})_{\perp},
\label{eq:2.32a}
\ee
\be
\partial_t\,\delta T + u_{\alpha}\partial^{\alpha} T = D_T\,\partial_{\alpha}
   \partial^{\alpha}\delta T
   + \frac{1}{\rho_0 T_0 c_p}\,\partial_{\alpha} q^{\alpha},
\label{eq:2.32b}
\ee
\ese
where $\rho_0$ is the average mass density, $c_p$ is the specific heat per mass
at constant pressure, and $D_T = \lambda/\rho_0\,c_p$ is the thermal
diffusivity. These bilinear equations can be solved by means of Fourier and
Laplace transformations. Focusing on static, or equal-time, correlations, one
finds, for example,
\bse
\label{eqs:2.33}
\be
\langle\left\vert\delta\rho(\bm{k})\right\vert^2\rangle = \rho T\left(
  \frac{\partial\rho}{\partial p}\right)_T + \frac{\rho T(\alpha_T{\hat{\bm k}}
    _{\perp}\cdot \bm{\nabla}T)^2}{D_T(\nu +D_T)\bm{k}^4}\ ,
\label{eq:2.33a}
\ee
\be
\langle[{\hat{\bm k}}_{\perp}\cdot {\bm g}({\bm k})]\,\delta\rho(-{\bm k})
   \rangle = \rho T\alpha_T\,\frac{\hat{\bm{k}}_{\perp}\cdot \bm{\nabla}T
             }{(\nu +D_T)\bm{k}^2}\ ,
\label{eq:2.33b}
\ee
where
\be
\alpha_T = -\frac{1}{\rho}\left(\frac{\partial\rho}{\partial T}\right)_p,
\label{eq:2.33c}
\ee
\ese
is the thermal expansion coefficient at constant pressure.

There are several remarkable aspects of these results. First, Eq.\
(\ref{eq:2.33a}) for the density correlations implies that the first term,
which also exists in equilibrium, is delta-correlated in real space, while the
second term decays as $\text{const.} - \vert\bm{x}\vert$ in three-dimensions
\cite{Schmitz_Cohen_1985, Ortiz_Cordon_Sengers_2001}. Equation\
(\ref{eq:2.33b}) shows that the transverse-momentum--density correlation
function decays as $ 1/\vert\bm{x}\vert$ in three-dimensions. Both of these
results show that spatial correlations in a nonequilibrium steady state exhibit
GSI. Second, the right-hand side of Eq.\ (\ref{eq:2.33b}) is essentially the
integrand of a LTT contribution to the heat conductivity $\lambda$
\cite{Hohenberg_Halperin_1977}. This demonstrates the close connection between
the LTTs in equilibrium time correlation functions and the spatial GSI of
equal-time correlation functions in nonequilibrium situations. As in the case
of the equilibrium LTTs, this is an example of indirect GSI.

\begin{figure}[t]
\includegraphics[width=6cm]{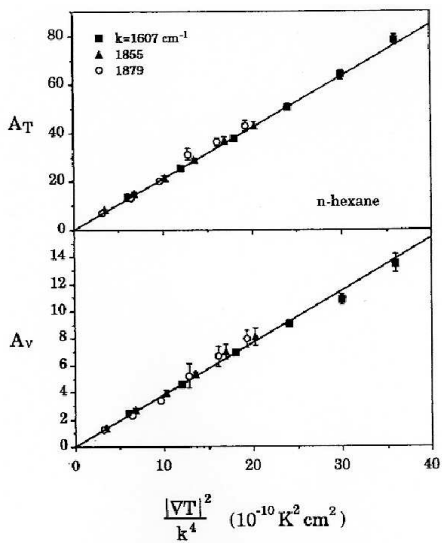}
\caption{\label{fig:4} Amplitudes $A_T$ and $A_{\rm v}$ of the nonequilibrium
   temperature and transverse-momentum (viscous) fluctuations in liquid hexane at
   $25^o\,{\rm C}$ as a function of $({\bm\nabla} T)^2/{\bf k}^4$. The symbols
   indicate experimental data. The solid lines represent the values predicted
   by Eqs.\ (\ref{eqs:2.34}). Notice the excellent agreement with no adjustable
   parameters. From \protect\textcite{Li_et_al_1994}.
 }
\end{figure}

These fluctuations can be directly measured by small-angle light-scattering
experiments. Specifically, the dynamic structure factor $S_{\rho\rho}(
\bm{k},t) = \langle\delta\rho(\bm{k},t)\delta\rho(-{\bm k},0)\rangle$, which is
proportional to the scattering cross section, in a nonequilibrium steady state
has the form \cite{Dorfman_Kirkpatrick_Sengers_1994}
\bse
\label{eqs:2.34}
\be
S_{\rho\rho}(\bm{k},t) = S_0 \left[(1+A_T) e^{-D_T\bm{k}^2 t} - A_{\nu} e^{-\nu
\bm{k}^2 t}\right],
\label{eq:2.34a}
\ee
where $S_0$ is the structure factor in equilibrium, and
\be
A_T = \frac{c_p\,\nu}{T D_T(\nu^2  -D_T^2)}\,\frac{({\hat{\bm
k}}_{\perp}\cdot{\bm\nabla}T)^2}{\bm{k}^4},
\label{eq:2.34b}
\ee
\be
A_{\nu} = \frac{c_p}{T(\nu^2 - D_T^2)}\,\frac{({\hat{\bm
k}}_{\perp}\cdot{\bm\nabla}T)}{\bm{k}^4}.
\label{eq:2.34c}
\ee
\ese
For $t=0$ one recovers the equal-time density correlation function, Eq.\
(\ref{eq:2.33a}). The amplitudes $A_{T}$ and $A_{\nu }$ are proportional to
$({\bm\nabla}T)^2/\bm{k}^4$, which has been verified by experiments, see Fig.\
\ref{fig:4}. Notice that the amplitude of the temperature fluctuations is
enhanced by a factor of a hundred compared to the scattering by an equilibrium
fluid. In real space, this strongly singular wave number dependence corresponds
to a decay as $\text{const.} - \vert{\bm x}\vert$ in three-dimensions.

\subsection{Quantum Systems}
\label{subsec:II.B}

The same mechanisms discussed above for classical systems also lead to
long-range correlations in space and time in quantum systems. The chief
distinctions between quantum systems (most interesting at $T=0$) \ and
classical systems are as follows.

(1) There are more soft, or gapless, modes in quantum systems. These extra
modes are due to a Goldstone mechanism that is absent at $T\neq 0$. For
example, particle-hole excitations across a Fermi surface are soft at $T=0$,
but acquire a mass at $T\neq 0$.

(2) As discussed in Sec.\ \ref{subsec:I.C}, there is a coupling between the
statics and the dynamics in quantum statistical mechanics that is absent in the
classical theory. In general, this means that long-ranged correlations in time
correlation functions imply long-ranged spatial correlations in equal-time
correlation functions, and thus in thermodynamic susceptibilities.

In this subsection we explain these points in some detail, and lay the
foundation for later sections where we discuss the importance of GSI for
understanding many quantum phase transitions. Our focus will be on itinerant
interacting electron systems. To motivate our discussions, and to make contact
with the classical results discussed in the preceding section, we first will
give some of the results showing GSI in zero (or very low) temperature electron
systems, both disordered and clean.\footnote{\label{fn:25} We will refer to
systems with and without quenched disorder as `disordered' and `clean',
respectively.} We choose to discuss the disordered case first, because in this
case the nonanalyticities that reflect the GSI are well known, although they
usually have not been thought of as examples of GSI. In noninteracting electron
systems they are usually referred to as ``weak-localization
effects'',\footnote{\label{fn:26} The term ``weak localization'' is
ill-defined, and differently applied by different authors. The most restrictive
meaning refers to the weakly nonmetallic weak-disorder regime in $d=2$, but we
use it to denote the nonanalytic dependence of observables on the frequency or
the wave number in noninteracting disordered electron systems in any
dimension.} and in interacting ones, as ``interaction effects'' or
``Altshuler-Aronov effects''; we will collectively refer to both classes as
``GSI effects''.\footnote{\label{fn:27} Occasionally both classes have been
collectively called ``weak-localization effects'' (e.g., in
\onlinecite{Kirkpatrick_Belitz_Sengers_2002}). Since most people associate
``weak localization'' with quenched disorder at least to some degree, this use
of the term tends to obscure the fact that precisely analogous effects are
found in clean systems.} The fact that there are analogous effects in clean
electron systems is much less well known. We then discuss the most important
soft modes in electron fluids, paying particular attention to those that exist
only at $T=0$. This is followed by an account of two distinct approaches to GSI
in quantum systems: The first approach is via explicit many-body calculations,
which is analogous to a generalized kinetic-theory approach to the classical
fluid. The second approach relies on the concept of clean and disordered
Fermi-liquid fixed points, and uses very general RG arguments. This is
analogous to the treatment of the classical Heisenberg ferromagnet discussed in
Sec.\ \ref{par:II.A.1.c}.

\subsubsection{Examples of generic scale invariance in itinerant electron
               systems}
\label{subsubsec:II.B.1}

\paragraph{Disordered systems in equilibrium}
\label{par:II.B.1.a}

In the entire metallic phase of disordered interacting electron systems,
various transport coefficients and thermodynamic quantities show nonanalytic
frequency and wave number dependencies that represent GSI. Since we are dealing
with a quantum system, there also are corresponding nonanalytic temperature
dependencies. For instance, for $2<d<4$ the electrical conductivity $\sigma$,
the specific heat coefficient $\gamma =C_V/T$, and the static spin
susceptibility $\chi_{\text{s}}$, as functions of the wave vector ${\bm k}$,
the frequency $\Omega$, and the temperature $T$, are given by (for reviews,
see, \onlinecite{Altshuler_Aronov_1984, Lee_Ramakrishnan_1985})
\bse
\label{eqs:2.35}
\bea
\sigma(\Omega\rightarrow 0,T=0)/\sigma (0,0) &=& 1
                             + c_d^{\sigma}\,\Omega^{(d-2)/2},\qquad\quad
\label{eq:2.35a}\\
\sigma(\Omega=0,T\to 0)/\sigma (0,0) &=& 1
                             + \tilde{c}_d^{\sigma}\,T^{(d-2)/2},
\label{eq:2.35b}\\
\gamma (T\rightarrow 0)/\gamma (0) &=& 1 + c_d^{\gamma}\,T^{(d-2)/2},
\label{eq:2.35c}\\
\chi_{\text{s}}(\bm{k}\rightarrow 0,T=0)/\chi_{\text{s}}(0,0) &=& 1 -
c_d^{\chi_s}\,
   \vert\bm{k}\vert^{d-2}.
\label{eq:2.35d}
\eea
\ese
Here the $c_d^{\sigma,\gamma,\chi_s}$ are positive coefficients that depend on
the disorder, the interaction strength,\footnote{\label{fn:28} Some of these
nonanalyticities, e.g., the one in Eq.\ (\ref{eq:2.35a}), are present even in
noninteracting electron systems (see, e.g., \onlinecite{Lee_Ramakrishnan_1985,
Kramer_MacKinnon_1993}), which are analogous to the classical Lorentz model
(see, e.g., \onlinecite{Hauge_1974}). We will be mostly interested in
interacting electron systems, which are analogous to classical fluids.} and the
dimensionality. In $d=2$ and $d=4$ the fractional powers in these equations are
replaced by integer powers times logarithms. In the time domain, Eq.\
(\ref{eq:2.35a}) implies that current correlations decay as $1/t^{d/2}$, while
Eq.\ (\ref{eq:2.35d}) implies that spatial spin correlations decay as
$1/r^{2(d-1)}$.

\begin{figure}[t]
\includegraphics[width=5cm]{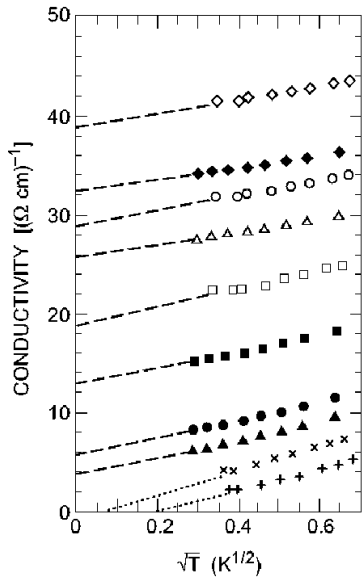}
\caption{\label{fig:5} Static low-$T$ conductivity of nine Si:B
   samples in a magnetic field plotted versus $\sqrt{T}$.
   After \textcite{Dai_Zhang_Sarachik_1992}.
 }
\end{figure}

All of the above effects are examples of indirect GSI, as will become clear
from the derivations given below. There are other examples of long-ranged
correlations in addition to the ones given above; the latter were chosen
because of their experimental relevance. For instance, an experiment showing
the $\sqrt{T}$ dependence of the conductivity expressed by Eq.\
(\ref{eq:2.35b}) is shown in Fig.\ \ref{fig:5}.
It is of interest to note, however, that not all obvious correlation functions
have nonanalyticities and exhibit GSI. For example, explicit calculations show
that the density susceptibility, or compressibility, has no $\vert{\bm
k}\vert^{d-2}$ nonanalyticity, in contrast to the spin susceptibility, Eq.\
(\ref{eq:2.35d}). This has been discussed by
\textcite{Belitz_Kirkpatrick_Vojta_2002}.

\paragraph{Clean systems in equilibrium}
\label{par:II.B.1.b}

For clean electronic systems, two interesting nonanalyticities that reflect GSI
are, for $1<d<3$ (for a recent overview, see,
\onlinecite{Chubukov_Maslov_2003b}),
\bse
\label{eqs:2.36}
\be
\hskip 62pt \gamma (T\rightarrow 0)/\gamma (0) = 1 + b_d^{\gamma}\,T^{d-1},
\label{eq:2.36a}
\ee
\vskip -22pt
\be
\chi_{\text{s}}(\bm{k}\to 0,T=0)/\chi_{\text{s}}(0,0) = 1
     + b_d^{\chi_{\text{s}}}\,\vert\bm{k}\vert^{d-1},
\label{eq:2.36b}
\ee
\ese
with the $b_d^{\gamma,\chi_{\text{s}}}$ positive, interaction dependent
coefficients. In $d=1$ and $d=3$ the integer powers in these equations become
multiplied by logarithms.

Let us explain the interesting sign difference between the nonanalytic
corrections to the spin susceptibility in the clean and dirty cases,
respectively, which will become important in Sec.\ \ref{sec:IV}. Quenched
disorder is known to increase the effective spin-triplet interaction strength
between the electrons \cite{Altshuler_Aronov_1984}. This effect is strongest at
zero wave number. $\chi_{\text{s}}(0,0)$ is therefore
enhanced,\footnote{\label{fn:29} Note that this represents an effect of
disorder that {\em enhances} the tendency towards ferromagnetism. It comes in
addition to the more basic, and more obvious, opposite effect due to the
dilution of the ferromagnet \cite{Grinstein_1985}. The important point for our
argument is that the former effect is wave number dependent, while the latter
is not.} but the enhancement effect decreases with increasing wave number,
which leads to the negative correction in Eq.\ (\ref{eq:2.35d}). In the clean
case, on the other hand, the generic soft modes represent fluctuations that
weaken the tendency towards ferromagnetism. This effect decreases
$\chi_{\text{s}}(0,0)$ compared to the Pauli value (more precisely, the Pauli
value with mean-field, or Hartree-Fock, corrections), and the wave number
dependent correction is positive.

\subsubsection{Soft modes in itinerant electron systems}
\label{subsubsec:II.B.2}

Let us now consider the soft modes that are responsible for the results listed
above. Apart from the soft modes due to the standard conservation laws, which
are the same as in the classical case, the most interesting soft modes in
itinerant electron systems are the Goldstone modes of a broken symmetry that is
characteristic of quantum systems. To understand these new modes, we consider
the formal action given by Eq.\ (\ref{eq:1.9b}) in more detail. Specifying the
Hamiltonian, we consider a model action
\bse
\label{eqs:2.37}
\be
S = S_0 + S_{\text{pot}} + S_{\text{int}},
\label{eq:2.37a}
\ee
with
\bea
S_0 &=& \int dx \sum_{\sigma} \bar{\psi}_{\sigma}(x)\left[-\partial_{\tau} +
       \bm{\nabla }^{2}/2m_{\text{e}} + \mu \right]\psi_{\sigma}(x).
\nonumber\\
\label{eq:2.37b}\\
S_{\text{pot}} &=& \int dx\ u({\bm x})\,\bar\psi_{\sigma}(x)\,\psi_{\sigma}(x).
\label{eq:2.37c}
\eea
\ese
Here $x\equiv (\bm{x,}\tau )$ and $\int dx=\int d\bm{x} \int_0^{\beta} d\tau$,
$m_{\text{e}}$ is the electron mass, $\mu $ is the chemical potential, and
$u(\bm{x)}$ is a one-body potential. $u$ can represent an electron-lattice
interaction, but for the physical applications we are interested in this would
not lead to qualitative deviations from the behavior of a simple `jellium'
model of a parabolic band, and we will therefore not pursue this possibility.
We will, however, want to treat the interaction of electrons with quenched
(i.e., static) random impurities \cite{Edwards_1958,
Abrikosov_Gorkov_Dzyaloshinski_1963}. In this case $u({\bm x})$ is a random
function. It is governed by a distribution that we take to be Gaussian with
zero mean, and a second moment
\be
\{u(\bm{x})u(\bm{y})\}_{\text{dis}} = U_0\,\delta (\bm{x-y})
         \equiv \frac{\delta(\bm{x-y})}{2\pi N_{\text{F}}\tau_{\text{rel}}}\ .
\label{eq:2.38}
\ee
Here $\{\ldots\}_{\text{dis}}$ denotes the disorder average,
$\tau_{\text{rel}}$ is the elastic relaxation time, and $N_{\text{F}}$ is the
density of states per spin at the Fermi surface. The factors in Eq.\
(\ref{eq:2.38}) are chosen such that disordered average Greens function has a
single-particle lifetime equal to $1/2\tau_{\text{rel}}$. We will consider both
clean systems, where $U_0 = 0$, and disordered ones, where $U_0 > 0$.

$S_{\text{int}}$ in Eq.\ (\ref{eq:2.37a}) represents the Coulomb potential, but
it is often advantageous to integrate out certain degrees of freedom to arrive
at an effective short-range interaction. For our present purposes it is not
necessary to specify the precise form of $S_{\text{int}}$; it will suffice to
postulate that the ground state of the interacting system is a Fermi liquid for
$U_0=0$, or a disordered Fermi liquid for $U_0\neq 0$.\footnote{\label{fn:30} A
disordered analog of Landau's Fermi-liquid theory (e.g.,
\onlinecite{Baym_Pethick_1991}) has been developed by
\textcite{Castellani_DiCastro_1985, Castellani_DiCastro_1986,
Castellani_Kotliar_Lee_1987, Castellani_et_al_1987, Castellani_et_al_1988}.}
However, for later reference we write down a popular model
\cite{Abrikosov_Gorkov_Dzyaloshinski_1963} that describes the self interaction
of the electron number density, $n(x)$, and the electron spin density,
$\bm{n}_{\,\text{s}}(x)$, via point-like, instantaneous interaction amplitudes
$\Gamma_{\text{s}}$ and $\Gamma_{\text{t}}$, respectively,
\bse
\label{eqs:2.39}
\be
S_{\text{int}} = \frac{-\Gamma_{\text{s}}}{2}\int dx\ n(x)\,n(x)
   + \frac{\Gamma_{\text{t}}}{2}\int dx\
   \bm{n}_{\,\text{s}}(x)\cdot\bm{n}_{\,\text{s}}(x).
\label{eq:2.39a}
\ee
In terms of fermionic fields, $n$ and $\bm{n}_{\text{s}}$ are given by
\bea
n(x) &=& \sum_{\sigma} {\bar\psi}_{\sigma}(x)\,\psi_{\sigma}(x),
\label{eq:2.39b}\\
\bm{n}_{\text{s}}(x) &=& \sum_{\sigma,\sigma'}{\bar\psi}_{\sigma}(x)\,
   \bm{\sigma}_{\sigma,\sigma'}\,\psi_{\sigma'}(x).
\label{eq:2.39c}
\eea
\ese
Here $\bm{\sigma} = (\sigma^x,\sigma^y,\sigma^z)$ represents the Pauli
matrices. An underlying repulsive Coulomb interaction leads to
$\Gamma_{\text{s}},\Gamma_{\text{t}} > 0$ in Eq.\ (\ref{eq:2.39a}).

It is useful to perform a Fourier representation from imaginary time to
fermionic Matsubara frequencies $\omega_n = 2\pi T(n+1/2)$,
\bse
\label{eqs:2.40}
\bea
\psi_{n,\sigma}(\bm{x}) &=& \sqrt{T}\int_0^{\beta}d\tau\
   e^{i\omega_n\tau}\psi_{\sigma}(x),
\label{eq:2.40a}\\
\bar{\psi}_{n,\sigma}(\bm{x}) &=& \sqrt{T}\int_0^{\beta}d\tau\
   e^{i\omega_n\tau}\bar{\psi}_{\sigma}(x).
\label{eq:2.40b}
\eea
\ese
The noninteracting part of the action can then be written,
\bea
{\tilde S}_0 &=& S_0 + S_{\text{pot}} = \int d\bm{x}
                 \sum_{\sigma,n}\bar{\psi}_{n,\sigma}(\bm{x})\,
                 \bigl[i\omega_n + \bm{\nabla}^{2}/2m_{\text{e}}
                 \nonumber\\
                 &&\hskip 70pt + \mu + u(\bm{x})\bigr]\,\psi_{n,\sigma}(\bm{x}).
\label{eq:2.41}
\eea
In the disordered case, it is further convenient to integrate out the quenched
disorder by means of the replica trick (\onlinecite{Edwards_Anderson_1975}; for
a pedagogical discussion of this technique, see, \onlinecite{Grinstein_1985}).
Accordingly, one introduces $N$ identical replicas of the system, labelled by
an index $\alpha$, and integrates out $u(\bm{x})$. $S_{\text{pot}}$ then gets
replaced by
\bea
S_{\text{dis}} &=& \frac{1}{4\pi N_{\text{F}}\tau_{\text{rel}}}
\sum_{\alpha,\beta=1}^{N} \int d{\bm x} \sum_{n,m}
{\bar\psi}_{n,\sigma}^{\alpha}({\bm x})\,\psi_{n,\sigma}^{\alpha}({\bm x})\,
\nonumber\\
&&\hskip 60pt\times{\bar\psi}_{m,\sigma'}^{\beta}({\bm x})\,
\psi_{m,\sigma'}^{\beta}({\bm x}),
\label{eq:2.42}
\eea
and physical quantities are obtained by letting $N\to 0 $ at the end of
calculations.

We now note two crucial features \cite{Wegner_Schaefer_1980}. First, for
$\omega_n = 0$, the action ${\tilde S}_0$ is invariant under transformations of
the fermionic fields that leave $\sum_n\bar{\psi}_n\psi_n$ invariant. That is,
${\tilde S}_0$ is invariant under a continuous rotation in frequency
space.\footnote{\label{fn:31} Because of the anticommuting nature of the
fermion fields, the Lie group in question is symplectic; for a model with $2M$
Matsubara frequencies it is $\text{Sp}\,(2M)$.} Second, this symmetry is
spontaneously broken symmetry whenever $N_{\text{F}}\neq 0$. To see this,
consider the order parameter
\bea
\bar{Q} &=& \lim_{\omega_n\to 0+}\left\langle\bar{\psi}_{n,\sigma}(\bm{x})\,
    \psi_{n,\sigma}(\bm{x})\right\rangle
\nonumber\\
&& -\lim_{\omega_n\to 0-}\left\langle\bar{\psi}_{n,\sigma}(\bm{x})\,
    \psi_{n,\sigma}(\bm{x})\right\rangle.
\label{eq:2.43}
\eea
$\bar{Q}$ is the difference between retarded and advanced Green functions, and
is proportional to $N_{\text{F}}$. The causal Green function has a cut along
the real axis for all frequencies where $\bar{Q}\propto N_{\text{F}}\neq 0$,
which breaks the symmetry between frequencies with positive and negative
imaginary parts, or between retarded and advanced degrees of freedom. Since
$S_{\text{int}}$ cannot explicitly break this symmetry,\footnote{\label{fn:32}
This follows generally from time translational invariance, and can be checked
explicitly for the interaction defined in Eq.\ (\ref{eq:2.39a}).} this will
result in soft modes according to Goldstone's theorem. Technically, these soft
modes are most conveniently discussed in terms of fluctuations of $4\times 4$
matrices that are isomorphic to bilinear products of fermionic fields
\cite{Efetov_Larkin_Khmelnitskii_1980},
\be
Q_{12}\cong \frac{i}{2}\left(
\begin{array}{cccc}
   -\psi_{1\uparrow }{\bar\psi}_{2\uparrow}  &
   -\psi_{1\uparrow}{\bar\psi}_{2\downarrow} &
   -\psi_{1\uparrow}\psi_{2\downarrow} &
   \ \ \psi_{1\uparrow}\psi_{2\uparrow}
\\
   -\psi_{1\downarrow}{\bar\psi}_{2\uparrow} &
   -\psi_{1\downarrow}{\bar\psi}_{2\downarrow} &
   -\psi_{1\downarrow}\psi_{2\downarrow} &
   \ \ \psi_{1\downarrow}\psi_{2\uparrow}
\\
   \ \ {\bar\psi}_{1\downarrow}{\bar\psi}_{2\uparrow} &
   \ \ {\bar\psi}_{1\downarrow}{\bar\psi}_{2\downarrow} &
   \ \ {\bar\psi}_{1\downarrow}\psi_{2\downarrow} &
   - {\bar\psi}_{1\downarrow}\psi_{2\uparrow}
\\
   -{\bar\psi}_{1\uparrow}{\bar\psi}_{2\uparrow} &
   -{\bar\psi}_{1\uparrow}{\bar\psi}_{2\downarrow} &
   -{\bar\psi}_{1\uparrow }\psi_{2\downarrow} &
   \ \ {\bar\psi}_{1\uparrow}\psi_{2\uparrow}
\end{array}\right)\ .
\label{eq:2.44}
\ee
Here all fields are understood to be at the position $\bm{x}$, and $1\equiv
(n_1,\alpha_1)$, etc., comprises both frequency labels $n$ and, for the
disordered case, replica labels $\alpha$. Since a nonzero frequency explicitly
breaks the symmetry, one expects, for $n_{1}n_{2}<0$,
\be
\lim_{\bm{k}\to 0}\left\langle Q_{n_{1}n_{2}}(\bm{k})\,Q_{n_{1}n_{2}}(-
\bm{k})\right\rangle\propto \frac{N_{\text{F}}}{\omega_{n_1}-\omega_{n_2}}\ .
\label{eq:2.45}
\ee
For $n_1n_2>0$, on the other hand, we expect the correlation function to
approach a finite constant in the limit of small wave numbers and frequencies.
That is, $Q_{nm}$ is soft if the frequencies $\omega_n$ and $\omega_m$ have
opposite signs, and massive if they have the same sign. For later reference,
let us introduce a notation that distinguishes between the soft and massive
components of $Q$. We write
\be
Q_{nm}(\bm{x}) = \begin{cases} q_{nm}(\bm{x}) & \text{if}\quad n>0, m<0\cr
                        q^{\dagger}_{nm}(\bm{x}) & \text{if}\quad n<0, m>0\cr
                        P_{nm}(\bm{x}) & \text{if}\quad nm>0
                 \end{cases}\ .
\label{eq:2.46}
\ee

Explicit calculations confirm these expectations \cite{Wegner_Schaefer_1980,
Efetov_Larkin_Khmelnitskii_1980, Belitz_Kirkpatrick_1997}. For technical
reasons, it is convenient to expand the $4\times 4$ matrix given by Eq.\
(\ref{eq:2.44}) in a spin-quaternion basis,
\be
Q_{12}(\bm{x}) = \sum_{r,i=0}^{3}(\tau_{r}\otimes s_{i})\,
         {_r^i Q}_{12}(\bm{x}),
\label{eq:2.47}
\ee
with $\tau_0 = s_0$ the $2\times 2$ unit matrix, and $\tau_j = -s_j =
-i\sigma_j$, ($j=1,2,3$). In this basis, $i=0$ and $i=1,2,3$ describe the
spin-singlet and spin-triplet degrees of freedom, respectively. $r=0,3$
corresponds to the particle-hole channel (i.e., products of $\bar{\psi}\psi$),
while $r=1,2$ describes the particle-particle channel (i.e., products
$\bar{\psi}\bar{\psi}$ or $\psi\psi$).

For small wave numbers and low frequencies, and for disordered noninteracting
electrons, one finds
\cite{Efetov_Larkin_Khmelnitskii_1980}\footnote{\label{fn:33} The
normalizations of these propagators used in the literature vary, depending on
how the $Q$-fields have been scaled. The normalization used in Eqs.\
(\ref{eqs:2.48}) corresponds to a dimensionless $Q({\bm x})$, in accord with
the usual convention in the nonlinear $\sigma$ model (see Eqs.\
(\ref{eqs:2.62}) below), while the $Q({\bm x})$ as defined by Eq.\
(\ref{eq:2.44}) is dimensionally a density of states.}
\bse
\label{eqs:2.48}
\be
\left\langle{_r^i q}_{12}(\bm{k})\,{_s^j q}_{34}(\bm{p})\right\rangle =
   \frac{G}{8}\,\delta(\bm{k}+\bm{p})\,\delta_{rs}\,\delta_{ij}\,\delta_{13}\,
   \delta_{24}\,{\cal D}_{n_{1}n_{2}}(\bm{k}),
\label{eq:2.48a}
\ee
with
\be
{\cal D}_{n_{1}n_{2}}(\bm{k}) = \frac{1}{\bm{k}^2
   + G H\vert\omega_{n_1}-\omega_{n_2}\vert}\qquad (\text{disordered})\ .
\label{eq:2.48b}
\ee
Here $H = \pi\,N_{\text{F}}/4$, and $G = 8/\pi\sigma_0 \propto
1/\tau_{\text{rel}}$, with $\sigma_0$ the conductivity in Boltzmann
approximation, see Eq.\ (\ref{eq:2.56}) below. $1/GH \equiv D$ is the electron
diffusion coefficient. In the low-disorder limit, $D =
v_{\text{F}}^{2}\tau_{\text{rel}}/d$, with $v_{\text{F}}$ the Fermi velocity.
Note that this structure is consistent with Eq.\ (\ref{eq:2.45}), which was
based on very the general arguments and Goldstone's theorem. For clean
noninteracting systems, Eq.\ (\ref{eq:2.48a}) is still valid, but $G =
2/N_{\text{F}}v_{\text{F}}$, and ${\cal D}$ is given by\footnote{\label{fn:34}
Equation (\ref{eq:2.48c}) is a schematic representation, which has the correct
scaling properties, of a more complicated function. In contrast, Eq.
(\ref{eq:2.48b}) represents the exact propagator in the limit of small
frequencies and wave numbers.}
\be
{\cal D}_{n_{1}n_{2}}(\bm{k}) = \frac{1}{\vert\bm{k}\vert +
                 G H\vert\omega_{n_1}-\omega_{n_2}\vert}\qquad
                 (\text{clean})\ .
\label{eq:2.48c}
\ee
\ese
$H$ is the same in either case. As expected, for disordered electrons ${\cal
D}$ is diffusive, while for clean electrons it is ballistic.

For interacting systems, clean or disordered, the results are structurally the
same \cite{Finkelstein_1983, Belitz_Kirkpatrick_1997}. The correlation
functions remain diffusive and ballistic, respectively, but they are no longer
the same in the spin-singlet ($i=0$) and spin-triplet ($i=1,2,3$) channels.
Rather, instead of $H$ in Eqs.\ (\ref{eq:2.48b}, \ref{eq:2.48c}), the
combinations $H + K_{\text{s}}$ and $H + K_{\text{t}}$, respectively, appear in
the two channels, with $K_{\text{s,t}} =
\pi\,N_{\text{F}}^2\Gamma_{\text{s,t}}/2$. We will not detailed correlation
functions for the interacting case; we just mention the physical significance
of these various coupling constants. $H$ determines the specific heat
coefficient $\gamma = C_V/T$ via $\gamma = 8\pi H/3$
\cite{Castellani_DiCastro_1986},\footnote{\label{fn:35} This reference
established the relation between $\gamma_V$ and $H$ for disordered systems
only. A later proof by means of Ward identities \cite{Castellani_et_al_1988}
can be generalized to apply to clean systems as well.} and can be interpreted
as a quasiparticle density of states \cite{Castellani_Kotliar_Lee_1987}.
$H+K_{\text{s}} = \pi\partial n/\partial\mu$ and $H+K_{\text{t}} =
\pi\chi_{\text{s}}$ are proportional to the thermodynamic density
susceptibility $\partial n/\partial\mu$ and the spin susceptibility
$\chi_{\text{s}}$, respectively \cite{Finkelstein_1984b,
Castellani_et_al_1984b, Castellani_et_al_1986}. Notice that $\partial
n/\partial\mu$, $\chi_{\text{s}}$, and $3\gamma/2\pi^2$ all are equal to
$N_{\text{F}}$ for noninteracting electrons, but differ for interacting
systems.

The above discussion assumed that the underlying continuous symmetry (see the
discussion above Eq.\ (\ref{eq:2.43})) is not broken explicitly. Such an
explicit symmetry breaking occurs, e.g., in the presence of a magnetic field,
magnetic impurities, or spin-orbit scattering
\cite{Efetov_Larkin_Khmelnitskii_1980, Finkelstein_1984a, Finkelstein_1984b,
Castellani_et_al_1984a}. In the presence of any of these symmetry breakers,
some of the channels classified by the spin and particle-hole indices $i$ and
$r$ of the matrix $^i_rQ$ become massive. For instance, magnetic impurities
lead to a mass in both the particle-particle channel ($r=1,2$) and the
spin-triplet channel ($i=1,2,3$). A summary of these effects has been given by
\textcite{Belitz_Kirkpatrick_1994}. The nonanalyticities in various observables
quoted in Sec.\ \ref{subsubsec:II.B.1} above are cut off accordingly, depending
on which soft channel they rely on.

\subsubsection{Generic scale invariance via explicit calculations}
\label{subsubsec:II.B.3}

One way to obtain the results quoted in Sec.\ \ref{subsubsec:II.B.1} is by
explicit many-body, or Feynman diagram, calculations to lowest order in a small
parameter \cite{Fetter_Walecka_1971, Abrikosov_Gorkov_Dzyaloshinski_1963}. To
illustrate this approach, we compute the electrical conductivity $\sigma$ in a
noninteracting, disordered, electronic system, to lowest nontrivial order in
the disorder about the Boltzmann value. The calculation starts with the Kubo
expression for $\sigma(\Omega)$ \cite{Kubo_1957},
\bse
\label{eqs:2.49}
\be
\sigma(\Omega) = \frac{i}{i\Omega_n}\left[\pi(\bm{q}=0,i\Omega_n)
                    + \frac{ne^2}{m}\right]_{i\Omega_n\rightarrow\Omega+i0},
\label{eq:2.49a}
\ee
with $\pi$ the longitudinal current correlation function,
\be
\pi(\bm{q},i\Omega_n) = \frac{-1}{{\bm q}^2}\int_{0}^{\beta }d\tau\
e^{i\Omega_n\tau}\,
   \left\langle T_{\tau}\,\bm{q}\cdot\hat{\!\!\bm{j}}(\bm{q},\tau)\ \bm{q}\cdot
                        \hat{\!\!\bm{j}}(\bm{q},0)\right\rangle.
\label{eq:2.49b}
\ee
Here $T_{\tau }$ is the imaginary time ordering operator, and $\
\hat{\!\!\bm{j}}$ is the current operator,
\be
\hat{\!\!\bm{j}}(\bm{q}) = \frac{1}{m}\sum_{\bm{k}}\bm{k}\
\hat{a}_{\bm{k}-\bm{q}/2}^{\dagger}\,\hat{a}_{\bm{k}+\bm{q}/2}\ ,
\label{eq:2.49c}
\ee
\ese
with $\hat{a}^{\dagger}$ and $\hat{a}$ electron creation and annihilation
operators, respectively. Wick's theorem can be used to evaluate time ordered
correlation functions such as the one in Eq.\ (\ref{eq:2.49b}).

\begin{figure}[t]
\includegraphics[width=7cm]{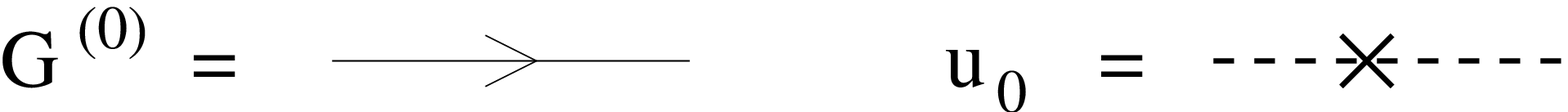}
\caption{\label{fig:6} Diagrammatic representation of the Green function
 $G^{(0)}$ and the impurity factor $u_0$.
 }
\end{figure}

\begin{figure}[t]
\includegraphics[width=7cm]{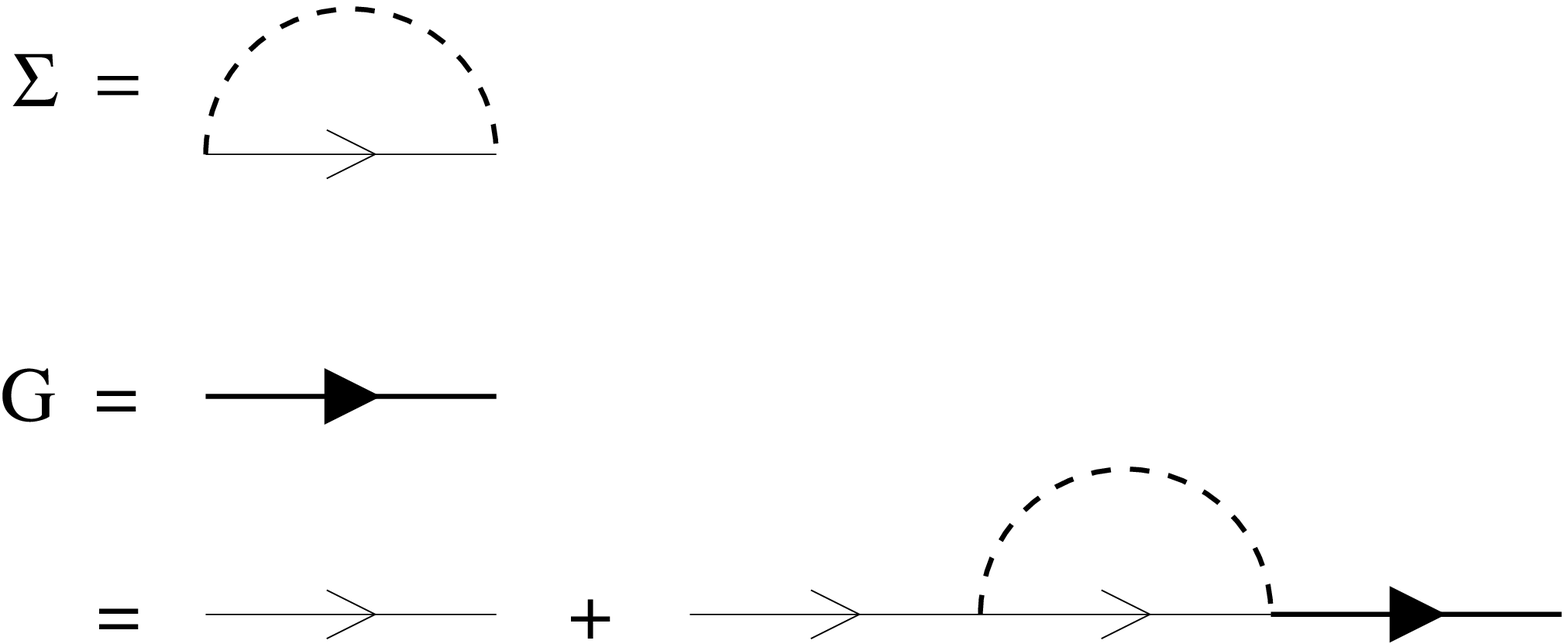}
\caption{\label{fig:7} The electronic self energy and the Green function in
   Born approximation.
}
\end{figure}

\begin{figure}[t]
\includegraphics[width=8cm]{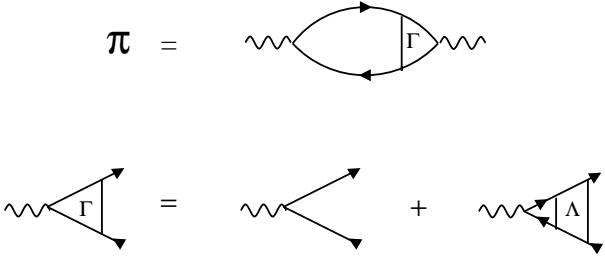}
\caption{\label{fig:8} Diagrammatic representation of the current correlation
   function, and of the relation between the functions $\Gamma$ and $\Lambda$.
 }
\end{figure}

For our present purposes, the small parameter for a perturbative treatment is
the impurity density $n_{i}$, or, alternatively, $1/k_{\text{F}}\ell$, with
$k_{\text{F}}$ the Fermi wave number, and $\ell$ the mean-free path between
electron-impurity collisions. The averaging over the positions of the
impurities can be performed using standard techniques \cite{Edwards_1958,
Abrikosov_Gorkov_Dzyaloshinski_1963}. The building blocks of the theory are,
the bare-electron Green function,
\be
G^{(0)}(\bm{q,}i\omega_n) = -\int_{0}^{\beta}d\tau\ e^{i\omega_n\tau}\,
   \left\langle T_{\tau}\hat{a}_{\bm{q}}(\tau)\,\hat{a}_{\bm{q}}^{\dagger}(0)
      \right\rangle_0\ ,
\label{eq:2.50}
\ee
and the impurity factor $U_0$, Eq.\ (\ref{eq:2.38}). The subscript $0$ in Eq.\
(\ref{eq:2.50}) indicates that both the average and the time dependence of
$\hat{a}$ are taken with the free-electron part of the Hamiltonian only.
Diagrammatically, we denote $G^{(0)}$ by a directed straight line, and $u_{0}$
by two broken lines (one for each factor of the impurity potential) and a cross
(for the factor of $n_{i}\sim 1/k_{\text{F}}\ell $); see Fig.\ \ref{fig:6}.
For our free-electron model one has
\be
G^{(0)}(\bm{q},i\omega_n) = (i\omega_n - \bm{q}^2/2m_{\text{e}} + \mu)^{-1}.
\label{eq:2.51}
\ee
The exact disorder-averaged Green function can be written in terms of a self
energy $\Sigma$ as
\bse
\label{eqs:2.52}
\be
G(\bm{q,}i\omega_n) = \left(i\omega_n - \bm{q}^2/2m_{\text{e}} + \mu + \Sigma
  (\bm{q},i\omega_n)\right)^{-1}.
\label{eq:2.52a}
\ee
The Born approximation for $\Sigma$ and for $G$ is shown diagrammatically in
Fig.\ \ref{fig:7}.
Analytically, it is given by
\be
\Sigma (\bm{q},i\omega_n) = \frac{i}{2\tau_{\text{rel}}}\ \sgn(\omega_n).
\label{eq:2.52b}
\ee
\ese
In this approximation, single-particle excitations thus decay exponentially
with a mean-free time $\tau_{\text{rel}}$. One commonly defines a
(longitudinal) current vertex function, $\Gamma$, by writing $\pi$ as
\bse
\label{eqs:2.53}
\bea
\pi(\bm{q},i\Omega_n) &=& \int_{\bm{p}}T\sum_{i\omega_n}
   \Gamma_0({\bm p},{\bm q})\,G(\bm{p}-\bm{q}/2,i\omega_n-i\Omega_n/2)
\nonumber\\
&&\hskip -55pt \times \, G(\bm{p}+{\bm q}/2,i\omega_n+i\Omega_n/2)\,
                         \,\Gamma(\bm{p},\bm{q};i\omega_n,i\Omega_n).
\label{eq:2.53a}
\eea
Here $\int_{\bm{p}} = \int d{\bm p}/(2\pi)^d$ in $d$-dimensions, and
\be
\Gamma_0({\bm p},{\bm q}) = {\bm p}\cdot{\bm q}/\vert{\bm q}\vert
\label{eq:2.53b}
\ee
\ese
is the bare current vertex. It is often convenient to express $\Gamma$ in terms
of another vertex function, $\Lambda$, via the equation
\bea
\Gamma(\bm{p},\bm{q};i\omega_n,i\Omega_n) &=& \Gamma_0({\bm p},{\bm q})
\nonumber\\
&& \hskip -80pt + \int_{\bm{k}}\Gamma_0({\bm k},{\bm
q})\,G(\bm{k}-\bm{q}/2,i\omega_n-i\Omega_n)
\nonumber\\
&&\hskip -60pt\times G(\bm{k}+\bm{q}/2,i\omega_n+i\Omega_n/2)\,
                    \Lambda(\bm{k},\bm{p},\bm{q};i\omega_n,i\Omega_n).
\nonumber\\
\label{eq:2.54}
\eea
Diagrammatically, these equations are illustrated in Fig.\ \ref{fig:8}.

\begin{figure}[t]
\includegraphics[width=8cm]{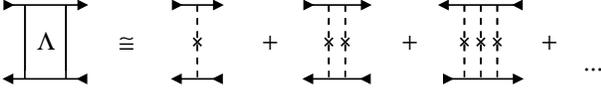}
\caption{\label{fig:9} The vertex function $\Lambda$ in Boltzmann
 approximation.
 }
\end{figure}

\begin{figure}[t]
\includegraphics[width=8cm]{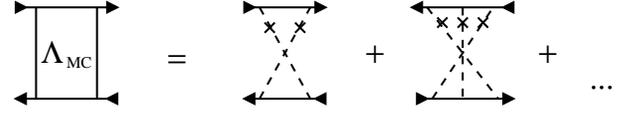}
\caption{\label{fig:10} Crossed-ladder contribution to $\Lambda$.
 }
\end{figure}

The Boltzmann approximation $\Lambda_{\text{B}}$ for $\Lambda$ (which is
equivalent to solving the Boltzmann equation for $\sigma$) is shown
diagrammatically in Fig.\ \ref{fig:9}.
Analytically, we find, for small $\Omega_n$ and ${\bm q}$,
\bea
\Lambda_{\text{B}}(\bm{k},\bm{p},\bm{q};i\omega_n,i\Omega_n)
        \Bigl\vert_{\bm{q}\to 0} &=&
                \Theta\left(-\omega_n(\omega_n + \Omega_n)\right)
\nonumber\\
&&\times \frac{4\pi N_{\text{F}}u_0^2}
 {D\bm{q}^{2}+\left\vert\Omega_n\right\vert}\ .
 \label{eq:2.55}
\eea
Using this in Eq.\ (\ref{eq:2.54}), and the result in Eqs.\ (\ref{eq:2.53a})
and (\ref{eq:2.49a}), the Boltzmann approximation for $\sigma$
is\footnote{\label{fn:36} Due to our point-like impurity potential, $\Gamma$ in
Boltzmann approximation is equal to $\Gamma_0$, and the conductivity is given
in terms of the single-particle relaxation time $\tau_{\text{rel}}$. For a more
general scattering potential, the vertex corrections in Eq.\ (\ref{eq:2.54})
are nonzero, and the Boltzmann conductivity is given by Eq.\ (\ref{eq:2.56})
with $\tau_{\text{rel}}$ replaced by the transport relaxation time
$\tau_{\text{tr}}$. See, \textcite{Abrikosov_Gorkov_Dzyaloshinski_1963} Sec.
39.2.}
\be
\sigma_0 = n\,e^2\tau_{\text{rel}}/m_{\text{e}}.
\label{eq:2.56}
\ee
There are, of course, an infinite number of corrections to $\sigma_{0}$ of
$O(1/k_{\text{F}}\ell)$ and higher. Of particular interest to us are the
contributions that couple to the electronic soft modes, Eqs.\ (\ref{eqs:2.48}).
As in the classical LTT calculation, Sec.\ \ref{par:II.A.2.b}, it is the
coupling of these modes to the current fluctuations that lead to the slow decay
of current correlations, and to a singular dependence of the conductivity on
the frequency. The relevant diagrams for this soft-mode contribution to
$\sigma$ are the so-called maximally crossed diagrams shown in Fig.\
\ref{fig:10} \cite{Gorkov_Larkin_Khmelnitskii_1979}.
For these contributions to $\Lambda$, which we denote by
$\Lambda_{\text{MC}}$, one finds
\bea
\Lambda_{\text{MC}}(\bm{k},\bm{p},\bm{q};i\omega_n,i\Omega_n)
   \Bigl\vert_{\bm{k}\approx\bm{p}} &=&
         \Theta\left(-\omega_n(\omega_n + \Omega_n)\right)
\nonumber\\
&&\hskip -30pt \times
      \frac{4\pi N_{\text{F}}u_0^2}
 {D(\bm{k}+\bm{p})^{2}+\left\vert\Omega_n
                                 \right\vert}\ .
\label{eq:2.57}
\eea
Notice that this is just the Boltzmann result, Eq.\ (\ref{eq:2.55}), with
$\bm{q}\to\bm{k}+\bm{p}$ \cite{Vollhardt_Woelfle_1980}. For small real
frequencies this contribution leads to the following result for the
conductivity \cite{Gorkov_Larkin_Khmelnitskii_1979},
\be
\sigma (\Omega )=\sigma _{0}\left\{ 1-\frac{1}{\pi N_{\text{F}}}\int_{\bm{q}}
\frac{1}{-i\Omega +D\bm{q}^{2}}\right\} .
\label{eq:2.58}
\ee
In $d=2$, this `correction' to $\sigma_0$ leads to a (negative) logarithmic
divergence. This reflects the fact that, at least for noninteracting electrons,
there is no metallic state in $d=2$ \cite{Abrahams_et_al_1979}. This phenomenon
(the ``weak localization'' proper, see footnote \ref{fn:26}) has generated a
huge body of literature (for reviews, see, \onlinecite{Bergmann_1984,
Lee_Ramakrishnan_1985, Kramer_MacKinnon_1993}). To the extent that it is caused
by the LTT not being integrable, this is analogous to the breakdown of
classical hydrodynamics in $d\leq 2$ (see Sec.\ \ref{par:II.A.2.b}), although
the physics behind it is quite different.

In $d>2$, Eq.\ (\ref{eq:2.58}) predicts that $\sigma(\Omega)$ is a nonanalytic
function of frequency, the small-frequency behavior for $2<d<4$ being
\bse
\label{eqs:2.59}
\be
\sigma(\Omega\to 0)/\sigma(0) = 1 + \text{const.}\times \Omega^{(d-2)/2},
\label{eq:2.59a}
\ee
With $\text{const.}>0$. This particular correction is one of several that
contribute to Eq.\ (\ref{eq:2.35a}); in the nomenclature explained at the
beginning of Sec.\ \ref{subsec:II.B} it is the weak-localization contribution.
Another one is the interaction contribution that was found by
\textcite{Altshuler_Aronov_Lee_1980}.\footnote{\label{fn:37} Historically, the
interaction effect predates weak localization. It was first discussed by
\textcite{Schmid_1974} for the electron inelastic lifetime, and by
\textcite{Altshuler_Aronov_1979} for the density of states, and a closely
related effect was found by \textcite{Brinkman_Engelsberg_1968b}.} In the time
domain this corresponds to a behavior of the current-current correlation
function, Eq.\ (\ref{eq:2.49b}), at long (real) times $t$,
\be
\pi(\bm{q}=0,t\rightarrow\infty)\sim -1/t^{d/2}.
\label{eq:2.59b}
\ee
\ese
Since the current is not conserved in our model, this is an example of indirect
GSI. Notice that the exponent is the same as for the corresponding LTT in a
classical fluid, Eq.\ (\ref{eq:2.31a}), but the sign is different. Physically,
this means that, in the quantum case, the leading LTTs decrease the diffusion
coefficient, while in the classical fluid case they increase it. This is
because the scattering in the electron-impurity model is due to static
impurities, while in the classical fluid it is due to the other fluid
particles.\footnote{\label{fn:38} This makes the electron-impurity model more
closely analogous to the classical Lorentz model, which shows a weaker
$1/t^{(d+2)/2}$ LTT (with a negative sign) than the classical fluid
\cite{Hauge_1974}. The mode-mode coupling effects are stronger in the quantum
system than in the classical one, which is why the electron model has a LTT
with the same strength as the classical fluid, but with the sign of the
classical Lorentz model.}

\subsubsection{Generic scale invariance via renormalization-group arguments}
\label{subsubsec:II.B.4}

The results given in Secs.\ \ref{subsubsec:II.B.1} and \ref{subsubsec:II.B.3}
can also be obtained independent of perturbation theory, by general RG and
scaling arguments which establish these LTTs as the exact leading behavior for
long times or small frequencies \cite{Belitz_Kirkpatrick_1997}. These authors
showed that in the disordered case, the GSI effects in $d>2$ can be understood
as corrections to scaling at a stable zero-temperature fixed point that
describes a disordered Fermi liquid. This also makes more explicit the
connection between weak-localization and Altshuler-Aronov effects on one hand,
and the Goldstone modes of the $O(N)$ nonlinear $\sigma$ model (Sec.\
\ref{subsubsec:II.A.1}), or the hydrodynamic theory of GSI (Sec.\
\ref{subsubsec:II.A.3}), on the other.\footnote{\label{fn:39} We note in
passing that it is possible to cast classical hydrodynamics in the form of a
field theory \cite{Martin_Siggia_Rose_1973}. This technique has not been
applied to the LTTs so far; doing so would make the analogy even closer.} This
is the first instance where we see a close, if not entirely obvious, analogy
between classical and quantum effects.

\paragraph{Disordered electron systems}
\label{par:II.B.4.a}

The long-wavelength and low-frequency excitations in a noninteracting
disordered electron systems can be described by a nonlinear $\sigma$ model
\cite{Wegner_1979, Wegner_Schaefer_1980, McKane_Stone_1981}. This has been
generalized to interacting systems by \textcite{Finkelstein_1983}; a derivation
in the spirit of \textcite{Wegner_Schaefer_1980} has been given by
\textcite{Belitz_Kirkpatrick_1997} and
\textcite{Belitz_Evers_Kirkpatrick_1998}. As for the simpler $\sigma$ model
derived and discussed in Sec.\ \ref{subsubsec:II.A.1}, the basic idea is to
construct an action solely in terms of the massless modes. The derivation of
the effective theory then becomes a two-step process. First, one needs to
identify the massless modes of the system, cf. Sec.\ \ref{subsubsec:II.B.2}.
Second, the microscopic theory, Eqs.\ (\ref{eqs:2.37}-\ref{eqs:2.39}), must be
transformed into an effective one that keeps only the soft modes, while all
other degrees of freedom are integrated out in some reasonable approximation.
The result of this procedure is a generalized nonlinear $\sigma$ model. Within
this theory the partition function can be written as an integral over a matrix
field ${\hat Q}$ that essentially contains the soft sectors of the field $Q$,
Eq.\ (\ref{eq:2.44}),
\be
Z=\int D[\hat{Q}]\ e^{{\cal A}[\hat{Q}]}.
\label{eq:2.60}
\ee
The effective action ${\cal A}$ is given by
\bse
\label{eqs:2.61}
\be
{\cal A} = \frac{-1}{2G}\int d\bm{x}\ \tr(\bm{\nabla }{\hat Q}(\bm{x}))^2
   + 2H\int d\bm{x}\ \tr(\Omega\, {\hat Q}(\bm{x})) + {\cal A}_{\text{int}}.
\label{eq:2.61a}
\ee
Here ${\cal A}_{\text{int}}\sim O(T\Gamma {\hat Q}^2)$ is the interaction part
of the action, i.e., $S_{\text{int}}$ from Eq.\ (\ref{eq:2.39a}) expressed in
terms of $Q$-matrices. Schematically, leaving out the detailed structure of the
frequency, replica, and spin-quaternion labels, it is of the
form\footnote{\label{fn:40} The complete expression \cite{Finkelstein_1983}
will not be needed here.}
\be
{\cal A}_{\text{int}} = T\,\Gamma \int d{\bm x}\ {\hat Q}(\bm{x})\,{\hat
          Q}(\bm{x}).
\label{eq:2.61b}
\ee
\ese
For our present purposes, $\Gamma$ can stand for either $\Gamma_{\text{s}}$ or
$\Gamma_{\text{t}}$. $\Omega$ is a diagonal matrix with fermionic Matsubara
frequencies as diagonal elements, and the coupling constants $G$ and $H$ were
defined after Eq.\ (\ref{eq:2.48b}). ${\hat Q}$ is subject to the constraints
\bse
\label{eqs:2.62}
\be
{\hat Q}^{2}(\bm{x}) = 1\ ,\ {\hat Q}^{\dagger} = {\hat Q}\ ,\
   \tr {\hat Q}(\bm{x}) = 1.
\label{eq:2.62a}
\ee
A standard way to enforce these constraints, analogous to Eqs.\
(\ref{eqs:2.6}), is to write ${\hat Q}$ as a block matrix in frequency space,
\be
{\hat Q} = \left(
\begin{array}{cc}
(1-qq^{\dagger })^{1/2} & q \\
q^{\dagger } & -(1-q^{\dagger }q)^{1/2}
\end{array}
\right) ,
\label{eq:2.62b}
\ee
\ese
where the matrix $q$ has elements $q_{nm}$ with $n\geqslant 0,m<0$.

Insight into the GSI effects in a disordered metal can be gained by an RG
analysis that focuses on a stable fixed point of the above generalized $\sigma$
model \cite{Belitz_Kirkpatrick_1997}. This fixed point provides an RG
description of the disordered Fermi liquid ground state (see footnote
\ref{fn:30}), in analogy to Shankar's RG description of a clean Fermi liquid
\cite{Shankar_1994}. The procedure in analogy to Sec.\ \ref{par:II.A.1.c}. We
again define the scale dimension of a length $L$ to be $[L] = -1$. The stable
disordered Fermi-liquid fixed point is characterized by the choice
\be
\left[q(\bm{x})\right] = -(d-2)/2
\label{eq:2.63}
\ee
for the scale dimension of the field $q$, which corresponds to diffusive
correlations of the $q$. This choice is consistent with what one expects for
the soft modes in a disordered metal, see Eq.\ (\ref{eq:2.48b}). In addition,
the scale dimension of the frequency or temperature, i.e., the dynamical
scaling exponent $z = \left[\Omega\right] = \left[T\right]$, is needed. In
order for the fixed point to be consistent with diffusion, the frequency must
scale as the square of the wave number, so we choose
\be
z = 2.
\label{eq:2.64}
\ee
Now we expand the action in powers of $q$. In a symbolic notation that leaves
out everything not needed for power counting purposes, we have
\bea
{\cal A} &=& -\frac{1}{G}\int d\bm{x}\ (\nabla q)^2 + H\int d\bm{x}\
\Omega\,q^2 + \Gamma\,T \int d\bm{x}\ q^2
\nonumber\\
&& +\ O(\nabla^2 q^4,\Omega\,q^4,T\,q^3).
\label{eq:2.65}
\eea
Power counting shows that all of the coupling constants in Eq.\ (\ref{eq:2.65})
have vanishing scale dimensions with respect to our disordered Fermi-liquid
fixed point,
\be
\left[G\right] = \left[H\right] = \left[\Gamma\right] = 0.
\label{eq:2.66}
\ee

Now consider the leading corrections to the fixed-point action, as indicated by
Eq.\ (\ref{eq:2.65}). Power counting shows that all of these terms are
irrelevant with respect to the disordered Fermi-liquid fixed point as long as
$d>2$. Furthermore, all of the terms that were neglected in deriving the
generalized nonlinear $\sigma$ model can be shown to be even more irrelevant
than the ones considered here. The conclusion from these arguments is that the
term given explicitly in Eq.\ (\ref{eq:2.65}) constitute a stable fixed-point
action, and that the leading irrelevant operators (which we denote collectively
by $u$) have scale dimensions
\be
\left[u\right] = -(d-2).
\label{eq:2.67}
\ee

These results can be used to derive the GSI effects from scaling arguments. We
first consider the dynamical conductivity $\sigma(\Omega)$. Its bare value is
proportional to $G$, and according to Eq.\ (\ref{eq:2.66}) its scale dimension
is zero. We therefore have the homogeneity law (cf. Sec.\ \ref{subsec:I.B})
\bse
\label{eqs:2.68}
\be
\sigma(\Omega,u) = \sigma(\Omega\, b^{z},u\,b^{-(d-2)}).
\label{eq:2.68a}
\ee
By putting $b=1/\Omega ^{1/z}$, and using $z=2$ as well as the fact that
$\sigma(1,x)$ is an analytic function of $x$, we find that the conductivity has
a singularity at zero frequency, or LTT, of the form
\be
\sigma(\Omega) = \sigma(\Omega=0) + \text{const.}\times\Omega^{(d-2)/2}.
\label{eq:2.68b}
\ee
\ese
That is, we recover Eqs.\ (\ref{eq:2.35a}) and (\ref{eq:2.59a}). The present
analysis proves (with the same caveats as in Sec.\ \ref{par:II.A.1.c}) that the
$\Omega^{(d-2)/2}$ is the exact leading nonanalytic behavior.

The specific heat coefficient $\gamma = C_{V}/T$ is proportional to the
quasiparticle density of states $H$, whose scale dimension vanishes according
to Eq.\ (\ref{eq:2.66}). We thus have
\be
\gamma(T,u) = \gamma(T\,b^z,u\,b^{-(d-2)}),
\label{eq:2.69}
\ee
which leads to the low-temperature behavior given by Eq.\ (\ref{eq:2.35b}).

Finally, we consider the wave vector dependent spin susceptibility $\chi
_{\text{s}}(\bm{k})$. $\chi_{\text{s}}$ is a two-particle correlation and thus
can be expressed in terms of a $Q$-$Q$ correlation function. The leading
correction to the finite Fermi-liquid value is obtained by replacing both of
the $Q$ by $q$. We thus have structurally $\chi_{\text{s}}\sim T\int d\bm{x}\
\langle q^{\dagger}\,q\rangle$, with scale dimension
$\left[\chi_{\text{s}}\right] = 0$. The corresponding homogeneity law is
\be
\chi_{\text{s}}(\bm{k},u) = \chi_{\text{s}}(\bm{k}\,b,u\,b^{-(d-2)}),
\label{eq:2.70}
\ee
which leads to a nonanalytic dependence on the wave number as given by Eq.\
(\ref{eq:2.35c}).

We conclude with an important caveat. As all scaling arguments, those presented
above are exact, but very formal. There is no guarantee that the prefactors of
the various nonanalyticities, i.e., the coefficients
$c_d^{\sigma,\gamma,\chi_{\text{s}}}$, are nonzero. Indeed, explicit
calculations show that both $ c_d^{\gamma}$ and $c_d^{\chi_{\text{s}}}$ are
nonzero only in the presence of electron-electron interactions (e.g.,
\onlinecite{Altshuler_Aronov_1984}). Further, the same arguments would lead to
the conclusion that the density susceptibility has the same form as Eq.\
(\ref{eq:2.70}) for the spin susceptibility, while perturbation theory finds
that the corresponding prefactor vanishes even in the presence of interactions,
see the discussion at the end of Sec.\ \ref{par:II.B.1.a}. The situation is
thus as follows. If explicit calculation show that a nonanalyticity has a
nonzero prefactor, then the scaling arguments can be used to show that the
perturbative result is indeed the leading nonanalytic behavior, which one can
never conclude from perturbation theory by itself. However, the scaling
arguments by themselves can never establish the existence of a particular
nonanalyticity.

\paragraph{Clean electron systems}
\label{par:II.B.4.b}

Structurally identical arguments can be made for clean electronic systems
\cite{Belitz_Kirkpatrick_1997}. There are two motivating factors. First, while
the weak-localization effects are clearly caused by disorder, disorder plays a
much less crucial role in the explicit calculations for the Altshuler-Aronov
effects that lead to the same type of nonanalyticities, and the scaling
arguments presented above do not at all depend on disorder in any obvious way.
Second, it is well known that in $d=1$, a perturbative expansion in powers of
the interaction breaks down due to logarithmic divergencies, and that this
divergence is related to the breakdown of Fermi-liquid theory in this dimension
\cite{Dzyaloshinski_Larkin_1971, Schulz_1995}. A natural question is, what
happens to these singularities for $d>1$? Explicit perturbation theory shows
that both clean and disordered interacting electronic systems have related
nonanalyticities that reflect the GSI in both systems, and that these
singularities are closely related to the terms that cause the breakdown of
Fermi liquid theory in $d=1$ \cite{Belitz_Kirkpatrick_Vojta_1997,
Chitov_Millis_2001, Chubukov_Maslov_2003a, Chubukov_Maslov_2003b,
Galitski_Das_Sarma_2003}.

The formal scaling arguments for these effects proceeds as in Sec.\
\ref{par:II.B.4.a} above. The only structural difference is that the two-point
vertex in the disordered case, $\Gamma _{\text{disordered}}^{(2)} =
\bm{k}^{2}+GH\Omega$, is replaced by $\Gamma_{\text{clean}}^{(2)} = \vert
\bm{k}\vert + GH\Omega$ in the clean case, and in the terms of higher order in
$q$, the $\nabla^2$ operator in the disordered case is effectively replaced by
a $\nabla$ operator. In terms of scale dimensions, the net result of these
substitutions, using the notation of the previous subsection, is
\bse
\label{eqs:2.71}
\bea
\left[q({\bm x})\right] &=& -(d-1)/2,
\label{eq:2.71a)}\\
z &=& 1,
\label{eq:2.71c}\\
\left[u\right] &=& -(d-1),
\label{eq:2.71d}
\eea
\ese
with $u$ denoting the least irrelevant variables. Note that technically the
difference between the disordered and clean case is that $(d-2)$ in the
disordered case is replaced by $(d-1)$ in the clean case.

Using the Eqs.\ (\ref{eqs:2.71}) and arguments identical to those used in the
disordered case, one obtains the results given by Eqs.\ (\ref{eqs:2.36}). As
for the disordered case, this reasoning shows that the Eqs.\ (\ref{eqs:2.36})
are exact.

\subsubsection{Another example of generic scale invariance: Disordered systems
               in a nonequilibrium steady state}
\label{subsubsec:II.B.5}

The electron-impurity model in a nonequilibrium steady state has been studied
\cite{Yoshimura_Kirkpatrick_1996} by means of many-body diagrammatic techniques
in conjunction with Zubarev's nonequilibrium statistical operator method
\cite{Zubarev_1974}. This work showed that effects analogous to those discussed
in Sec.\ \ref{subsubsec:II.A.3} for classical nonequilibrium fluids also exist
in disordered electron systems at $T=0$, and that quantum GSI effects due to
nonequilibrium conditions are stronger, i.e., of longer range, than in
classical systems.

The particular model studied was that defined by Eqs. (\ref{eqs:2.37}) and
(\ref{eq:2.38}), without the electron-electron interaction term
$S_{\text{int}}$, but in the presence of a chemical-potential gradient
${\bm\nabla}\mu$. Consider the nonequilibrium part of the electronic structure
factor
\be
C_{2}(\bm{x},\bm{y}) = \{\langle\delta n(\bm{x})\delta
n(\bm{y})\rangle\}_{\text{dis}},
\label{eq:2.72}
\ee
where $\langle\ldots\rangle$ denotes a nonequilibrium thermal average. As one
might expect, the calculation leads to a result that is analogous to the one in
the classical fluid, Eqs. (\ref{eqs:2.34}), viz.,
\be
C_{2}(\bm{k}) = \frac{N_{\text{F}}\mu\,\tau_{\text{rel}}}{6d\,\pi
(D\bm{k}^{2})^{2}}\,\left[25({\bm\nabla} \mu )^{2}-12({\hat{\bm{k}}}\cdot
{\bm\nabla} \mu )^{2}\right].
\label{eq:2.73}
\ee
As in the classical fluid, this corresponds to a decay of correlations in real
space given by $\text{const.} - \vert\bm{x}\vert$ in three-dimensions.

In comparison, for a classical Lorentz gas (\onlinecite{Hauge_1974}, see also
foonotes \ref{fn:28} and \ref{fn:38}) one finds instead
\be
C_{2}(\bm{k})\sim ({\bm\nabla} \mu )^{2}/\bm{k}^{2}.
\label{eq:2.74}
\ee
That is, in both quantum and classical cases $C_{2}$ exhibits GSI, but the
effect is stronger in the quantum case due to the existence of more soft modes.

\subsubsection{Quantum Griffiths phenomena: power laws from rare
regions in disordered systems}
\label{subsubsecII.B.6}

In this section we discuss another mechanism for long-range correlations in
time in disordered quantum systems, the so-called quantum Griffiths phenomena.
In contrast to the examples above, the relevant soft modes are local in space.

Griffiths phenomena are nonperturbative effects of rare strong-disorder
fluctuations. They were first discovered in the context of classical phase
transitions in quenched disordered systems \cite{Griffiths_1969}. Griffiths
phenomena can be understood as follows: In general, impurities will decrease
the critical temperature $T_c$ from its clean value $T_c^0$. In the temperature
region $T_c<T<T_c^0$ the system does not display global order, but in an
infinite system one will find arbitrarily large regions that are devoid of
impurities, and hence show local order, with a small but nonzero probability
that usually decreases exponentially with the size of the region. These static
disorder fluctuations are known as ``rare regions'', and the order parameter
fluctuations induced by them belong to a class of excitations known as ``local
moments''; sometimes they are also referred to as ``instantons''. Since they
are weakly coupled to one another, and flipping them requires changing the
order parameter in a whole region, the rare regions have very slow dynamics.
\textcite{Griffiths_1969} was the first to show that they lead to a nonanalytic
free energy everywhere in the region $T_c<T<T_c^0$, which is known as the
Griffiths phase, or, more appropriately, the Griffiths region. In generic
classical systems, the contribution of the Griffiths singularities to
thermodynamic (equilibrium) observables is very weak since the singularity in
the free energy is only an essential one. In contrast, the consequences for the
dynamics are much more severe, with the rare regions dominating the behavior
for long times \cite{Randeria_Sethna_Palmer_1985, Dhar_Randeria_Sethna_1988,
Bray_1988}. In the classical McCoy-Wu model, where the disorder along a certain
direction is correlated, the Griffiths singularities have drastic effects even
for the statics \cite{McCoy_Wu_1968, McCoy_Wu_1969}.

At quantum phase transitions, the quenched disorder is perfectly correlated in
one of the relevant dimensions, {\it viz.}, the imaginary time dimension.
Therefore, the quantum Griffiths effects are generically as strong as in the
classical McCoy-Wu model \cite{Rieger_Young_1996, Thill_Huse_1995}. One of the
most obvious realizations of quantum Griffiths phenomena can be found in random
quantum Ising models,
\be
H = \sum_{\langle i,j \rangle} J_{ij}\, \sigma_i^z \sigma_j^z + \sum_i h_i\,
    \sigma_i^x,
\label{eq:2.75}
\ee
with Pauli matrices $\sigma^x$ and $\sigma^z$, and the summation runs over
nearest neighbors on a lattice. Both the Ising couplings $J_{ij}$ and the
transverse fields $h_i$ are independent random variables. This model has been
studied extensively in $d=1$ \cite{Fisher_1995, Young_Rieger_1996, Young_1997,
Igloi_Rieger_1998, Igloi_Juhasz_Rieger_1999} and $d=2$ \cite{Pich_et_al_1998,
Motrunich_et_al_2000, Rieger_Young_1996}, and the results are expected to
qualitatively hold in $d=3$ as well \cite{Motrunich_et_al_2000}. Let us
consider a system which is globally in the disordered phase, with a rare region
of volume $V_R$ that is locally in the ordered phase. The probability density
for such a region to occur is exponentially small, $p \propto \exp(-a V_R)$,
with $a$ a disorder dependent coefficient. However, since the volume dependence
of the local energy gap $\Delta$ is also exponential \cite{Young_Rieger_1996},
$\Delta \sim \exp(-b V_R)$, the probability density $P(\Delta)$ for finding a
gap of size $\Delta$ varies as $\Delta$ to a power. One thus has
\cite{Young_1997}
\be
P(\Delta) = p\,(V_R)\,\vert dV_R/d\Delta\vert \propto \Delta^{a/b-1}.
\label{eq:2.76}
\ee
Thus, we obtain a power-law density of low-energy excitations. Notice that the
exponent $a/b$ (which in the literature is often called an inverse dynamical
scaling exponent, and denoted by $1/z$) is a continuous function of the
disorder strength.

Many results follow from this. For instance, a region with a local energy gap
$\Delta$ has a local spin susceptibility that decays exponentially in the limit
of long imaginary times, $\chi_{\text loc}(\tau\to\infty) \propto
\exp(-\Delta\tau)$. Averaging by means of $P$, Eq.\ (\ref{eq:2.76}), yields
\bse
\label{eqs:2.77}
\be
\chi_{\text loc}^{\text av}(\tau\to\infty) \propto \tau^{-a/b}.
\label{eq:2.77a}
\ee
The temperature dependence of the static average susceptibility is then
\be
\chi_{\text loc}^{\text av}(T) = \int_0^{1/T}d\tau\ \chi_{\text loc}^{\text
av}(\tau)
   \propto T^{a/b-1}.
\label{eq:2.77b}
\ee
If $a<b$, the local zero-temperature susceptibility diverges, even though the
system is globally still in the disordered phase. The typical correlation
function falls off much faster than the average one; for $d=1$,
\textcite{Young_1997} has found a stretched-exponential behavior
\be
\chi_{\text loc}^{\text typ} (\tau\to\infty) \propto \exp\left( -c\,
\tau^{-a/(a+b)} \right)
\ee
\ese
with $c$ another coefficient. The large difference between the average and the
typical values is characteristic of very broad probability distributions of the
corresponding observables.

\subsubsection{Quantum long-time tails from detailed balance}
\label{subsubsec:II.B.7}

Finally, we discuss a very simple source of long time tails, or temporal GSI,
which is operative at zero temperature only. To be specific, consider a model
with spin diffusion. The imaginary, or dissipative, part of the dynamical spin
susceptibility as a function of wave number and real frequency is
\be
\chi^{{\lower 5pt\hbox{\,$''$}}}({\bm k},\Omega) = \chi({\bm k})\ \frac{D{\bm
k}^2\,\Omega}
     {\Omega^2 + (D{\bm k}^2)^2}\ ,
\label{eq:2.78}
\ee
with $D$ the spin diffusion coefficient and $\chi({\bm k})$ the static spin
susceptibility. $\chi^{{\lower 5pt\hbox{\,$''$}}}({\bm k},\Omega)$ is an
analytic function of frequency. Upon Fourier transformation, this yields an
exponential decay of correlations in real time, $\chi({\bm k},t)\propto
\exp(-D{\bm k}^2 t)$. Now consider the dynamic structure factor, $S({\bm
k},\Omega )$, and the symmetrized fluctuation function, $\varphi ({\bm
k},\Omega)$, which are defined by \cite{Forster_1975}
\bse
\label{eqs:2.79}
\bea
\hskip -30pt S({\bm k},\Omega) &=& \int d{\bm x}\int_{-\infty}^{\infty} dt\
e^{-i{\bm k}\cdot{\bm x} + i\Omega
t} \bigl\langle {\hat M}({\bm x},t)\,{\hat M}({\bm 0},0)\bigr\rangle, \nonumber\\
\label{eq:2.79a}\\
\varphi ({\bm k},\Omega) &=& \frac{1}{2}\int d{\bm x}\int_{-\infty}^{\infty}
dt\ e^{-i {\bm k}\cdot{\bm x} + i\Omega t} \bigl\langle {\hat M}({\bm
x},t)\,{\hat M}({\bm 0},0)
\nonumber\\
\hskip 20pt&&   + {\hat M}({\bm 0},0)\,{\hat M}({\bm x},t)\bigr\rangle,
\label{eq:2.79b}
\eea
\ese
where ${\hat M}$ is the magnetization operator. (We neglect the vector nature
of ${\hat M}$ for simplicity.) The fluctuation-dissipation theorem
(\onlinecite{Callen_Welton_1951}, see also \onlinecite{Forster_1975}) relates
these two correlation functions to $\chi^{''}$ via
\bse
\label{eqs:2.80}
\bea
S({\bm k},\Omega) &=& \frac{2}{1 - e^{-\beta\Omega}}\ \chi^{{\lower
5pt\hbox{\,$''$}}}({\bm k},\Omega),
\label{eq:2.80a}\\
\varphi({\bm k},\Omega) &=& \coth (\beta\Omega/2)\ \chi^{{\lower
5pt\hbox{\,$''$}}}({\bm k},\Omega).
\label{eq:2.80b}
\eea
\ese
For any nonzero temperature, these correlations are again analytic functions of
the frequency, and their Fourier transforms therefore decay exponentially in
time. However, in the zero-temperature limit, $\beta \to \infty$ (and for ${\bm
k}\neq 0$), both of these correlation functions are not analytic at $\Omega
=0$; namely, they are proportional to $\Omega\,\Theta(\Omega)$. This
nonanalyticity leads to a power-law decay in real time (see, e.g.,
\onlinecite{Lighthill_1958}),
\be
S({\bm k},t\to \infty )\propto \varphi ({\bm k},t\to \infty )\propto 1/t^2.
\label{eq:2.81}
\ee
Physically, the nonanalyticities in the Eqs.\ (\ref{eqs:2.80}) at $T=0$ reflect
the absence of excitations in the ground state. This is contained in the
detailed balance relation
\be
S({\bm k},-\Omega) = e^{-\beta\Omega}\,S({\bm k},\Omega),
\label{eq:2.82}
\ee
which implies that $S({\bm k},\Omega)$ at $T=0$ must vanish for $\Omega<0$.
Note that these long-time tails are present at all wave numbers, not just in
the long-wavelength limit. More generally, the above considerations imply that
$S$ and $\varphi$ will display LTTs for any zero-temperature system. For low
but nonzero temperature there will be a pre-asymptotic LTT followed by an
exponential asymptotic decay.

For later reference we also consider the spin susceptibility at imaginary
Matsubara frequencies $i\Omega_n = i2\pi Tn$,
\be
\chi({\bm k},i\Omega_n) = \int_{-\infty}^{\infty} \frac{d\omega}{\pi}\
\frac{\chi^{{\lower 5pt\hbox{\,$''$}}}({\bm k},\omega)}{\omega - i\Omega_n} =
\chi({\bm k})\,\frac{D{\bm k}^2}{\vert\Omega_n\vert + D{\bm k}^2}\ ,
\label{eq:2.83}
\ee
which is a nonanalytic function of $\Omega_n$. Accordingly, the imaginary time
correlation function $\chi({\bm k},\tau) = T\sum_n e^{-i\Omega_n\tau} \chi({\bm
k},i\Omega_n)$ at $T=0$ has a LTT for large imaginary time $\tau$,
\be
\chi({\bm k},\tau\to\infty) \propto 1/\tau^2.
\label{eq:2.84}
\ee
More generally, any imaginary time correlation function has a long imaginary
time tail if the corresponding causal function is nonanalytic at zero
frequency.

\section{Influence of Generic Scale Invariance on Classical Critical Behavior}
\label{sec:III}

We now combine the concept of GSI with critical phenomena. Given that critical
phenomena are driven by soft modes, one expects that soft modes connected to
GSI, provided they couple to the order-parameter fluctuations, will influence
the critical behavior. This is indeed the case in both classical and quantum
systems. In this section we will discuss two classical examples. The first one
deals with the nematic--smectic-A transition in liquid crystals, where the
generic soft modes are the Goldstone modes of the nematic phase.
\cite{Halperin_Lubensky_Ma_1974, Chen_Lubensky_Nelson_1978} This turns out to
be closely related to the normal metal-superconductor transition, where the
generic soft modes are virtual photons. The second one considers phase
separation in a binary liquid subject to shear, and the generic soft modes are
due to the nonequilibrium situation \cite{Onuki_Kawasaki_1979}. Another
classical example, which we will not dicuss here, involves the ferromagnetic
transition in compressible magnets \cite{Bergman_Halperin_1976,Aharony_1976}.
Quantum phase transitions will be discussed in Sec.\ \ref{sec:IV} below.

\subsection{The nematic--smectic-A transition in liquid crystals}
\label{subsec:III.A}

Nematic liquid crystals consist of rod-shaped molecules. In the nematic phase,
there is directional order, i.e., the rod axes tend to be parallel to one
another, while the centers of gravity of the molecules do not show long-range
order \cite{DeGennes_Prost_1993}. In the smectic-A phase, the molecules are
additionally arranged in layers. Within each layer, the rod axes are on average
aligned perpendicular to the layer, but the molecules still form a
two-dimensional liquid. In either phase, there are Goldstone modes associated
with the broken rotational invariance that is characteristic of the directional
order. It turns out that these Goldstone modes couple to the order parameter
for the smectic-A order, and have an important influence on the critical
behavior. Let us now discuss this effect, following
\textcite{Halperin_Lubensky_Ma_1974}, \textcite{Chen_Lubensky_Nelson_1978}, and
\textcite{DeGennes_Prost_1993}.

\subsubsection{Action, and analogy with superconductors}
\label{subsubsec:III.A.1}

The layers of the smectic-A phase are described as a density modulation in,
say, the $z$-direction,
\be
\rho({\bm x}) = \rho_0 + \rho_1({\bm x})\cos\left(q\,z + \varphi({\bm
x})\right).
\label{eq:3.1}
\ee
In the smectic-A phase, the fluctuating amplitude $\rho_1({\bm x})$ has a
nonzero mean value, $q$ is the wave number associated with the smectic layer
spacing, and $\varphi$ is a phase. It is convenient to combine $\rho_1$ with
$\varphi$ to form a complex order parameter (see \textcite{DeGennes_Prost_1993}
and references therein)
\be
\psi({\bm x}) = \rho_1({\bm x})\,e^{i\varphi({\bm x})}.
\label{eq:3.2}
\ee
One piece of the action for the smectic-A order then takes the form
\bse
\label{eqs:3.3}
\be
S_{\text{A}}[\psi] = \int d{\bm x}\ \left\{r_0\,\vert\psi({\bm x})\vert^2
   + c\,\vert{\bm\nabla}\psi({\bm x})\vert^2
   + u_0\,\vert\psi({\bm x})\vert^4\right\}.
\label{eq:3.3a}
\ee
This is the LGW functional for a complex scalar field, or, equivalently, the
$O(2)$ version of Eq.\ (\ref{eq:2.3}).\footnote{\label{fn:41} The gradient
terms parallel and perpendicular to the layers have different coefficients, but
the action can be made isotropic by an appropriate real-space scale change.}
$r_0$ and $u_0$ are bare parameters that will be renormalized later. This
action is incomplete. One must add to it, (1) a term describing the
fluctuations of the nematic order parameter, and (2) a term describing the
coupling between the two. The nematic rods are described by a unit vector
(called the ``director'') ${\bm n}$, and the director fluctuations are given by
${\bm A}({\bm x}) = {\bm n}({\bm x}) - {\hat z}$. To linear order in the
fluctuations, ${\bm n}^2 = 1$ enforces ${\bm A} = (A_x,A_y,0)$. The Gaussian
contribution of ${\bm A}$ to the action takes the form of three
gradient-squared terms \cite{DeGennes_Prost_1993}, viz.,
\bea
S_{\text{N}}[{\bm A}] &=& \int d{\bm x}\ \Bigl\{K_1\,[{\bm\nabla}\cdot{\bm
A}]^2
     + K_2\,\left[{\hat z}\cdot({\bm\nabla}\times{\bm A}({\bm x}))\right]^2
\nonumber\\
&&\hskip 50pt +\, K_3\left[\partial{\bm A}({\bm x})/\partial z\right]^2\Bigr\},
\label{eq:3.3b}
\eea
where $K_1$, $K_2$, and $K_3$ are related to elastic constants. The coupling
between $\psi$ and ${\bm A}$ can be shown to be adequately represented by
replacing the ${\bm\nabla}\psi$ in Eq.\ (\ref{eq:3.3a}) by
\be
{\bm\nabla}\psi({\bm x}) \rightarrow [{\bm\nabla} - iq{\bm A}({\bm
x})]\psi({\bm x}).
\label{eq:3.3c}
\ee
\ese

We see that the coupling between the order parameter and the director
fluctuations takes the same form as the one between the order parameter and the
vector potential in Eq.\ (\ref{eq:2.15a}). It is important to realize, however,
that the action given by Eqs. (\ref{eqs:3.3}), in contrast to Eqs.\
(\ref{eqs:2.15}), is not invariant under local gauge transformations due to the
additional gradient terms in Eq.\ (\ref{eq:3.3b}). This has profound
consequences for the ordered phase; for instance, it leads, via a
Landau-Peierls instability, to the absence of true long-range order in a
smectic (\textcite{Lubensky_1983}, see also footnote \ref{fn:42} below).
However, it has been shown that a simplified action obtained by neglecting
$K_1$ and putting $K_2 = K_3 = 1/8\pi\mu$ has the same critical properties
\cite{Lubensky_1983}. Although the full action has properties that are in
general very different from those of the gauge theory given in Eq.\
(\ref{eq:2.15a}), the former thus reduces to the latter for the purposes of
determining the critical behavior. Keeping in mind that ${\bm A}$ lies in the
$xy$-plane, an appropriate effective action therefore is
\bea
S_{\rm NA}[\psi,{\bm A}]\hskip -2pt &=& \hskip -2pt \int d{\bm x}\
\Bigl\{r_0\,\vert\psi({\bm x})\vert^2
   + c\,\left\vert [\nabla - iq{\bm A}({\bm x})]\psi({\bm x})\right\vert^2
   \nonumber\\
&&\hskip 0pt +\, u_0\,\vert\psi({\bm x})\vert^4 +
\frac{1}{8\pi\mu}\,\left[{\bm\nabla}\times
       {\bm A}({\bm x})\right]^2\Bigr\}.
\label{eq:3.4}
\eea
As we have discussed in Sec.\ \ref{subsubsec:II.A.4}, this action formally {\em
is} gauge invariant.\footnote{\label{fn:42} It is important to realize,
however, that the only physical gauge is still the one where $A_z=0$. ``Gauge
transformations'' that take one to, e.g., the Coulomb gauge used in Sec.\
\ref{subsubsec:II.A.4}, lead to a transformed order parameter which no longer
represents a physical observable. For instance, the transformed order parameter
exhibits true long-range order in the smectic phase, whereas, as mentioned in
the text, the original one does not \cite{DeGennes_Prost_1993, Lubensky_1983}.
Another way to say this is that the phase of the order parameter in a smectic
is observable, while in a superconductor, it is not.} We also recognize it as
the action of a spin-singlet superconductor, with $\psi$ the superconducting
order parameter, ${\bm A}$ the vector potential, $q$ the Cooper pair charge,
and $\mu$ the magnetic permeability (see, e.g., \onlinecite{DeGennes_1989}).
The nematic--smectic-A transition can thus be mapped onto a
superconductor-normal metal transition, even though the generic soft modes in
the superconducting case, viz., the photons, have a very different origin than
the director fluctuations: The latter are the Goldstone modes of a broken
symmetry, while the former are the result of gauge invariance, as was discussed
in Sec.\ \ref{sec:II}. This remarkable analogy \cite{Halperin_Lubensky_1974} is
a good example of the versatility of effective field theories. This becomes
even more apparent upon the realization that the action (\ref{eq:3.4}) is
closely related to the scalar electrodynamics problem studied by
\textcite{Coleman_Weinberg_1973}, which demonstrated a mechanism for the
spontaneous generation of mass in particle physics.

We make on final remark to put this section into context. In the disordered
phase we have generic long-ranged correlations, or GSI,  from the gauge field
fluctations, cf. Eq.\ (\ref{eq:2.17a}). In the following subsections we will
see how these generic long-ranged correlations modify the critical behavior
compared to the one that would result from $S_{\text{A}}$, Eq.\
(\ref{eq:3.3a}), alone. In particular, see Eq.\ (\ref{eq:3.7}) below, and the
discussion following it.

\subsubsection{Gaussian approximation}
\label{subsubsec:III.A.2}

The simplest way to deal with the action $S_{\text{NA}}$ is to treat it in
Gaussian approximation, that is, to neglect the nonlinear coupling between
$\bm{A}$ and $\psi$. The $\psi^4$ term needs to be kept for stability reasons,
of course. In this approximation the  model reduces to an XY-model, which makes
the liquid crystal analogous to a superfluid rather than to a superconductor,
and it predicts a continuous transition in the XY-universality class
\cite{DeGennes_1972}. As we know from Sec.\ \ref{subsubsec:II.A.4}, and will
see again below, this Gaussian approximation misses some crucial aspects of the
full model.

\subsubsection{Renormalized mean-field theory}
\label{subsubsec:III.A.3}

The action (\ref{eq:3.4}) is characterized by two length
scales.\footnote{\label{fn:43} The full action for the liquid crystal problem
contains many more length scales, and hence more cases than the type-I and
type-II of the superconductor need to be distinguished. These have not been
fully classified.} In a superconducting language, they are, the coherence
length
\bse
\label{eqs:3.5}
\be
\xi = \sqrt{c/\vert r\vert},
\label{eq:3.5a}
\ee
and the London penetration depth
\be
\lambda \equiv k_{\lambda}^{-1}= \sqrt{1/8\pi\mu c q^2\langle\vert\psi({\bm
x})\vert^2\rangle}.
\label{eq:3.5b}
\ee
\ese
Here $r$ is the renormalized distance from the critical point, see below. The
ratio $\kappa = \lambda/\xi = 1/k_{\lambda}\xi$ is the Landau-Ginzburg
parameter. Notice that $\kappa$ is independent of $r$, since
$\langle\vert\psi\vert\rangle \propto \sqrt{\vert r\vert}$. $\kappa < 1$
characterizes type-I superconductors, where the order parameter coherence
length is larger than the penetration depth. In a magnetic field, these
materials form a Meissner phase that completely expels the magnetic
flux.\footnote{\label{fn:44} To avoid misunderstandings, we stress that the
conclusions drawn below are valid in the absence of an external magnetic field,
as the vector potential ${\bm A}$ can describe spontaneous electromagnetic
fluctuations.} In the extreme type-I limit, $\kappa\ll 1$, order-parameter
fluctuations are negligible, $\psi({\bm x})\approx\langle\psi({\bm
x})\rangle\equiv \psi$, and the part of the action that depends on the vector
potential takes the form
\begin{equation*}
\frac{1}{8\pi\mu}\int d{\bm x}\ \left[k_{\lambda}^2\,{\bm A}^2({\bm x})
      + \left({\bm\nabla}\times{\bm A}({\bm x})\right)^2\right].
\end{equation*}
We see that in the normal conducting phase, the transverse photons are soft, as
they should be according to Sec.\ \ref{par:II.A.4.a} (these are the generic
soft modes), while in the superconducting phase they acquire a mass
proportional to $k_{\lambda}\propto\psi$, cf. Sec.\ \ref{par:II.A.4.b}.
Choosing a gauge fixing term, Eq.\ (\ref{eq:2.16}), with
$\eta=1$\footnote{\label{fn:45} Notice that the renormalized mean-field theory
neglects all fluctuations of the field $\psi$, in contrast to the Gaussian
theory discussed in Sec.\ \ref{par:II.A.4.b}. In particular, no particular
gauge needs to be chosen in order to eliminate the Higgs field. One should also
keep in mind that the full action for the liquid crystal, which is not gauge
invariant due to the additional gradient terms in Eq.\ (\ref{eq:3.3b}), has a
different soft-mode spectrum in the ordered phase than the superconductor.} one
finds for the ${\bm A}$-propagator
\be
\langle A_{\alpha}({\bm x})\,A_{\beta}({\bm y})\rangle = \delta_{\alpha\beta}\,
   \delta({\bm x}-{\bm y})\ \frac{4\pi\mu}{k_{\lambda}^2 + {\bm\nabla}^2}\ .
\label{eq:3.6}
\ee

Since the vector potential no longer couples to any fluctuating fields, it can
be integrated out to obtain an action entirely in terms of the superconducting
order parameter \cite{Halperin_Lubensky_Ma_1974},
\bea
S_{\text{eff}}[\psi] &=& \ln\int D[{\bm A}]\ e^{-S_{\text{NA}}[\psi,{\bm A}]}
\nonumber\\
&&\hskip -40pt = V\left(r_0\vert\psi\vert^2 + u_0\vert\psi\vert^4\right) -
\frac{1}{2}\,\Tr\ln\left({\bm\nabla}^2 + k_{\lambda}^2\right).
\label{eq:3.7}
\eea
Here $S_{\text{NA}}$ is given by Eq.\ (\ref{eq:3.4}), $V$ is the system volume,
and we have neglected a constant contribution to the action. The above
procedure is exact to the extent that fluctuations of $\psi$ can be neglected;
otherwise, it is an approximation. It produces an action entirely in terms of
$\psi$, but, since one has integrated out a soft mode, the price one pays is
that this action is nonlocal.\footnote{\label{fn:46} More precisely, the mode
that has been integrated out has a small mass (for small $\vert\psi\vert$) that
is given by the order parameter.} The effect of the nonlocality is most
conveniently studied by expanding the equation of state in powers of the order
parameter $\psi$, and the free energy density $f = T\,S_{\text{eff}}/V$ can be
obtained by integrating the result order by order. The details have been given
by \textcite{Chen_Lubensky_Nelson_1978}. The result for the leading terms in
three-dimensions is
\bse
\label{eqs:3.8}
\be
f/T = r\vert\psi\vert^2 - v_3\vert\psi\vert^3
               + u\vert\psi\vert^4 - h\,\psi\qquad (d=3).
\label{eq:3.8a}
\ee
Here we have added a field $h$ conjugate to the order parameter. $r$ and $u$
are given by $r_0$ and $u_0$ with additive corrections proportional to $\mu
q^2$ and $(\mu q^2)^2$, respectively. These changes do not affect the behavior
of the theory. However, $v_3>0$ is a positive coupling constant proportional to
$\sqrt{\mu q^2}$ that drives the transition first order. Notice that the new
term generated by the generic soft modes is nonanalytic in the order parameter;
such a term can never appear in a local Landau expansion. This provides a
rather extreme example of generic soft modes influencing critical behavior: The
critical point is destroyed, and a first-order transition takes place instead.
This phenomenon is known as a fluctuation-induced first-order transition in
condensed matter physics,\footnote{\label{fn:47} Notice that the fluctuations
in question are not the order-parameter fluctuations, but rather the generic
soft modes.} and as the Coleman-Weinberg mechanism in high-energy physics.
Interestingly, it is not the soft mode {\it per se} that produces the
first-order transition; it is the fact that a nonzero order parameter gives the
soft mode a mass. This is an example of a more general principle that has been
explored by \textcite{Belitz_Kirkpatrick_Vojta_2002}. For later reference we
also quote the free energy density in four dimensions
\cite{Chen_Lubensky_Nelson_1978}, although this case is not of physical
relevance for liquid crystals,
\be
f/T = r\vert\psi\vert^2 + v_4\vert\psi\vert^4\ln\vert\psi\vert +
u\vert\psi\vert^4 - h\,\psi\qquad (d=4),
\label{eq:3.8b}
\ee
\ese
with $v_4>0$.

Note that the free energy functionals given by Eqs.\ (\ref{eqs:3.8}) are
nonanalytic in $\psi$, and that this nonanalyticity has nothing to do with the
phase transition. Indeed, they denote the proper free energy for the order
parameter caused by a conjugate field far from the transition. Physically, the
nonanalyticity reflects GSI in the disordered phase, and it is directly related
to a nonanalytic wave number dependence of the order parameter susceptibility
$\chi_{\psi}$. This becomes obvious if one remembers that in any mean-field
theory, $r = 1 - \Gamma\chi_{\psi}$, with $\Gamma$ the appropriate interaction
coupling constant, and that $\psi$ scales like a wave number far from the phase
transition, as can be seen from Eqs.\ (\ref{eq:3.5b}, \ref{eq:3.7}). In a
scaling sense, Eq.\ (\ref{eq:3.8a}) thus corresponds to a wave number
dependence of $\chi_{\psi}$ given by
\bse
\label{eqs:3.8'}
\be
\chi_{\psi}({\bm k})/\chi_{\psi}(0) = 1 + b^{\chi_{\psi}}\vert{\bm k}\vert,
\label{eq:3.8'a}
\ee
with $b^{\chi_{\psi}}>0$ a positive coefficient. Notice that this is analogous
to Eqs.\ (\ref{eq:2.36b}, \ref{eq:2.35d}), although $\chi_{\psi}$ is not
observable for the superconductor, and related to an observable susceptibility
in a complicated way for the liquid crystal. In Sec.\ \ref{sec:IV} we will
discuss examples where the magnetic susceptibility, which is directly
measurable, plays an analogous role. In real space, this behavior corresponds
to
\be
\chi_{\text{s}}(r>0,\vert{\bm x}\vert\to\infty )\propto 1/\vert{\bm x}\vert^4.
\label{eq:3.8'b}
\ee
\ese
This manifestation of GSI is ultimately responsible for the failure of the
Gaussian theory to correctly describe the phase transition.

The conclusion that the nematic--smectic-A transition and the BCS
superconducting transition are of first order is inevitable if order-parameter
fluctuations are negligible. Quantitatively, the effect turns out to be
unobservably small in superconductors, as the first-order nature of the
transition is predicted to manifest itself only within about one $\mu$K from
the transition temperature. In liquid crystals, on the other hand, the numbers
are much more favorable \cite{Halperin_Lubensky_Ma_1974}. After a long time of
confusion, careful experiments have indeed confirmed the first-order nature of
the transition in a variety of liquid crystals
\cite{Anisimov_et_al_1990,Lelidis_2001, Yethiraj_Mukhopadhyay_Bechhoefer_2002}.

\subsubsection{Effect of order-parameter fluctuations}
\label{subsubsec:III.A.4}

Interesting and technically difficult questions arise when order-parameter
fluctuations are taken into account. First-order transitions are often thought
of as unaffected by order-parameter fluctuations, almost so by definition,
since they preempt a fluctuation-driven critical point. However, if the
first-order transition takes place within the critical region of an unrealized
second-order transition, then the critical fluctuations associated with the
latter can destabilize the mechanism that drives the transition first order.
The resulting transition can then be continuous, but described by a fixed point
different from the one that destabilizes the fluctuation-induced first-order
transition. An explicit example for such a ``fluctuation-induced second-order
transition'' was given by \textcite{Fucito_Parisi_1981}. How this works
technically becomes clear once one realizes that, within a RG treatment, a
fluctuation-induced first-order transition comes about by the quartic coupling
constant $u$ flowing to negative values. Critical fluctuations can weaken this
tendency, and lead to a positive fixed point value of $u$ after all. We will
discuss an explicit example of this happening in Sec.\ \ref{subsec:IV.A} below.

For the superconducting/nematic--smectic-A transition problem, a RG analysis to
one-loop order has been performed by \textcite{Chen_Lubensky_Nelson_1978}. To
first order in $\epsilon = 4-d$, these authors found no critical fixed point
and concluded that the transition is always of first order, both for
superconductors and for liquid crystals, and for both type-I and type-II
materials. However, later work by \textcite{Dasgupta_Halperin_1981}, prompted
by mounting experimental evidence that the transition is of second order in
type-II materials, concluded that, as one enters into the type-II region, the
first-order transition becomes weaker and weaker until the transition reverts
to second order in the so-called inverted XY-universality class. This has been
confirmed by a sizeable body of numerical and analytical evidence (see, e.g.,
\onlinecite{Herbut_Yethiraj_Bechhoefer_2001}, and references therein). Why the
perturbative RG does not show this fixed point is not quite clear.

\subsection{Critical behavior in classical fluids}
\label{subsec:III.B}

\begin{figure}[b]
\includegraphics[width=4.0cm]{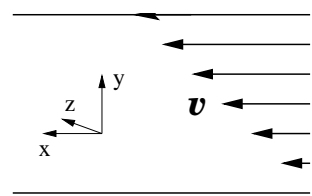}
\caption{\label{fig:11} Velocity field in a fluid subject to a steady, plane
 Couette flow.}
\end{figure}

Our second classical example deals with the critical behavior of a classical
fluid, both in equilibrium \cite{Kawasaki_1976, Hohenberg_Halperin_1977}, and
in a nonequilibrium situation created by applying a constant rate of shear
\cite{Onuki_Kawasaki_1979}. Let us consider a fluid confined between two
parallel plates subject to a constant rate $s$ of shear, see Fig.\
\ref{fig:11}.
The average local velocity is given by
\be
{\bm v}({\bm x}) = s\,y\,{\bm e}_x,
\label{eq:3.9}
\ee
with ${\bm e}_x$ a unit vector in $x$-direction. This externally induced
velocity has to be added to the Navier-Stokes equations (\ref{eqs:2.21}). Close
to the liquid-gas critical point, these equations can be replaced by the
following set of simpler equations \cite{Kawasaki_1970, Onuki_Kawasaki_1979},
\bse
\label{eqs:3.10}
\bea
\frac{\partial\psi}{\partial t} &=& -s\,y\frac{\partial}{\partial x}\,\psi
   -\rho\,\partial_{\alpha}\,\psi\,u^{\alpha}
   + \frac{\lambda}{2}\,\partial_{\alpha}\partial^{\alpha}
      \frac{\delta}{\delta\psi}\,S[\psi] + \theta,
\nonumber\\
\label{eq:3.10a}\\
\frac{\partial u^{\alpha}}{\partial t} &=&
-\frac{\rho}{2}\left(\psi\,\partial^{\alpha}
   \frac{\delta}{\delta\psi}\,S[\psi]\right)_{\perp}
      + \eta\,\partial_{\beta}\partial^{\beta}\,u^{\alpha} +
      \zeta_{\perp}^{\alpha}.
      \nonumber\\
\label{eq:3.10b}
\eea
Here the order parameter $\psi$ represents the specific entropy, $\bm{u}$ is
the deviation of the local velocity from its average $\bm{v}$, and $\lambda$
and $\eta$ are the bare thermal conductivity and the bare shear viscosity as in
Eqs.\ (\ref{eqs:2.21}).\footnote{\label{fn:48} Here we discuss a pure fluid for
simplicity. For experimental purposes, binary fluids are preferred for
technical reasons. This changes the interpretation of the various quantities in
Eqs.\ (\ref{eqs:3.10}), but the physics remains the same.} ${\bm v}_{\perp}$
denotes the transverse part of a vector $\bm{v}$. $S$ is a classical action, or
free energy functional, given by
\be
S[\psi] = \int d{\bm x}\ \left[r\,\psi^2 + c\,({\bm\nabla}\psi)^2
   + u\,\psi^4 \right],
\label{eq:3.10c}
\ee
and $\theta$ and ${\bm\zeta}$ are Langevin forces that obey
\bea
\langle\theta({\bm x},t)\,\theta({\bm x}',t')\rangle &=& -2\lambda{\bm\nabla}^2
   \delta({\bm x}-{\bm x}')\,\delta(t-t'),
\nonumber\\
\label{eq:3.10d}\\
\langle\zeta_{\alpha}({\bm x},t)\,\zeta_{\beta}({\bm x}',t')\rangle &=&
   -2\delta_{\alpha\beta}\,\eta{\bm\nabla}^2
   \delta({\bm x}-{\bm x}')\,\delta(t-t').
\nonumber\\
\label{eq:3.10e}
\eea
\ese

In the absence of shear, $s=0$, Eqs.\ (\ref{eqs:3.10}) describe Model H of
\textcite{Hohenberg_Halperin_1977}. It is a time-dependent Ginzburg-Landau
theory for the conserved order parameter $\psi$ coupled to the conserved
auxiliary field $\bm{u}$. Since the shear modes are soft, this coupling
influences the critical behavior. We will now discuss this influence, first in
equilibrium, and then for a system with $s\neq 0$.

\subsubsection{Critical dynamics in equilibrium fluids}
\label{subsubsec:III.B.1}

Let us return to the discussion of long-time tails in Sec.\ \ref{subsec:II.B}.
There, we considered corrections to the bare kinematic viscosity $\nu$ in the
GSI region far from criticality, and mentioned that analogous results hold for
other transport coefficients. Of particular interest for the critical dynamics
near the liquid-gas critical point is a contribution to the correction to the
thermal conductivity, $\delta\lambda$, from the coupling of the
transverse-velocity fluctuations to the entropy fluctuations described by Eqs.\
(\ref{eqs:3.10}) for $s=0$. Before performing the wave number integral, this
particular contribution, which we denote by $\delta\lambda_{\perp s}$, is
\cite{Kawasaki_1976, Hohenberg_Halperin_1977}
\be
\delta\lambda_{\perp s}({\bf k},t) = \frac{1}{\rho^2}\sum_{\bf p}\chi({\bf p})
   \sum_{i}({\hat{\bf k}}\cdot{\hat{\bf p}}_{\perp}^{(i)})^2\,
      e^{-(\nu{\bf p}_{-}^2 + D_T{\bf p}_{+}^2)\,t}\quad.
\label{eq:3.11}
\ee
Here $D_T = \lambda/\rho\,c_p$ is the thermal diffusivity in terms of $c_p$,
the specific heat at constant pressure. $\chi$ is the order parameter
susceptibility for the phase transition, where we have anticipated that near
the critical point we will need the momentum-dependent $\chi$, and ${\bf
p}_{\pm} = {\bf p}\pm{\bf k}/2$. Setting ${\bf k}=0$ and carrying out the
momentum integral leads to a $t^{-d/2}$ LTT. The correction to the thermal
conductivity is obtained by integrating $\delta\lambda (t)$ over all times, as
in the case of the kinematic viscosity.\footnote{\label{fn:49} These results
imply that for $d\leq 2$, conventional hydrodynamics does not exist. Indeed, it
is now known that for these dimensions the hydrodynamic equations are nonlocal
in space and time. For a discussion of this topic, see,
\textcite{Forster_Nelson_Stephen_1977} and references therein. In smectic
liquid crystal phases this effect is stronger, since some susceptibilities
behave, for certain directions in wave vector space, as $1/{\bf k}^4$,
amplifying the LTT effect. This causes a breakdown of local hydrodynamics for
all dimensions $d\leq 5$ \cite{Mazenko_Ramaswamy_Toner_1983}.}

By examining Eq.\ (\ref{eq:3.11}) one easily identifies a mechanism by which
the LTT effects can become even stronger. Consider a system with long-range
static correlations, for instance due to Goldstone modes, or due to the
vicinity of a continuous phase transition. In either case, some
susceptibilities, e.g. the $\chi$ in Eq.\ (\ref{eq:3.11}), become long-ranged,
amplifying the LTT effect. Before we discuss the realization of this scenario
in the vicinity of a phase transition we make one last point concerning the
LTTs. So far we have stressed the leading tails that decay as $t^{-d/2}$, but
there are numerous subleading LTTs as well. Most of them are uninteresting, but
one becomes important near the critical point, via the mechanism discussed in
the last paragraph. According to Eq.\ (\ref{eq:3.11}), a central quantity for
determining the critical contribution to $\lambda$ is the shear viscosity
$\eta$ (which enters $\nu$, see the definition of $\nu$ after Eq.\
(\ref{eq:2.24}).). It turns out that the contribution to $\eta$ that is
dominant near the critical point is a subleading LTT away from criticality. It
involves a coupling of two heat or entropy modes and is given by
\cite{Kawasaki_1976}
\bea
\delta\eta ({\bf k},t)&=&\frac{A}{{\bf k}^2}\sum_{\bf p}\chi ({\bf p})\,
   \chi ({\bf k}-{\bf p})\left(\frac{1}{\chi({\bf p})}-\frac{1}{\chi({\bf k}
      -{\bf p})}\right)^2
\nonumber\\
&&\times ({\hat{\bf k}}_{\perp}\cdot{\bf p})^2\,e^{-D_T[{\bf p}^2
       + ({\bf k}-{\bf p})^2)]t}\quad,
\label{eq:3.12}
\eea
with $A$ a constant. Away from the critical point, this LTT decays as
$t^{-(d/2+2)}$, so in fact it is a next-to-next leading LTT. Nevertheless, it
is the dominant mode-coupling contribution to $\eta$ near the critical point
because of the two factors of $\chi $ in the numerator of Eq.\ (\ref{eq:3.12}).

Near continuous phase transitions, fluctuations grow and ultimately diverge at
the critical point. For the liquid-gas critical point the order parameter is
the difference between the density and the critical density, and the divergent
fluctuations are the density fluctuations as described by the density
susceptibility. The susceptibility $\chi$ in Eq.\ (\ref{eq:3.12}) is
proportional to this divergent susceptibility. In the Ornstein-Zernike
approximation it is given by Eqs.\ (\ref{eqs:1.2}). Away from the critical
point, $r\neq 0$, $\chi$ decays exponentially in real space, see Eq.\
(\ref{eq:1.2b}).

Carrying out the time integral, the leading singular contribution to the
static, wave number-dependent thermal conductivity is
\cite{Hohenberg_Halperin_1977}
\be
\delta\lambda ({\bf k}) = \frac{1}{\rho^2}\sum_{\bf p}\chi ({\bf p})\,
   \frac{\sum_{i}({\hat{\bf k}}\cdot{\hat{\bf p}}_{\perp}^{(i)})^2}
      {\nu{\bf p}_{-}^2 + D_T{\bf p}_{+}^2}\,.
\label{eq:3.14}
\ee
Using Eq.\ (\ref{eq:1.2a}) in this equation we see that the homogeneous thermal
conductivity is infinite at the critical point for all $d\leq 4$, diverging as
$\vert r\vert^{-(4-d)/2}$. This is result of the amplification of the LTT by
the critical fluctuations.

By the same mechanism, we see that Eq.\ (\ref{eq:1.2a}) leads to a
logarithmically singular contribution to $\delta\eta$, if we take into account
that Eq.\ (\ref{eq:3.14}) implies that at the critical point, $D_T({\bf k})
\sim \vert{\bf k}\vert^{d-2}$.

\begin{figure}[t]
\includegraphics[width=7.0cm]{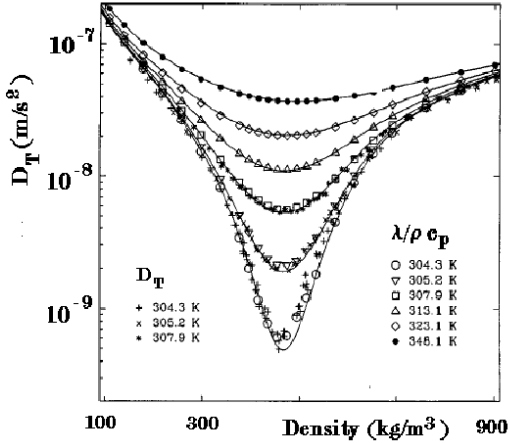}
\vskip 5mm
\caption{The thermal diffusivity $D_T = \lambda/\rho\,c_p$ of carbon dioxide
 in the critical region as a function of density at various temperatures
 ($T_{\rm c} = 304.12\,{\rm K}$). The symbols indicate experimental data for
 $D_T$ measured directly, and for $\lambda/\rho\,c_p$ deduced from
 thermal-conductivity data. The solid curves represent
 values calculated from the mode-coupling theory. After
 \protect\textcite{Luettmer-Strathmann_Sengers_Olchowy_1995}.}
 \vskip 0mm
\label{fig:12}
\end{figure}

The above conclusions result from one-loop calculations that use the
Ornstein-Zernike susceptibility. To go beyond these approximations it is
necessary to, (1) use the correct scaling form for the susceptibility, and, (2)
use either a self-consistent mode-coupling theory \cite{Kadanoff_Swift_1968} or
a RG approach\cite{Forster_Nelson_Stephen_1977, Hohenberg_Halperin_1977} to
improve on the one-loop approximation. The result is that the thermal
conductivity diverges as
\bse
\label{eqs:3.15}
\be
\lambda\propto\vert r\vert^{-0.57}\quad,
\label{eq:3.15a}
\ee
and that the thermal diffusivity vanishes as
\be
D_T\propto\vert r\vert^{0.67}\quad.
\label{eq:3.15b}
\ee
\ese
This is in good agreement with experimental results, as shown in Fig.\
\ref{fig:12}, which compares experimental and theoretical results for the
thermal diffusivity of carbon dioxide in the critical region.

All of the above results are obtained by considering the effects on the
dynamics (i.e, on transport coefficients) of the static critical behavior, as
expressed by the susceptibility $\chi$, which has been determined
independently. There is no feedback of the critical dynamics on the statics.
This is an example of the missing coupling between the statics and the dynamics
in classical equilibrium systems that we alluded to in the Introduction. As we
will see, this changes in a nonequilibrium situation.

\subsubsection{Critical dynamics in a fluid under shear}
\label{subsubsec:III.B.2}

Let us now consider Eqs.\ (\ref{eqs:3.10}) in the presence of shear, $s\neq 0$.
In this situation the entire fluid can still be at its critical temperature and
density. One therefore still expects a sharp phase transition (in an infinitely
large system), albeit a nonequilibrium one. This is in contrast to driving the
system out of equilibrium by means of, e.g., a temperature gradient. Inspecting
Eq.\ (\ref{eq:3.10a}), we make the following observations. (1) Shear will have
a tendency to make the order parameter susceptibility less soft. This is
obvious since the operator $y\,\partial/\partial x$ in Eq.\ (\ref{eq:3.10a}),
which scales like a mass by naive power counting, enters additively to the
diffusive operator ${\bm\nabla}^2$. In addition, the susceptibility will become
anisotropic. (2) $1/s$ sets a new time scale in the problem, which needs to be
compared to the relaxation time $\tau$. One therefore expects equilibrium
critical behavior in the weak-shear region $s\tau<1$, and a crossover to
different behavior in the strong-shear region $s\tau>1$. Alternatively, one can
define a characteristic wave number $k_s$ by $\tau(k_s) = 1/s$. The
strong-shear region is then given by $k_s\xi>1$, with $\xi$ the order parameter
correlation length.

It follows from point (1) above that the upper critical dimension cannot be
greater than the one in the equilibrium case, which is equal to $4$. At least
for $d>4$, the nonlinearities in the stochastic equations (\ref{eqs:3.10}) must
therefore be irrelevant. In the linearized equations, $\psi$ and $\bm{u}$
decouple, and the equation for the former becomes
\be
\frac{\partial\psi}{\partial t} = -s\,y\frac{\partial}{\partial x}\,\psi
   + \lambda\,{\bm\nabla}^2(r - {\bm\nabla}^2)\,\psi + \theta.
\label{eq:3.16}
\ee
The static order parameter susceptibility,
\be
\chi_{\psi}(\bm{k}) = \frac{1}{T}\int \frac{d\Omega}{2\pi}\int dt\
   e^{i\Omega t}\,\langle\psi_{\bm{k}}(t)\,\psi_{-\bm{k}}(0)\rangle,
\label{eq:3.17}
\ee
can be obtained from Eq.\ (\ref{eq:3.16}) Fourier transforming and using Eq.\
(\ref{eq:3.10d}). This yields the following differential equation for
$\chi_{\psi}$,
\be
\left[\lambda\,{\bm k}^2(r + {\bm k}^2) -
            \frac{1}{2}\,s\,k_x\,\frac{\partial}{\partial k_y}\right]
   \chi_{\psi}(\bm{k}) = \lambda\,{\bm k}^2.
\label{eq:3.18}
\ee
In the strong-shear region, the solution of this equation is adequately
represented by \cite{Onuki_Kawasaki_1979}
\bse
\label{eqs:3.19}
\be
\chi_{\psi}(\bm{k}) = 1/(r + {\text{const.}}\times\,k_s^{8/5}\vert
k_x\vert^{2/5} + {\bm k}^2),
\label{eq:3.19a}
\ee
with the constant of $O(1)$. This is the generalization of the Ornstein-Zernike
susceptibility, Eq.\ (\ref{eq:1.2a}), to a sheared system. (Remember that
$\chi$ in Eq.\ (\ref{eq:3.14}) is proportional to $\chi_{\psi}$.) As expected,
$\chi_{\psi}(\bm{k})$ is strongly anisotropic, and less soft than for $s=0$.
The nonequilibrium situation induces long-ranged static order parameter
correlations even away from criticality that manifest themselves in the
nonanalytic wave number dependence of $\chi_{\psi}$. In real space in $d=3$,
one finds power-law correlations at $r\neq 0$ \cite{Onuki_Kawasaki_1979},
\be
\chi_{\psi}(r>0,\vert{\bm x}\vert\to\infty)\propto 1/\vert x\vert^{7/5}
\label{eq:3.19b}
\ee
\ese
for $\vert {\bm x}_{\perp}\vert \ll k_s^{-1} \ll \vert x\vert$, where ${\bm x}
= {\bm x}_{\perp} + x\,{\hat{\bm e}}_x$. For $\vert{\bm x}_{\perp}\vert \gg
\vert x\vert$, on the other hand, $\chi_{\psi}$ decays exponentially. That is,
$\chi _{\psi }$ exhibits GSI (and extreme anisotropy). At criticality, $r=0$,
$\chi_{\psi}$ is of even longer range \cite{Onuki_Kawasaki_1979}. If we use Eq.
(\ref{eq:3.19a}) for $r=0$ in Eq.\ (\ref{eq:3.14}), we find
\bea
\delta\lambda(\bm{k}=0) &\propto& \int_0^{k_s} dp\ p^{\,d-3}\,k_s^{-8/5}\,\vert
     p_x\vert^{-2/5}
\nonumber\\
&\propto& k_s^{-8/5}\int_0^{k_s} dp\ p^{\,d-17/5}.
\label{eq:3.20}
\eea
In contrast to the equilibrium situation, where $\delta\lambda(\bm{k}=0)$
diverged for all $d\leq 4$, we see that in the presence of shear the divergence
occurs only for $d\leq 12/5$.

The above result suggests that the critical behavior of the thermal
conductivity $\lambda$ is mean-field like, i.e., given by the time-dependent
Ginzburg-Landau theory, for $d>12/5$, and in particular in $d=3$. The physical
reason is that the long-range order parameter correlations stabilize the
mean-field critical behavior. It is not obvious that this conclusion is
correct, though, as it has been derived by using the approximation
(\ref{eq:3.19a}), which neglected the nonlinearities in the equations of
motion. \textcite{Onuki_Kawasaki_1979} have performed a RG analysis which shows
that the simple argument given above is indeed correct. In the presence of
shear, the upper critical dimensionality is $d_c^{\,+} = 12/5$, and for $d >
d_c^{\,+}$ there is a new simple critical fixed point where $\chi_{\psi}$ and
$\delta\lambda$ are given by Eqs.\ (\ref{eq:3.19a}) and (\ref{eq:3.20}),
respectively. The RG treatment shows that the flow equations for $\chi_{\psi}$
and the transport coefficients are coupled. This means that, contrary to the
equilibrium case, it is {\em not} possible to solve for the static critical
behavior independently of the dynamics. \textcite{Onuki_Kawasaki_1979} also
considered the equation of state, which they found to be of mean-field form for
$d>12/5$. In particular, the critical exponent $\beta$ has its mean-field value
$\beta=1/2$.

\begin{figure}[t]
\includegraphics[width=7.0cm]{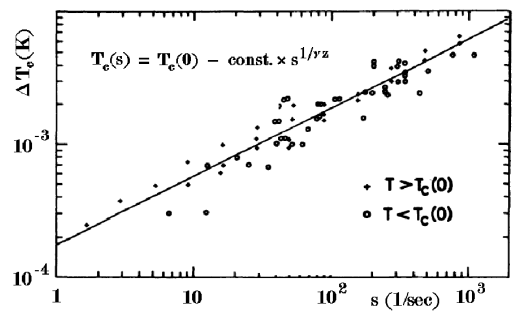}
\vskip 5mm
\caption{Critical temperature versus shear in the binary fluid
 cyclohexane-aniline. Symbols represent the experimental data, and the solid
 line corresponds to the theoretical exponent $1/\nu z \approx 0.53$. After
 \protect\textcite{Beysens_Gbadamassi_1980}.}
 \vskip 0mm
\label{fig:13}
\end{figure}

Another theoretical prediction regards the suppression of the critical
temperature by the shear. Consider the inverse correlation length, $\xi^{-1}$,
as a function of $r$ and $s$. Since $s$ is an inverse time, it is expected to
scale with the dynamical critical exponent $z$. Dynamical scaling then predicts
the relation
\bea
\xi^{-1}(r,s) &=& b^{-1}\xi^{-1}(r\,b^{1/\nu},s\,b^z)
\nonumber\\
              &=& r^{\nu}\xi^{-1}(1,s/r^{\nu z}),
\label{eq:3.21}
\eea
with $b$ an arbitrary scale factor, and $\nu$ and $z$ the correlation length
exponent and the dynamical critical exponent, respectively, at the equilibrium
transition. For $s\neq 0$, the inverse correlation length thus vanishes not at
$r=0$, but rather at $r={\text{const.}}\times s^{1/\nu z}$. The shear
dependence of the critical temperature, for small shear, is therefore
\be
T_{\text{c}}(s) = T_{\text{c}}(0) - \text{const.}\times s^{1/\nu z}.
\label{eq:3.22}
\ee
At the equilibrium transition, $\nu\approx 0.63$, and $z\approx 3$ in $d=3$
\cite{Hohenberg_Halperin_1977}. Many of these results have been confirmed
experimentally in light-scattering experiments on binary fluids
\cite{Beysens_Gbadamassi_1980}. For the $T_{\text{c}}$-suppression, the
experimental results are shown in Fig.\ \ref{fig:13} together with the
theoretical prediction.

The classical nonequilibrium phase transition described above is very much
analogous to the equilibrium quantum phase transition in a ferromagnet with
quenched disorder, which we will discuss in Sec.\ \ref{subsec:IV.B}. In that
case, too, the order parameter susceptibility reflects long-ranged
correlations, although they are due to generic soft modes rather than a
nonequilibrium situation. This in turn leads to the stabilization of a simple
Gaussian critical fixed point, much like in the nonequilibrium classical fluid.
An important difference is that in the quantum ferromagnetic case, the equation
of state is {\em not} mean-field like, see Eq.\ (\ref{eq:4.24b}) below. The
reason for this difference is not known.

\section{Influence of Generic Scale Invariance on Quantum Critical Behavior}
\label{sec:IV}

We now turn to quantum systems. As we have discussed before, the only
qualitative difference compared to the classical case is the coupling between
the statics and the dynamics in quantum mechanics, which leads to, e.g.,
long-time tails affecting the critical behavior of thermodynamic quantities
even in equilibrium. Otherwise, the phenomena are qualitatively very much
analogous. However, there is an important quantitative difference. In any
itinerant electron system, there are generic soft modes, viz., the
particle-hole excitations that can be understood as Goldstone modes as
explained in Sec.\ \ref{subsubsec:II.B.2}, which are soft {\em only} at zero
temperature. Our first example for which these particle-hole excitations play
the role of the generic soft modes is the case of the ferromagnetic transition
in metals at $T=0$. We will treat clean and disordered systems separately,
since they turn out to behave quite differently, and are qualitatively
remarkably similar to the classical nematic--smectic-A transition discussed in
Sec.\ \ref{subsec:III.A}, and to the critical dynamics of a classical fluid
under shear, Sec. \ref{subsubsec:III.B.2}, respectively. Another example is the
transition from a normal metal to a BCS superconductor $T=0$, which we cover in
Sec. \ref{subsec:IV.C}. The case of the quantum antiferromagnetic transition,
which is very different and at this point rather incompletely understood, is
discussed in Sec.\ \ref{subsec:IV.D}.

\subsection{Quantum Ferromagnetic Transition in Clean Systems}
\label{subsec:IV.A}

Let us consider a clean itinerant quantum Heisenberg ferromagnet as our first
example of a quantum phase transition that is influenced by generic soft modes.
In Sec.\ \ref{subsec:IV.B} we will discuss the effects of quenched disorder and
also give some additional technical details.

Our exposition does not follow historical lines. The first detailed theory of
the quantum ferromagnetic transition was given in an influential paper by
\textcite{Hertz_1976}, who derived an effective action from a microscopic model
and concluded that the critical behavior is mean-field like in all dimensions
$d>1$. As we will show, this conclusion was the result of an approximation that
ignores all but the most basic aspects of the generic soft modes.

\subsubsection{Soft-mode action for itinerant quantum ferromagnets}
\label{subsubsec:IV.A.1}

The order parameter is the fluctuating magnetization field ${\bm M}(x)$ (with
$x = ({\bm x},\tau)$ the space--imaginary-time coordinate as in Sec.\
\ref{subsubsec:II.B.2}) whose expectation value is proportional to the
magnetization $m$. The generic soft modes in this system are the diffusive
particle-hole excitations. They were parameterized in Sec.\
\ref{subsubsec:II.B.2} in terms of the soft sector $q$ of the matrix field $Q$,
see Eqs.\ (\ref{eq:2.46}, \ref{eq:2.47}), with the corresponding propagator
given by Eq.\ (\ref{eq:2.48c}). The coupling between the two is provided by the
fact that the magnetization couples linearly to the spin density, which can be
expressed in terms of the $Q$. The action will thus consist of a part that
depends only on the magnetization, a part that depends only on the generic soft
modes, and a coupling between the two,
\be
{\cal A}[\bm{M},Q] = {\cal A}_M + {\cal A}_q + {\cal A}_{M,q}
\label{eq:4.1}
\ee
${\cal A}_M$ is a static, local, LGW functional for the magnetization
fluctuations. It can be chosen static because, as we will see in Sec.\
\ref{subsubsec:IV.A.2} below, the low-frequency dynamical part will be
generated by the coupling to the ballistic modes. We thus write
\be
{\cal A}_M[\bm{M}] = \int dx\ \left[{\bm M}(x)\,(r_0 - c{\bm\nabla}^2)\,{\bm
M}(x) + u_0\,{\bm M}^4(x)\right].
\label{eq:4.2}
\ee
${\cal A}_q$ must yield the ballistic propagator of the soft modes, Eqs.\
(\ref{eqs:2.48}). The Gaussian part of the fermionic action will therefore have
the form
\bse
\label{eqs:4.3}
\be
{\cal A}_{q}^{(2)} \hskip -1pt = \hskip -1pt \frac{-4}{G}\!\int\!\hskip -1pt
       d{\bm x}\,d{\bm y}\!\hskip -1pt\sum_{1,2,3,4} \sum_{r,i}
         {_r^i q}_{12}({\bm x})\,{^i\Gamma}_{12,34}^{(2)}({\bm x}-{\bm y})\,
          {_r^i q}_{34}({\bm y}).
\label{eq:4.3a}
\ee
The particle-particle degrees of freedom are irrelevant for this problem, and
we can therefore restrict the sum over the index $r$ to $r=0,3$. The vertex
function $\Gamma^{(2)}$ is most easily written in momentum space,
\bea
{^i\Gamma}_{12,34}^{(2)}({\bm k}) &=& \delta_{13}\,\delta_{24}\,
    \Gamma^{(2,0)}_{12}({\bm k})
    + \delta_{1-3,2-4}\,\delta_{i0}\,2\pi T G K_{\text{s}}
\nonumber\\
&& +\,\delta_{1-3,2-4}\,(1-\delta_{i0})\,2\pi T G {\tilde K}_{\text{t}}.
\label{eq:4.3b}
\eea
with
\be
\Gamma^{(2,0)}_{12}({\bm k}) = \vert{\bm k}\vert + GH\Omega_{1-2}
   \quad.
\label{eq:4.3c}
\ee
\ese
The inverse of $\Gamma^{(2,0)}$ yields the noninteracting propagator, Eqs.\
(\ref{eq:2.48a}, \ref{eq:2.48c}), and the inverse of $\Gamma^{(2)}$, its
generalization to interacting systems. $K_{\text{s}}$ is the spin-singlet
interaction amplitude defined after Eq.\ (\ref{eq:2.48c}), and ${\tilde
K}_{\text{t}}$ is a `residual' spin-triplet interaction.\footnote{\label{fn:50}
One might argue that the spin-triplet degrees of freedom are included in
${\bm{M}}$, so also including them in ${\cal A}_q$ constitutes double counting.
To see that this is not true, imagine deriving the effective action from a
microscopic model, e.g., Eqs. (\ref{eqs:2.37}-\ref{eqs:2.39}). This introduces
${\bm{M}}$ by decoupling the spin-triplet interaction term in Eq.\
(\ref{eq:2.39a}) by means of a Hubbard-Stratonovich transformation
\cite{Hubbard_1959, Stratonovich_1957}, which leaves spin-triplet degrees of
freedom in the noninteracting part of the action $S_0$, and hence in ${\cal
A}_q^{(2)}$.\\ A weaker version of the same objection is that ${\tilde
K}_{\text{t}}$ should be zero, since it represents the interaction that has
been decoupled. However, this argument ignores the fact that a spin-triplet
interaction will be generated from the spin-singlet one under renormalization,
even if there is none in the bare action. The only restriction on ${\tilde
K}_{\text{t}}$ is therefore that it must not be so large as to induce a
ferromagnetic instability in the electronic ``reference system'' described by
${\cal A}_q$ in the absence of a coupling to ${\bm M}$.}

${\cal A}_{M,q}$ originates from a term ${\cal A}_{M-Q}$ that couples ${\bm M}$
and $Q$. Such a term must be present since in the presence of a magnetization
the fermionic spin density will couple linearly to it. Using Eq.\
(\ref{eq:2.44}) to express the spin density in terms of $Q$, we thus obtain
\bse
\label{eqs:4.4}
\bea
{\cal A}_{M-Q} &=& 2c_1\sqrt{T}\int d{\bm x}\sum_n\sum_{i=1}^3
             M_n^{i}({\bm x})
\nonumber\\
&&\times\sum_{r=0,3}(-1)^{r/2}\sum_{m}\tr[(\tau_{r}\otimes s_{i})\,
   Q_{m,m+n}({\bm x})]\ ,
\nonumber\\
\label{eq:4.4a}
\eea
with a model-dependent coefficient $c_1$. Defining a symmetrized magnetization
field by
\be
b_{12}({\bm x}) = \sum_{i,r}\left(\tau_r\otimes s_i\right)\,
                  {^i_rb}_{12}({\bm x})\quad,
\label{eq:4.4b}
\ee
with components
\be
{_r^i b}_{12}({\bm x}) = (-)^{r/2}\sum_{n}\delta_{n,n_1-n_2}\left[
   M_n^i({\bm x})
- (-)^{r} M_{-n}^i({\bm x})\right],
\label{eq:4.4c}
\ee
allows to rewrite Eq.\ (\ref{eq:4.4a}) in a more compact form,
\be
{\cal A}_{M-Q} = c_1\sqrt{T}\int d{\bm x}\
                 \tr\left(b({\bm x})\,Q({\bm x})\right)\quad.
\label{eq:4.4d}
\ee
\ese
Using Eq.\ (\ref{eq:2.46}) in Eq.\ (\ref{eq:4.4a}) or (\ref{eq:4.4d}), and
integrating out the massive $P$-fluctuations, obviously leads to a series of
terms coupling ${\bm M}$ and $q$, ${\bm M}$ and $q^2$, etc. We thus obtain
${\cal A}_{M,q}$ in form of a series
\bse
\label{eqs:4.5}
\be
{\cal A}_{M,q} = {\cal A}_{M-q} + {\cal A}_{M-q^2} + \ldots
\label{eq:4.5a}
\ee
The first term in this series is obtained by just replacing $Q$ by $q$ and
$q^{\dagger}$, respectively, in Eq.\ (\ref{eq:4.4d}),
\be\
A_{M-q} = 8c_1T^{1/2}\sum_{12}\int d{\bm x}\sum_r\sum_{i=1}^3
            {_r^i b}_{12}({\bm x})\,{_r^i q}_{12}({\bm x}).
\label{eq:4.5b}
\ee
The next term in this expansion has an overall structure
\be
{\cal A}_{M-q^2} \propto c_2\sqrt{T} \int d{\bm x}\ \tr \left(b({\bm
x})\,q({\bm x})\,
   q^{\dagger}({\bm x})\right),
\label{eq:4.5c}
\ee
\ese
with $c_2$ another positive constant. The detailed structure
\cite{Kirkpatrick_Belitz_2002} can be obtained by integrating out the
$P$-fluctuations in tree approximation, in analogy to the derivation of the
nonlinear $\sigma$ model in the disordered case \cite{Wegner_Schaefer_1980,
McKane_Stone_1981}. Terms of higher order in $q$ in this expansion will turn
out to be irrelevant for determining the behavior at the quantum phase
transition.

Let us pause here and compare our soft-mode action with the one for the
classical liquid-crystal transition in Sec.\ \ref{subsubsec:III.A.1}. The
order-parameter part of the action, Eqs.\ (\ref{eq:3.3a}) and (\ref{eq:4.2}),
respectively, is a $\phi^4$-theory in either case, and the generic soft modes
are described by a Gaussian action, Eqs.\ (\ref{eq:3.3b}) and (\ref{eq:4.3a}),
respectively. Finally, in either case there is a direct coupling between the
order parameter and the generic soft modes. In the case of the liquid crystal,
this coupling is between the square of the order parameter and the square of
the soft-mode field, while in the case of the magnet the order parameter
couples linearly to all powers of the soft-mode field, but this does not have
major physical consequences, as we will see. Apart from this, the main
difference is that in the quantum case the fields depend on time or frequency
in addition to position or wave vector, which leads to a more complicated
detailed structure of the terms in the action, the coupling terms in
particular. Given these structural similarities, it is natural to analyze the
quantum ferromagnetic transition in analogy to the liquid crystal one, Secs.\
\ref{subsubsec:III.A.2} and \ref{subsubsec:III.A.3}.

\subsubsection{Gaussian approximation}
\label{subsubsec:IV.A.2}

A strict Gaussian approximation neglects the term $A_{M-q^2}$, Eq.\
(\ref{eq:4.5c}). One can then integrate out the intrinsic soft modes $q$ to
obtain an action \cite{Hertz_1976}
\be
{\cal A}_{\text{H}}[{\bm M}] = \sum_k {\bm M}(k)\,{\cal M}_n^{-1}({\bm
k})\,{\bm M}(-k) + O({\bm M}^4),
\label{eq:4.6}
\ee
where $k = ({\bm k},\Omega_n)$ is a momentum-frequency four-vector, $d =
4Gc_1^2/\pi$, and ${\cal M}^{-1}$ is the inverse of the paramagnon propagator
\bse
\label{eqs:4.7}
\be
{\cal M}_n({\bm k}) = \frac{1}{r_0 + c{\bm k}^2 +
d\vert\Omega_n\vert/\left(\vert{\bm k}\vert + GH\vert\Omega_n\vert\right)}\ .
\label{eq:4.7a}
\ee
In the context of the full action, ${\cal M}$ gives the Gaussian ${\bm M}$
propagator,
\be
\langle M_n^i({\bm k})\,M_m^j({\bm p})\rangle = \delta_{{\bm k},-{\bm p}}\,
   \delta_{n,-m}\,\delta_{ij}\,\frac{1}{2}\,{\cal M}_n({\bm k}).
\label{eq:4.7b}
\ee
\ese
${\cal M}$ and the fermionic propagator ${\cal D}$, Eq.\ (\ref{eq:2.48c}), with
the latter suitably generalized to allow for the interaction amplitudes
$K_{\text{s}}$ and ${\tilde K}_{\text{t}}$, are the Gaussian propagators of the
coupled field theory.\footnote{\label{fn:51} There also are mixed propagators
$\langle b\,q\rangle$, but they do not enter any diagrams that are important
for the phase transition.} The $\vert\Omega_n\vert/\vert{\bm k}\vert$ structure
of the dynamical piece of ${\cal M}$ is characteristic of the itinerant
electron degrees of freedom that couple to the magnetic ones; it also manifests
itself, e.g., in the frequency dependence of the Lindhard function (see, e.g.,
\onlinecite{Pines_Nozieres_1989}). Technically, it results from the inverse of
the vertex $\Gamma^{(2)}$, Eq.\ (\ref{eq:4.3b}), which gets multiplied by a
frequency due to frequency restrictions inherent in the definition of the $q$.
This illustrates how the coupling to the generic soft modes generates the
dynamics of the order parameter, see the remark above Eq.\ (\ref{eq:4.2}).

${\cal A}_{\text{H}}$ is Hertz's action, which predicts a continuous transition
with mean-field critical behavior for all dimensions $d>1$. From our experience
with the analogous classical transitions in Sec.\ \ref{subsec:III.A} above we
suspect that the Gaussian approximation yields qualitatively incorrect results.
Indeed, internal inconsistencies of Hertz's results have been discussed by
\textcite{Sachdev_1994} and \textcite{Dzero_Gorkov_2003}. More explicitly, the
instability of Hertz's fixed point can be shown formally by means of arguments
analogous to those we will present in Sec.\ \ref{par:IV.B.4.b} below for the
case of disordered magnets. In what follows, we show that the transition can be
either of first order, or of second order with non-mean-field critical
behavior.

\subsubsection{Renormalized mean-field theory}
\label{subsubsec:IV.A.3}

As a first step to improve upon the Gaussian approximation, it is natural to
construct a renormalized mean-field theory in analogy to Sec.\
\ref{subsubsec:III.A.3}. The relevant length scales are, the magnetic coherence
length
\bse
\label{eqs:4.8}
\be
\xi = \sqrt{c/\vert r\vert},
\label{eq:4.8a}
\ee
and a length
\be
\lambda = 1/G\,H^2\,c_2\,\sqrt{T}\langle M_n^3(\bm{x})\rangle
        = 1/\Lambda\,c_2\,\,m,
\label{eq:4.8b}
\ee
with $m$ is the magnetization. $\Lambda = G\sqrt{\pi H^3/8}$ depends on
microscopic length and energy scales only and ensures that $m$ is dimensionally
an inverse volume. In a magnetic phase, $\lambda$ determines the mass in the
transverse spin-triplet $q$-propagator, just like the London penetration depth
determines the mass of the transverse photon in a superconductor, or the mass
of the director fluctuations in a smectic-A phase, see Eq.\ (\ref{eq:3.6}). The
Ginzburg-Landau parameter,
\be
\kappa = \lambda/\xi \propto \sqrt{u}/c_2\sqrt{c},
\label{eq:4.8c}
\ee
\ese
is again independent of $m$ or $\vert r\vert$.

Let us now neglect the fluctuation of the magnetic order parameter, i.e., we
put
\be
M_n^i({\bm x}) \approx \delta_{i3}\,\delta_{n0}\,m\,\Lambda/G\,H^2\,\sqrt{T}.
\label{eq:4.9}
\ee
In the limit $\lambda\ll\xi$ this approximation becomes exact; in general, its
validity will need to be investigated. $q$ can then be integrated out, in exact
analogy to the treatment of $\bm{A}$ in Sec.\ \ref{subsubsec:III.A.3}. The
result for the free energy density $f$, in a magnetic field $h$, and with $f_0
= f(m=0)$, is
\bse
\label{eqs:4.10}
\bea
f &=& f_0 + r_0\,m^2 + u_0\,m^4 - h\,m
\nonumber\\
&&      + \frac{2}{V}\sum_{{\bm k} < \Lambda}
      T\sum_{n}\ln N({\bm k},\Omega_n;m).
\label{eq:4.10a}
\eea
$\Lambda$ is an ultraviolet momentum cutoff, and
\bea
N({\bm k},\Omega_n;m) &=& 16\,c_2^2\,G^4{\tilde K}_{\text{t}}^2\, m^2\,
                          \Omega_n^2\hskip 50pt
\nonumber\\
&&\hskip -70pt + \left(\vert{\bm k}\vert + GH\Omega_n\right)^2
     \left[\vert{\bm k}\vert + G(H + {\tilde K}_{\text{t}})\Omega_n\right]^2.
\label{eq:4.10b}
\eea
\ese
Minimizing $f$ with respect to the magnetization gives the equation of state.

The integral in Eq.\ (\ref{eq:4.10a}) has been analyzed by
\textcite{Kirkpatrick_Belitz_2002}. In $d=3$, the equation of state was found
to take the form
\bse
\label{eqs:4.11}
\bea
h &=& 2\,r\,m + 4\,v\,m^3\ln (m^2 + T^2)
\nonumber\\
&& +\, m^3\,\left(4u_0 + 2v\,\frac{m^2}{m^2 + T^2}\right).
\label{eq:4.11a}
\eea
For the free energy density at $T=h=0$ this implies
\cite{Belitz_Kirkpatrick_Vojta_1999}
\be
f = f_0 + r\,m^2 + v_3\,m^4\ln m^2 + u\,m^4
\label{eq:4.11b}
\ee
in $d=3$, and
\be
f = f_0 + r\,m^2 - v_d\,m^{d+1} + u\,m^4
\label{eq:4.11c}
\ee
\ese
in generic dimensions. In these equations, $f$, $m$, and $T$ are measured in
terms of suitable microscopic quantities such that $r$, $v_d$, and $u$ are all
dimensionless. $r$ and $u$ are given by $r_0$ and $u_0$, respectively, plus
additive renormalizations from the soft modes, in analogy to the classical
example of renormalized mean-field theory in Sec.\ \ref{subsubsec:III.A.3}.
$v_d>0$ is quadratic in $c_1^2$, so in strongly correlated systems $v_d$ is
larger than in weakly correlated ones. A comparison with Eq.\ (\ref{eq:3.8b})
shows that the free energy for the quantum ferromagnet in $d=3$ is precisely
analogous to that of the classical liquid crystal in $d=4$ in the same
approximation.

This free energy functional is nonanalytic in $m$, in analogy to Eqs.\
(\ref{eqs:3.8}) in our classical example, and the same discussion applies.
Namely, the nonanalyticity reflects GSI in the paramagnetic phase, and it is
directly related to the nonanalytic wave number dependence of the spin
susceptibility, Eq.\ (\ref{eq:2.36b}). In real space, the latter corresponds to
\be
\chi_{\text{s}}(r>0,\vert{\bm x}\vert\to\infty )\propto 1/\vert{\bm
x}\vert^{2d-1}.
\label{eq:4.12}
\ee
This manifestation of GSI is the ultimate physical reason behind the failure of
Hertz theory to correctly describe the quantum critical behavior.

Note that the derivation of the above mean-field result for the free energy,
Eqs.\ (\ref{eq:4.11b}, \ref{eq:4.11c}), implicity assumes that the nonanalytic
wave number dependence of $\chi_{\text{s}}$ given by Eq.\ (\ref{eq:2.36b}) or
Eq.\ (\ref{eq:4.12}), which is exact in the paramagnetic phase, continues to
hold in the critical region. This is a nontrivial assumption. In Sec.\
\ref{subsubsec:IV.A.4} below we will discuss how critical fluctuations can
modify this result. This point has recently been addressed by
\textcite{Chubukov_Pepin_Rech_2003}.

The phase diagram predicted by these equations in $d=3$ is shown, in a more
general context, in the first panel of Fig.\ \ref{fig:19} below. There is a
tricritical point at
\bse
\label{eqs:4.13}
\be
T = T_{\text{tc}} = e^{-u/2v}.
\label{eq:4.13a}
\ee
At $T=0$, there is a first-order phase transition at $r = r_1$, with the
magnetization changing discontinuously from zero to a value $m_1$. One finds
\be
r_1 = v\,m_1^{\ 2}\quad,\quad m_1 = e^{-(1+u/v)/2},
\label{eq:4.13b}
\ee
\ese

In $d=2$, there is no finite-temperature magnetic phase transition. However, at
zero temperature there is a QPT, which is predicted by the Eqs.\
(\ref{eqs:4.10}) to be discontinuous. In $d>3$ the nonanalytic terms produced
by the soft modes are subleading, and the transition is described by ordinary
mean-field theory. The generalized mean-field theory thus suggests an upper
critical dimension $d_{\text{c}}^{\,+}=3$. As we will see in the next
subsection, a more sophisticated analysis confirms this result.

\subsubsection{Effect of order-parameter fluctuations}
\label{subsubsec:IV.A.4}

The renormalized mean-field theory predicts that the quantum ferromagnetic
transition in clean systems is {\em always} of first order. As we have
discussed above, this conclusion is certainly valid in the limit
$\lambda\ll\xi$. In general, however, the order-parameter fluctuations need to
be taken into account. This can be done by a systematic RG analysis of the
action, Eq.\ (\ref{eq:4.1}). In contrast to the liquid crystal case, it turns
out that a one-loop calculation predicts, under certain conditions, a critical
fixed point that corresponds to a second-order transition.

The fixed point found by \textcite{Kirkpatrick_Belitz_2002} has the property
that $G$ is marginal. To one-loop order, in $d=3$, and for $r=0$, the coupling
constants $u$, $H$, $c_2$, and $c$ obey the flow
equations\footnote{\label{fn:52} At this level of detail of the discussion it
is not obvious that there are two time scales in the problem, one related to
the order-parameter fluctuations, and the other to the fermionic degrees of
freedom. As a consequence, $c_2$, which couples the two, does not have a unique
scale dimension and can be either marginal or irrelevant, depending on the
context. The flow equation (\ref{eq:4.14c}) applies to its irrelevant
incarnation. We will discuss this point in more detail in Sec.\
\ref{subsubsec:IV.B.4} below.}
\bse
\label{eqs:4.14}
\bea
\frac{du}{d\ln b} &=& -2u - A_u\,c_2^2/H,
\label{eq:4.14a}\\
\frac{dH}{d\ln b} &=& A_H/c\quad,\quad \frac{dc}{d\ln b} = -A_c/H,
\label{eq:4.14b}\\
\frac{dc_2}{d\ln b} &=& -c_2,
\label{eq:4.14c}
\eea
\ese
with $b$ the RG length rescaling factor, and $A_u$, $A_H$, and $A_c$ positive
constants. In a purely perturbative treatment, $c_2$ and $H$ on the right-hand
side of Eq.\ (\ref{eq:4.14a}) are constants, and $u$ inevitably becomes
negative at large scales ($b\to\infty$). A negative $u$ is usually interpreted
as signalizing a first-order transition;\footnote{\label{fn:53} While this
interpretation of $u$ flowing to negative values is not always correct, in the
current case it is bolstered by the renormalized mean-field theory.} this is
the RG version of the conclusion that the transition is always of first order.
However, this perturbative argument is not consistent since, at the same level
of the analysis, Eq.\ (\ref{eq:4.14b}) predicts a singular specific-heat
coefficient $H$, which couples back into the coefficients of the Landau theory.
This feedback is taken in to account by solving the one-loop equations
(\ref{eqs:4.14}) self-consistently, whereby the one-loop correction to the
$u$-flow equation decreases with increasing scale, due a combination of $c_2$
being irrelevant and $H$ increasing under renormalization. The conclusion that
$u$ flows to negative values is therefore no longer inevitable. Indeed, a
solution of the above flow equations \cite{Kirkpatrick_Belitz_2002} shows that
\bea
u(b\to\infty) &=& \left[u_0 - (A_u/A_c)(c_{2,0})^2\,c_0\right]\,b^{-2}
\nonumber\\
&\propto& (\kappa^2 - \kappa_{\text{c}}^2)\,b^{-2}.
\label{eq:4.15}
\eea
with $u_0 = u(b=1)$, etc., the bare coupling constants, $\kappa$ the
Ginzburg-Landau parameter from Eq.\ (\ref{eq:4.8c}), and $\kappa_{\text{c}}
\propto \sqrt{A_u/A_c}$. We see that the RG analysis predicts an asymptotically
negative value of $u$, and hence a fluctuation-induced first-order transition,
for small values of the Ginzburg-Landau parameter, $\kappa <
\kappa_{\text{c}}$, in agreement with the renormalized mean-field theory, and
in analogy with the liquid-crystal case. For $\kappa > \kappa_{\text{c}}$,
however, $u$ stays positive and one finds a critical fixed point, in contrast
to the liquid-crystal case. As mentioned in Sec.\ \ref{subsubsec:III.A.4}
above, the mechanism for this ``fluctuation-induced second-order transition''
is very similar to the one discussed by \textcite{Fucito_Parisi_1981} for a
classical system. The critical behavior at this transition will be summarized
in the next section.

\begin{figure}[t]
\includegraphics[width=6cm]{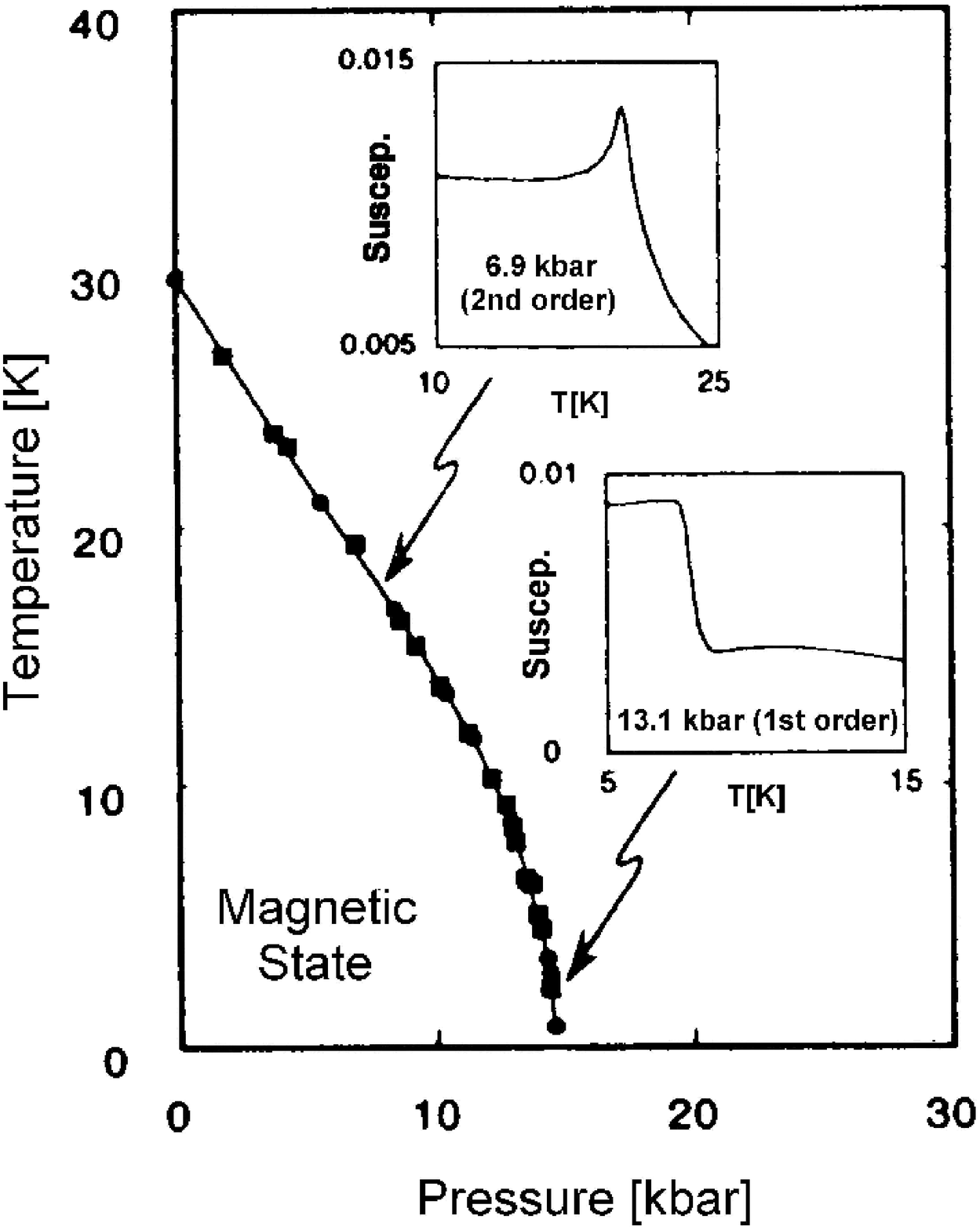}
\caption{\label{fig:13a} Phase diagram of MnSi as measured by
   \protect{\textcite{Pfleiderer_et_al_1997}}. The insets show the behavior of
   the susceptibility close to the transition point. From
   \protect{\textcite{us_hh_fm_1999}}.}
\end{figure}

\begin{figure}[t]
\includegraphics[width=8cm]{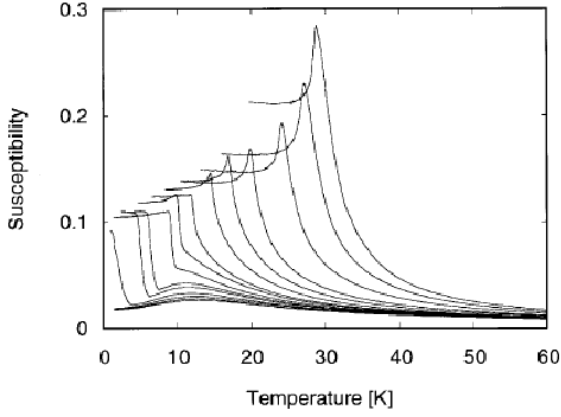}
\caption{\label{fig:14} Magnetic susceptibility (in SI units) versus
   temperature for MnSi at
   pressures corresponding to the data points in Fig.\ \ref{fig:13a}. Pressure
   values for the curves from right to left are, 1.80, 3.80, 6.90, 8.60, 10.15,
   11.25, 12.15, 13.45, 13.90, 14.45, 15.20, 15.70, and 16.10 kbar. The change
   from a sharply peaked susceptibility consistent with a second order
   transition to a discontinuous one as expected for a first-order transition
   is apparent. From \protect{\textcite{Pfleiderer_et_al_1997}}.
   }
\end{figure}

\begin{figure}[t]
\includegraphics[width=8cm]{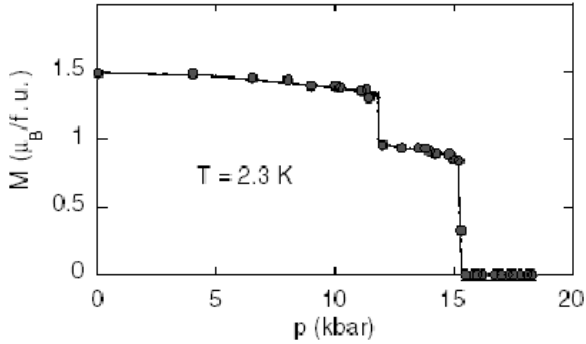}
\caption{\label{fig:15} Magnetization versus hydrostatic pressure at
$T=2.3{\text{K}}$ for UGe$_2$. The intial first-order transition at $p\approx
15 {\text{kbar}}$ is followed by a metamagnetic transition at $p\approx 12
{\text{kbar}}$. After \protect{\textcite{Pfleiderer_Huxley_2002}}.
   }
\end{figure}

There is experimental evidence that the quantum ferromagnetic transition in
clean systems is of first order in some materials, but continuous in others,
consistent with the theoretical picture given above.
\textcite{Pfleiderer_et_al_1997} have reported a continuous transition in MnSi
at moderate hydrostatic pressures corresponding to relatively high values of
$T_c$, while the transition becomes first order if $T_c$ is driven to low
values by higher values of the pressure, see Figs.\ \ref{fig:13a} and
\ref{fig:14}.\footnote{\label{fn:54} Strictly speaking, MnSi is not a
ferromagnet, but rather a weak helimagnet. Local remnants of the helical order
have been speculated to be responsible for the non-Fermi-liquid properties
observed in the paramagnetic phase \cite{Doiron-Leyraud_et_al_2003,
Pfleiderer_et_al_2004}. The influence of the helical order on the quantum
critical behavior has been studied by \textcite{Vojta_Sknepnek_2001}.}
Qualitatively the same magnetic phase diagram has been observed in UGe$_2$
\cite{Huxley_et_al_2001}, see Fig.\ \ref{fig:15}.
In ZrZn$_2$, on the other hand, the magnetic transition is of second order down
to the lowest temperatures observed \cite{Pfleiderer_et_al_2001}. The latter
two materials are of particular interest since they also have a phase of
coexistent ferromagnetism and superconductivity. The superconductivity is very
vulnerable to disorder, which testifies to the fact that the samples in
question are very clean. Another material where a continuous transition is
observed down to very low temperatures is Ni$_x$Pd$_{1-x}$
\cite{Nicklas_et_al_1999}. While the substitutional nature of this system
inevitably introduces some disorder, the quantum phase transition occurs at low
Ni concentration ($x\approx 0.02$), and disorder is believed to play a minor
role at the transition.

While the soft-mode theory is consistent with these observations, it is not the
only possible explanation. Another obvious possibility are band-structure
effects, which can change the sign of the coefficient $u$ in an ordinary Landau
theory, Eq.\ (\ref{eq:4.10a}) without the last term, and thus lead to
first-order transitions in some materials, and second-order transitions in
others. Indeed, band structure effects have been proposed to be the source of
the first-order transition, the superconductivity, and an observed metamagnetic
transition within the ferromagnetic phase
\cite{Sandeman_Lonzarich_Schofield_2003, Pfleiderer_Huxley_2002}.

\subsubsection{Critical behavior at the continuous transition}
\label{subsubsec:IV.A.5}

\paragraph{Critical behavior in $d=3$}
\label{par:IV.A.5.a}

A solution of the RG flow equations for the case $\kappa > \kappa_{\text{c}}$
\cite{Kirkpatrick_Belitz_2002} yields a wave number dependent coefficient $c$
in the paramagnon propagator, Eq.\ (\ref{eq:4.7b}),
\be
c({\bm k}\rightarrow 0) \propto (\ln 1/\vert{\bm k}\vert)^{-1/26}\quad,
\label{eq:4.16}
\ee
where the exponent is determined by the ratio $A_H/A_c = 27$. Expressing the
logarithmic corrections to power-law scaling in terms of scale-dependent
exponents, the critical exponent $\eta$, which describes the wave number
dependence of the paramagnon propagator at criticality, is therefore given by
\bse
\label{eqs:4.17}
\be
\eta = \frac{-1}{26}\,\ln\ln b/\ln b\quad.
\label{eq:4.17a}
\ee
The parameters $r$ and $d$ in the paramagnon propagator are not renormalized.
The correlation length exponent $\nu$, the susceptibility exponent $\gamma$,
and the dynamical exponent $z$ can therefore be directly read off Eqs.\
(\ref{eq:4.7b}),
\be
\nu = 1/(2-\eta)\quad,\quad z=3-\eta\quad,\quad\gamma = 1\quad.
\label{eq:4.17b}
\ee
The order parameter exponents $\beta$ and $\delta$ can be obtained from scaling
arguments for the free energy. The result is
\be
\beta = 1/2\quad,\quad\delta = 3\quad.
\label{eq:4.17c}
\ee
Finally, one can generalize the definition of the specific heat exponent
$\alpha$ familiar from thermal phase transitions by defining
$C_{\text{V}}\propto T^{-\alpha}$ at criticality. One finds
\be
\alpha = -1 + (\ln\ln b/\ln b - \eta)/z \quad.
\label{eq:4.17d}
\ee
\ese
The result for $\eta$ is valid to leading logarithmic accuracy; the values of
$\gamma$, $\beta$, and $\delta$, as well as the relations between $\eta$ and
$\nu$, $z$, and $\alpha$, respectively, are exact.

The theory thus predicts the critical behavior in $d=3$ to be mean-field like
with logarithmic corrections. Although the fixed point found by
\textcite{Hertz_1976} is unstable, it is so only marginally. Hertz's results,
and their extension to nonzero temperatures by \textcite{Millis_1993}, should
therefore apply apart from corrections that are too small to be detectable with
current experimental accuracies.

\begin{figure}[t]
\includegraphics[width=8cm]{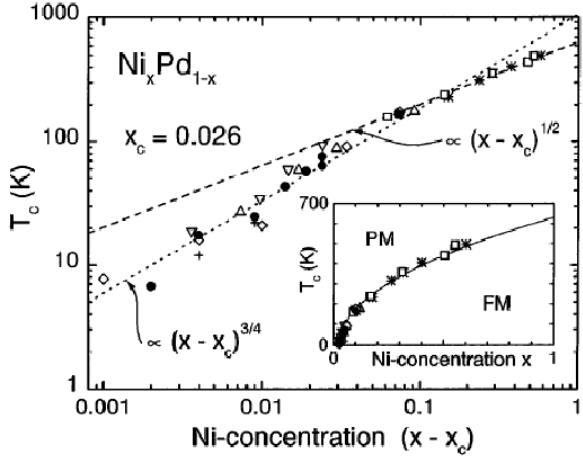}
\caption{\label{fig:16} Phase diagram of Ni$_x$Pd$_{1-x}$. Equation
   (\ref{eq:4.19}) is obeyed over an $x$-range of two decades close to the
   transition. From \protect{\textcite{Nicklas_et_al_1999}}.
   }
\end{figure}

\begin{figure}[t]
\includegraphics[width=5cm]{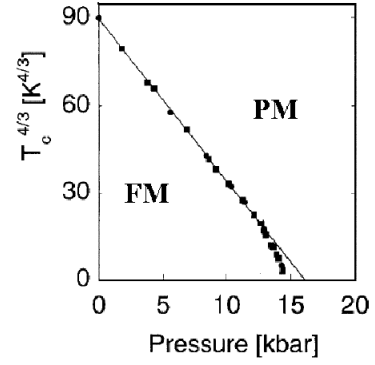}
\caption{\label{fig:17} Phase diagram of MnSi. These are the same data as in
    Fig.\ \ref{fig:13a}, scaled to display the relation given in
    Eq.\ (\ref{eq:4.19}). The tricritical point separating second and first
    order transitions coincides with the point where the scaling breaks down.
    After \protect{\textcite{Pfleiderer_et_al_1997}}.
   }
\end{figure}

Although the critical behavior has not been probed directly experimentally, a
combination of various exponents determines the shape of the phase diagram at
low temperatures. To see this, consider a homogeneity law for the
magnetization, which can be obtained by differentiating Eq.\ (\ref{eq:1.10})
twice with respect to $h$, and putting $h=0$:
\bse
\label{eqs:4.18}
\be
m(r,T,u) = b^{-d-z+2y_h}\,m(r\,b^{1/\nu},T\,b^z,u\,b^{[u]}).
\label{eq:4.18a}
\ee
Here we have included the coefficient $u$ of the quartic term in the free
energy, even though its scale dimension $[u]<0$ is negative, and $u$ is thus
irrelevant. The reason is that $u$ is a dangerous irrelevant variable with
respect to the magnetization \cite{Ma_1976, Fisher_1983}, which influences the
critical behavior of $m$. To see this, consider the equation of state, Eq.\
(\ref{eq:4.11a}). In the absence of the soft-mode corrections, one has
$m\propto \sqrt{-r/u}$. $m$ thus depends on $r/u$, and the homogeneity law can
be written
\bea
m(r/u,T) &=& b^{-d-z+2y_h}\,m\left((r/u)\,b^{1/\nu - [u]},T\,b^z\right)\hskip
30pt
\nonumber\\
         &=& (r/u)^{\beta}\,m\left(1,T\,(r/u)^{-\nu z/(1-\nu [u])}\right),
\label{eq:4.18b}
\eea
\ese
where $\beta = \nu (d+z-2y_h)/(1-\nu [u])$. The relation between $T_{\text{c}}$
and $r$ is now obtained from the requirement that $m$ must have a zero for $T =
T_{\text{c}}$. This yields $r = {\text{const.}}\times T^{(1-\nu [u])/\nu z}$
\cite{Millis_1993, Sachdev_1997}. $[u] =  4-d-z$, as can be seen from Eq.\
(\ref{eq:4.2}) by power counting, and neglecting the logarithmic corrections to
scaling we find from Eqs.\ (\ref{eqs:4.17}) \cite{Millis_1993}
\be
T_{\text{c}} \propto r^{3/4}.
\label{eq:4.19}
\ee
This agrees well with the phase diagram measured in Ni$_x$Pd$_{1-x}$, and also
with the portion of the phase diagram in MnSi where the transition is
continuous, see Figs.\ \ref{fig:16}, \ref{fig:17}.

\paragraph{Critical behavior in $d\neq 3$}
\label{par:IV.A.5.b}

In $d>3$, the RG analysis shows that Hertz's fixed point is stable, and all
exponents have their mean-field values. This confirms the suggestion of the
renormalized mean-field theory that the upper critical dimension is
$d_c^{\,+}=3$.

In $d<3$, the critical behavior can be studied by means of an expansion in
$\epsilon = 3-d$. \textcite{Kirkpatrick_Belitz_2002} have found a fixed point
where $G$, $H$, $c$, and $c_1$ are marginal. $c_2$ can again be either marginal
or irrelevant, depending on the context. A one-loop calculation of the exponent
$\eta$ and the specific heat exponent $\alpha$ yields
\be
\eta =  -\epsilon/26 \quad,\quad \alpha = -d/(3-\eta).
\label{eq:4.20}
\ee
$\nu$, $z$, $\gamma$, $\beta$, and $\delta$ are still given by Eqs.\
(\ref{eq:4.17b}, \ref{eq:4.17c}).

\subsection{Quantum Ferromagnetic Transition in Disordered Systems}
\label{subsec:IV.B}

If one adds quenched disorder to the problem considered in the previous
subsection, the action changes relatively little. The nature of the generic
soft modes changes; they are now diffusive rather than ballistic in nature, but
this change by itself will clearly only change the upper critical
dimensionality. Apart from this, the disorder needs to be averaged over, e.g.
by means of the replica trick. It turns out that the disorder average leads to
some terms in perturbation theory having a sign that is opposite from the
corresponding one in the clean case. For instance, the one-loop correction to
the quartic coupling constant $u$ is positive, and as a result the phase
transition is always of second order. Treating the generic soft modes in tree
approximation, \textcite{Hertz_1976} had concluded that the critical behavior
is mean-field like for all $d>0$. This turns out to not be true, for reasons
analogous to those that invalidate Hertz theory in the clean case. It turns
out, however, that the critical behavior can still be determined exactly for
all dimensions $d>2$, although it is not mean-field like. Historically, this
extension of Hertz's theory for the disordered case was developed earlier than
the corresponding treatment of the clean case \cite{Kirkpatrick_Belitz_1996,
Belitz_et_al_2001a, Belitz_et_al_2001b}. Here we explain the structure of this
theory and summarize its results.

\subsubsection{Soft-mode action}
\label{subsubsec:IV.B.1}

The action as given in Sec.\ \ref{subsubsec:IV.A.1} remains valid, with two
modifications. First, all fields carry replica indices that need to be summed
over. Specifically, ${\bm M}$ carries one replica index, while $Q$ and $q$
carry two replica indices, as explained in connection with Eq.\
(\ref{eq:2.44}). The symmetrized magnetization field $b$ is diagonal in its two
replica indices. Second, the fermionic vertex function $\Gamma^{(2)}$ changes;
Eqs.\ (\ref{eq:4.3b}) and (\ref{eq:4.3c}) get replaced by
\bse
\label{eqs:4.21}
\bea
{^i\Gamma}^{(2)}_{12,34}({\bm k}) &=&
\delta_{13}\,\delta_{24}\,\Gamma_{12}^{(2,0)}({\bm k}) +
\delta_{1-3,2-4}\,\delta_{\alpha_1\alpha_2}\,\delta_{\alpha_1\alpha_3}
\nonumber\\
&&\times 2\pi TG\,\left[\delta_{i0}K_{\text{s}} + (1-\delta_{i0}){\tilde
K}_{\text{t}}\right].
\label{eq:4.21a}
\eea
with replica indices $\alpha_i$, and
\be
\Gamma_{12}^{(2,0)}({\bm k}) = {\bm k}^2 + GH\Omega_{n_1-n_2}.
\label{eq:4.21b}
\ee
\ese
All other terms in the action remain unchanged, except for the addition of
replica indices.

It is obvious from simple physical consideration that the quenched disorder
must lead to additional terms in the action. For instance, consider the bare
distance from the critical point, $r$ in Eq.\ (\ref{eq:4.2}), a random function
of space, and integrate out this `random mass' with respect to some
distribution function. This will generate terms of higher order in ${\bm M}$,
starting at $O({\bm M}^4)$, that have a different imaginary-time structure than
the $u\,{\bm M}^4$ term in Eq.\ (\ref{eq:4.2}). Such random-mass terms do
indeed get generated by the basic disorder term in an underlying microscopic
action, Eq.\ (\ref{eq:2.37c}) or (\ref{eq:2.42}), and thus need to be added to
the effective action. However, they turn out to be irrelevant for the critical
behavior in all dimensions except $d=4$, and we therefore neglect them.

\subsubsection{Gaussian approximation}
\label{subsubsec:IV.B.2}

Keeping only terms that are bilinear in the fields yields the Gaussian
approximation, as in the clean case. The paramagnon propagator now reads
\be
{\cal M}_n({\bm k}) = \frac{1}{r_0 + c\,{\bm k}^2 +
d\,\vert\Omega_n\vert/\left({\bm k}^2 + GH\vert\Omega_n\vert\right)}\,.
\label{eq:4.22}
\ee
Formally this is the same as Eq.\ (\ref{eq:4.7a}) with $\vert{\bm k}\vert$
replaced by ${\bm k^2}$, but the coefficient $G$ has a different physical
interpretation as was explained in connection with Eqs.\ (\ref{eqs:2.48}). The
basic fermionic propagator is given by Eq.\ (\ref{eq:2.48b}). As in the clean
case, there is a mixed propagator that turns out not to be important for the
determination of the critical behavior.

\subsubsection{Renormalized mean-field theory}
\label{subsubsec:IV.B.3}

A renormalized mean-field theory can be constructed in exact analogy to the
treatment of the clean case in Sec.\ \ref{subsubsec:IV.A.3}
\cite{Sessions_Belitz_2003}. The free energy in this approximation is again
given by Eq.\ (\ref{eq:4.10a}), but the quantity $N$ in Eq.\ (\ref{eq:4.10b})
is replaced by
\be
N({\bm k},\Omega_n;m) = \frac{\left({\bm k}^2 + G(H + {\tilde
K}_{\text{t}})\Omega_n\right)^2 + 32\sqrt{\pi}c_2G^2m^2}{\left({\bm k}^2 +
GH\Omega_n\right)^2 + 32\sqrt{\pi}c_2G^2m^2}\,.
\label{eq:4.23}
\ee
An analysis of the integral in Eq.\ (\ref{eq:4.10a}) with this expression for
$N$ yields a free energy at $T=0$ of the form
\bse
\label{eqs:4.24}
\be
f = f_0 + r\,m^2 + w\,m^{(d+2)/2} + u\,m^4 - h\,,
\label{eq:4.24a}
\ee
or, equivalently, an equation of state
\be
h = r\,m + w\,m^{d/2} + u\,m^3.
\label{eq:4.24b}
\ee
\ese
In Eqs.\ (\ref{eqs:4.24}), $w>0$ is a positive coefficient that is proportional
to the disorder, and $r$ and $u$ again represent additive renormalizations of
$r_0$ and $u_0$, respectively.

Regarding the nonanalytic dependence of $f$ on $m$, the same comments apply as
in the clean case, Sec.\ \ref{subsubsec:IV.A.3}, and an assumption analogous to
the one mentioned after Eq.\ (\ref{eq:4.12}) has been made. As far as the GSI
is concerned, the only difference is that the decay of the spin susceptibility
in real space far from the transition is given, instead of Eq.\
(\ref{eq:4.12}), by
\be
\chi_{\text{s}}(r>0,\vert{\bm x}\vert\to\infty )\propto 1/\vert{\bm
x}\vert^{2d-2},
\label{eq:4.25}
\ee
which is the Fourier transform of Eq.\ (\ref{eq:2.35d}). Also, the
magnetization now scales as a wave number squared, which explains the exponent
of the nonanalyticity in Eq.\ (\ref{eq:4.24a}).

Despite these similarities, there is a crucial sign difference between the
nonanalytic term in the clean case, Eqs.\ (\ref{eq:4.11b}, \ref{eq:4.11c}), and
in the one in Eq.\ (\ref{eqs:4.24}). Although this is not of much significance
from a GSI point of view, it leads to the quantum phase transition to be of
first order in the clean case (at least at the level of the renormalized
mean-field theory), while it is of second order in the disordered case.
Physically, this can be understood by means of arguments very similar to those
that explained the sign difference between Eqs.\ (\ref{eq:2.35d}) and
(\ref{eq:2.36b}), respectively. In the clean case, the nonanalyticity is
produced by fluctuation effects, represented by the generic soft modes, that
weaken the tendency towards ferromagnetism. Accordingly, the constant
contribution from the integral over $\ln N$ in Eq.\ (\ref{eq:4.10a}), which
changes $r_0$ to $r$, is positive. The appearance of a nonzero magnetization
gives the soft modes a mass, and hence weakens these fluctuations, so the
leading $m$-dependent correction to the constant contribution is negative. In
the disordered case, on the other hand, the quenched disorder leads to
diffusive motion of the electrons, which is much slower than the ballistic
dynamics in the clean case. This increases the effective interaction strength
between the electrons, which in turn favors ferromagnetism. Consequently, $r$
is smaller than $r_0$. A nonzero magnetization again weakens this effect, and
therefore the leading nonanalytic $m$-dependence of the free energy is
positive.

\begin{figure}[t]
\includegraphics[width=6cm]{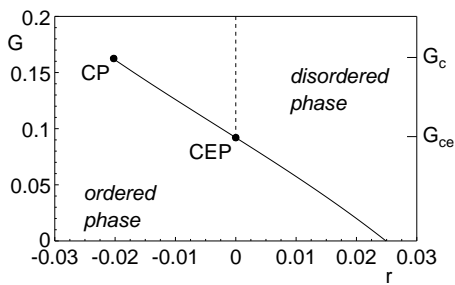}
\caption{\label{fig:18} Phase diagram in the $G$-$r$ plane resulting
    from Eq.\ (\ref{eq:4.26}) at
    $T=0$ for $u=1$, $v=0.5$, $A = 0.5$, $B_T = B_G = 1$,
    showing a second-order transition (dashed line), a first-order transition
    (solid line), a critical end point (CEP), and a critical point (CP).
    From \protect{\textcite{Belitz_Kirkpatrick_Vojta_1999}}.
   }
\end{figure}

\begin{figure*}[t]
\includegraphics[width=16cm]{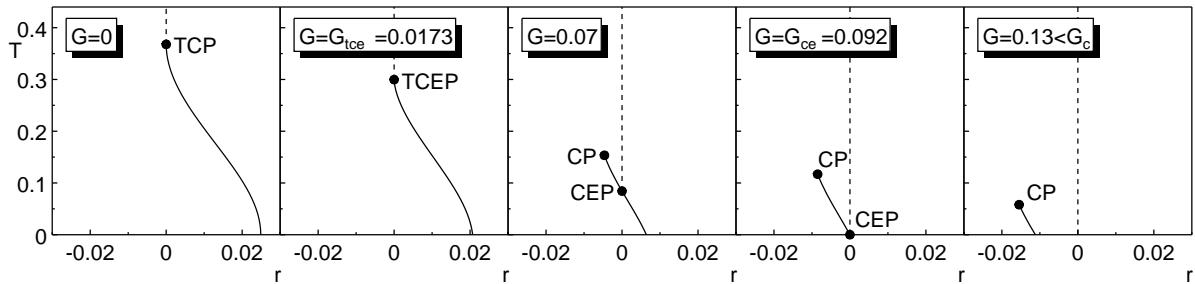}
\caption{\label{fig:19} Phase diagrams in the $T$-$r$ plane
    resulting from Eq.\ (\ref{eq:4.26}) for $u = B_G =1$,
    $v = B_T = A = 0.5$. TCP and TCEP denote a tricritical point and a
    tricritical endpoint, respectively. All other symbols are the same as in
    Fig.\ \ref{fig:18}.
    From \protect{\textcite{Belitz_Kirkpatrick_Vojta_1999}}.
   }
\end{figure*}

A more general form of the free energy in renormalized mean-field
approximation, which is valid for $T\geq 0$ and allows the zero-disorder limit
to be taken, has been considered by \textcite{Belitz_Kirkpatrick_Vojta_1999}.
The free energy density discussed by these authors takes the form
\bea
f &=& f_0 + r\,m^2 + (G/\epsilon_{\text{F}})\,A\,m^4\,\left[m^2 + (B_T
T)^2\right]^{-3/4}
\nonumber\\
&&+ v\,m^4\,\ln\left[m^2 + (T + B_G G)^2\right] + u\,m^4
   + O(m^6).
\nonumber\\
\label{eq:4.26}
\eea
Here $B_T$ and $B_G$ are dimensionless parameters that measure the relative
strengths of the temperature and the disorder dependence, respectively, in the
two nonanalytic terms, and $A$ is a measure of how strongly correlated the
system is. They are all expected to be of order unity. For $G=0$ the resulting
equation of state reduces to Eq.\ (\ref{eq:4.11a}). For sufficiently large
disorder $G$, the logarithmic term is unimportant, and at $T=0$ one recovers
Eqs.\ (\ref{eqs:4.24}). This form of the free energy thus interpolates between
the clean and disordered cases. It displays a rich phenomenology, as shown in
Figs.\ \ref{fig:18}, \ref{fig:19}. Notice that, for a disorder larger than a
threshold value, they predict a second magnetic transition inside the
ferromagnetic phase that is always of first order.

The critical exponents $\beta = 2/(d-2)$ and $\delta = d/2$ (for $2<d<6$) can
be read off Eq.\ (\ref{eq:4.24b}), but a determination of the remaining
exponents requires the consideration of order-parameter fluctuations. We will
discuss this in the next subsection, where we will also see that the critical
behavior predicted by the renormalized mean-field theory is very close to the
exact one.

\subsubsection{Effect of order-parameter fluctuations}
\label{subsubsec:IV.B.4}

The effects of order-parameter fluctuations on the disordered ferromagnetic
quantum phase transition have been studied at various levels.
\textcite{Hertz_1976} and \textcite{Millis_1993} considered order-parameter
fluctuations, but integrated out the generic soft modes in a tree approximation
that neglects all fermionic loops. \textcite{Kirkpatrick_Belitz_1996} kept
fermionic loops, but still integrated out the fermions, which led to a nonlocal
field theory in terms of the order parameter only, which they analyzed by means
of power counting. \textcite{Belitz_et_al_2001a, Belitz_et_al_2001b} kept all
of the soft modes explicitly, and on equal footing, in analogy to the theory
for clean systems discussed in Sec.\ \ref{subsec:IV.A}, and to the classical
theories covered in Sec.\ \ref{sec:III}. This theory contains the previous ones
as approximations, and we will sketch it in the remainder of this section.

\paragraph{Coupled field theory, and power counting}
\label{par:IV.B.4.a}

In contrast to the clean case, Sec.\ \ref{subsubsec:IV.B.1}, for disordered
magnets the fermionic part of the soft-mode action is known in closed form,
namely, the nonlinear $\sigma$ model, Eqs.\ (\ref{eqs:2.61}, \ref{eqs:2.62}),
with no spin-triplet interaction. The order-parameter part of the action is
again given by Eq.\ (\ref{eq:4.2}) with a replicated ${\bm M}$ field as
explained in Sec.\ \ref{subsubsec:IV.B.1}, and the coupling between the two is
given by Eq.\ (\ref{eq:4.4a}), with $Q$ replaced by ${\hat Q}$ and replica
indices added as appropriate. This action can be written down on general
principles, or it can be derived from the complete nonlinear $\sigma$ model,
Eqs.\ (\ref{eqs:2.61}), by performing a Hubbard-Stratonovich decoupling
\cite{Hubbard_1959, Stratonovich_1957} of the spin-triplet interaction and
adding an ${\bm M}^4$ term.\footnote{\label{fn:55} The quartic and higher terms
in ${\bm M}$ originate from massive modes in the underlying miscroscopic
action, which have been dropped in deriving the nonlinear $\sigma$ model.}
Since the nonlinear $\sigma$ model has been derived from a microscopic action,
this also provides a derivation of the final effective action from a
microscopic starting point, if that is desired.

We now expand this effective action in powers of ${\bm M}$ and $q$. In a
schematic notation that omits everything not necessary for power counting, we
have
\bse
\label{eqs:4.27}
\be
{\cal A}_{\rm eff}[{\bm M},q] = {\cal A}_M + {\cal A}_q + {\cal A}_{M,q}.
\label{eq:4.27a}
\ee
with
\bea
{\cal A}_M &=& -\int d{\bm x}\ M\,\left[r + c_{d-2}\,\partial_{\bm x}^{d-2} +
c\,\partial_{\bm x}^2\right]\,M
\nonumber\\
&&\hskip 70pt   + O(\partial_{\bm x}^4\, M^2, M^4),
\label{eq:4.27b}
\eea
\bea
{\cal A}_q &=& -\frac{1}{G}\int d{\bm x}\,(\partial_{\bm x} q)^2
     + H\int d{\bm x}\,\Omega\,q^2 + K_s\,T\int d{\bm x}\, q^2
\nonumber\\
&& \hskip -0pt - \frac{1}{G_4}\int d{\bm x}\,\partial_{\bm x}^2\, q^4
   + H_4\int d{\bm x}\,\Omega\,q^4
\nonumber\\
&& \hskip 50pt + O(T q^3, \partial_{\bm x}^2\, q^6,
      \Omega\,q^6),
\label{eq:4.27c}
\eea
\bea
{\cal A}_{M,q} &=& \sqrt{T}\,c_1 \int d{\bm x}\,M\,q + \sqrt{T}\,c_2 \int d{\bm
x}\,M\,q^2
\nonumber\\
   && \hskip 50pt + O(\sqrt{T}Mq^4)\quad.
\label{eq:4.27d}
\eea
\ese
Here the fields are understood to be functions of position, frequencies,
replica labels. Only quantities that carry a scale dimension are shown.
Accordingly, frequency and replica sums have been omitted, but appropriate
powers of the temperature are shown. All of the terms except the second one in
${\cal A}_M$ can be obtained by combining Eqs.\ (\ref{eq:4.2}, \ref{eq:4.3a},
\ref{eqs:4.4}, \ref{eqs:4.5}) and Eqs.\ (\ref{eqs:4.21}). The bare value of the
coefficient $c_{d-2}$ is zero, but we will see that such a term is generated
under renormalization.

For a power-counting analysis of this effective action, we assign scale
dimensions $[L] = -1$ and $[T] = [\Omega] = -z$ to lengths and temperatures or
frequencies, respectively. It is important to note that $z$ is unlikely to have
a unique value: Since the critical paramagnon propagator will in general have a
frequency-wave number relation that is different from that of the fermionic $q$
propagator, we expect at least two different time scales in the problem, hence
two different values of $z$. This needs to be taken in to account in the power
counting analysis.

\paragraph{Hertz's fixed point}
\label{par:IV.B.4.b}

\begin{figure}[t]
\includegraphics[width=4cm]{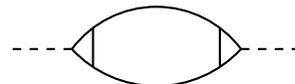}
\caption{\label{fig:20} One-loop renormalization of the $M^2$ vertex. Dashed
    lines denote $M$ fields, solid lines denote $q$ propagators, and the vertices
    carry a factor of $c_2$ each.
   }
\end{figure}

The fixed point identified by \textcite{Hertz_1976} is recovered from this
formalism by the coefficients $c$ and $c_1$ to be dimensionless. The field $q$
is expected to be diffusive, and we therefore choose its scale dimension to be
$[q] = (d-2)/2$, in accord with Eqs.\ (\ref{eq:2.63}). From ${\cal A}_M$ we
then obtain $[M] = (d-2)/2$ and $[t]=2$. $r$ is thus relevant, as expected, and
the correlation length exponent is $\nu = 1/[r] = 1/2$. $c_{d-2}$ is also
relevant, of course, but since its bare value is zero we ignore it for now. The
first term in ${\cal A}_{M,q}$ produces the dynamical part of the paramagnon,
as was demonstrated in Sec.\ \ref{subsubsec:IV.A.2}. The scale dimension of the
$\sqrt{T}$ prefactor thus yields the paramagnon or critical dynamical exponent
$z_{\text{c}} = 4$. The frequency and the temperature in the second and third
term, respectively, in ${\cal A}_q$ carry the fermionic time scale $z_q = 2$
that is consistent with diffusion. $G$, $H$, and $K_{\text{s}}$ are then all
marginal, and all higher order terms in ${\cal A}_q$ are irrelevant. For the
scale dimension of $c_2$ one finds
\be
[c_2] = -(d+z-6)/2.
\label{eq:4.28}
\ee
Due to the existence of two time scales mentioned above, one needs to
distinguish between two different variants of $c_2$, which differ with respect
to the value of $z$ that enters their scale dimension. With $z=z_{\text{c}}=4$,
$c_2 \equiv c_2^-$ is irrelevant for all $d>2$. However, with $z=z_q=2$, $c_2
\equiv c_2^+$ is relevant with respect to the putative fixed point for $d<4$.
This instability of Hertz's fixed point is indeed realized. Consider, for
instance, the renormalization of the ${\bm M}^2$ vertex by means of the diagram
shown in Fig.\ \ref{fig:20}. Since it is a pure fermion loop, the factors of
$\sqrt{T}$ which come with the vertices, and which are absorbed into the loop
integral over the frequency, carry indeed the fermionic time scale, and the
relevant coupling constant is $c_2^+$. We note that this is the same
perturbative contribution that leads to the nonanalytic wave number dependence
of the spin susceptibility deep in the paramagnetic phase, Eq.\
(\ref{eq:2.35d}). The term with coupling constant $c_{d-2}$ is thus generated
from the $c_2$ coupling. It is convenient to explicitly add this term to the
bare action, as we have done in Eq.\ (\ref{eq:4.27b}), although the physics it
represents is already contained in the $c_2$ coupling term. Indeed, with
respect to Hertz's fixed point, $(c_2^+)^2$ and $c_{d-2}$ have the same scale
dimension, $[c_{d-2}] = 2[c_2^+] = 4-d$.

One thus finds that the generic soft modes render Hertz's fixed point unstable,
and this first becomes apparent at one-loop order. This instability, although
not its source, can also be deduced from the fact that the mean-field value for
the correlation length exponent $\nu = 1/2$, violates, for all $d<4$, the
rigorous bound provided by the Harris criterion \cite{Harris_1974,
Chayes_et_al_1986} for systems with quenched disorder, $\nu\geq 2/d$.

\paragraph{A marginally unstable fixed point}
\label{par:IV.B.4.c}

The behavior of the renormalized mean-field theory is recovered from the RG if
one (1) takes the $c_{d-2}$ term into account, and (2) drops the requirement
that $c_2$ be marginal. We thus require only that $c_1$ be marginal, which
implies $[M] = 1 + (d-z)/2$, and that $[q] = (d-2)/2$ and $z_q = 2$ as before.
At this point it is useful to consider an analogy between the present problem
and that of a classical fluid under shear discussed in Sec.\
\ref{subsubsec:III.B.2}. In the latter case, long-ranged correlations between
order-parameter fluctuations stabilized mean-field critical behavior in
dimensions lower than the usual upper critical dimension. In the current
problem, the $c_{d-2}\partial_{\bm x}^{d-2}$ term in the action represents
long-range order parameter correlations that decay as $\vert{\bm
x}\vert^{-2(d-1)}$. By analogy with the classical fluid example, it is
therefore natural to expect the stabilization of a Gaussian fixed point where
$c_{d-2}$ is marginal in the critical paramagnon. This implies $[M]=1$, and
hence a critical time scale and an exponent $\eta$ characterized by
\bse
\label{eqs:4.29}
\be
z_c = 4 - \eta = d.
\label{eq:4.29a}
\ee
$r$ is relevant with $[r]=d-2$, which corresponds to a correlation length
exponent
\be
\nu = 1/(d-2).
\label{eq:4.29b}
\ee
\ese
In contrast to the situation at Hertz's fixed point, this respects the Harris
criterion.

Equation (\ref{eq:4.29a}) implies that the magnetic susceptibility at
criticality decays in real space according to
\be
\chi_{\text{s}}(r=0,\vert{\bm x}\vert\to\infty )\propto 1/\vert{\bm x}\vert^2.
\label{eq:4.30}
\ee
In the relevant dimensionality range, $d>2$, these correlations are of longer
range than the power-law correlations in the paramagnetic phase, Eq.\
(\ref{eq:4.25}). We see that, as in the case of the classical fluid, Sec.\
\ref{subsubsec:III.B.2}, the long-ranged correlations or GSI in the disordered
phase influence the critical behavior, and lead to correlations with an even
longer range at the critical point.

The results given above hold for $2<d<4$. For $d>4$, $\eta=0$ and $\nu=1/2$
have their mean-field values, and $z_c=4$. The exponents $\beta$ and $\delta$
can be obtained either from scaling arguments for the free energy
\cite{Belitz_et_al_2001b}, or from repeating the above counting arguments in
the ferromagnetic phase \cite{Sessions_Belitz_2003}. By either method one finds
the same result as from Eq.\ (\ref{eq:4.24b}), viz.,
\be
\beta = 2/(d-2)\quad,\quad\delta=d/2.
\label{eq:4.31}
\ee
These relations hold for $2<d<6$, while for $d>6$ one has the mean-field values
$\beta=1/2$, $\delta=3$. A detailed analysis reveals that Hertz's fixed point
is actually stable for all $d>4$, but for $4<d<6$ the coefficient $w$ in Eq.\
(\ref{eq:4.24b}) acts as a dangerous irrelevant variable with respect to the
magnetization, which explains why $\beta$ and $\delta$ lock into their
mean-field values only for $d>6$.

What remains to be done is to investigate the stability of the Gaussian fixed
point with the critical behavior described above. It turns out that it is
marginally unstable, and that the exact critical behavior is given by the power
laws given above with additional logarithmic corrections to scaling. We give
these results next.

\paragraph{Exact critical behavior}
\label{par:IV.B.4.d}

The results of the previous subsection characterize a Gaussian fixed point that
was described by the theory given by \textcite{Kirkpatrick_Belitz_1996}. To
check its stability, one needs to consider the scale dimensions of the
remaining terms in the action. $c_2$ has a scale dimension
\be
[c_2] = 1 - z/2
\label{eq:4.32}
\ee
with respect to the Gaussian fixed point, which implies that $c_2^+$ (i.e.,
$c_2$ with $z=z_q$) is marginal. This was to be expected, since $c_{d-2}$,
which describes the same physics as $c_2$, is also marginal. However, it
implies that the non-Gaussian $c_2$ term must be kept as part of the
fixed-point action. All other terms are nominally irrelevant. However, it turns
out that, for subtle reasons again related to the existence of multiple time
scales, the terms of order $q^4$ in ${\cal A}_q$ can also be effectively
marginal and must be kept as part of the fixed point action. For all $d>2$, the
latter then consists of the terms shown explicitly in Eqs.\
(\ref{eqs:4.27}).\footnote{\label{fn:56} A more conventional approach would be
to not include the $c_{d-2}$ term in the action, but rather have its effects
included in a renormalization of $c$ while working in $d=4-\epsilon$. This can
be done, but the loop expansion in this case does not lead to a controllable
expansion in powers of $\epsilon$ since $c_2$ remains strictly marginal order
by order. Nevertheless, such a procedure yields interesting technical insights
\cite{Rollbuehler_et_al_2004}.}

Since the fixed point action is not Gaussian, the exact critical behavior
cannot be obtained simply by power counting. However, it turns out that the
problem can still be solved exactly, by means of an infinite resummation of
perturbation theory \cite{Belitz_et_al_2001b}. For $2<d<4$, the result consists
of logarithmic corrections to the exponents derived in the previous subsection.
For instance, the correlation length depends on $r$ via
\be
\xi \propto r^{-1}\,e^{-{\text{const.}}\times (\ln\ln(1/r))^2},
\label{eq:4.33}
\ee
to leading logarithmic accuracy. Notice that the correction to the power law
varies more slowly than any power of $r$, but faster than any power of $\ln r$.
This behavior is conveniently expressed in terms of exponents that are scale
dependent via a function $g$ whose asymptotic behavior for large arguments is
\be
g(x \gg 1) \approx \left[2\ln(d/2)/\pi\right]^{-1/2}\,
                e^{[\ln (c(d)\,x)]^2/2\ln(d/2)}.
\label{eq:4.34}
\ee
The dimensionality dependent coefficient $c(d)$ goes to zero for $d\to 4$, and
to a constant for $d\to 2$. In the case of $\nu$, Eq.\ (\ref{eq:4.33})
corresponds to
\bse
\label{eqs:4.35}
\be
1/\nu = d - 2 + \ln g(\ln b)/\ln b.
\label{eq:4.35a}
\ee
For the other exponents one obtains
\bea
z_{\text{c}} &=& 4 - \eta = d + \ln g(\ln b)/\ln b,
\label{eq:4.35b}\\
\delta &=& -\alpha/2d = z_{\text{c}}/2,
\label{eq:4.35c}\\
\beta &=& 2\nu\quad,\quad\gamma = 1,
\label{eq:4.35d}\\
z_q &=& 2 + \ln g(\ln b)/\ln b,
\label{eq:4.35e}
\eea
\ese
where the specific heat exponent $\alpha$ is defined as in the context of Eq.\
(\ref{eq:4.17d}).

The critical behavior of various transport coefficients and relaxation rates,
as well as the tunneling density of states, have also been determined
\cite{Belitz_et_al_2000, Belitz_et_al_2001b}. In particular, these authors
showed that the quasiparticle properties at criticality are those of a marginal
Fermi liquid \cite{Varma_et_al_1989}. Here we just quote the result for the
electrical conductivity. In $d=3$ at criticality, one finds a nonanalytic
temperature dependence in addition to a noncritical background term,
\be
\sigma(T\to 0) = \sigma(T=0) +
{\text{const.}}\times\left[T\,g\left(\frac{1}{3}\,
   \ln (\epsilon_{\text{F}}/T)\right)\right]^{1/3},
\label{eq:4.36}
\ee
with $g(x)$ from Eq.\ (\ref{eq:4.34}).

\begin{figure}[t]
\includegraphics[width=8cm]{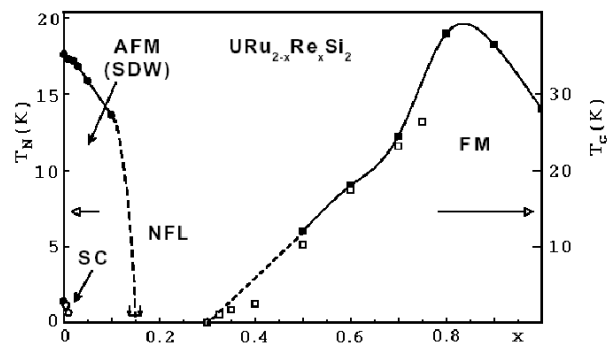}
\caption{\label{fig:21} Phase diagram of URuReSi, showing ferromagnetic (FM),
   antiferromagnetic (AFM), superconducting (SC), and non-Fermi liquid (NFL)
   phases. The different symbols refer to different methods for determining the
   phase boundaries. After \textcite{Bauer_2002}.
}
\end{figure}

Experimentally, there are few systematic studies of the influence of quenched
disorder on the ferromagnetic quantum phase transition. URu$_{2-x}$Re$_x$Si$_2$
shows a rich phase diagram in the $T$-$x$ plane, with phases displaying
ferromagnetic, antiferromagnetic, and superconducting order in addition to a
paramagnetic non-Fermi-liquid (``strange metal'') phase
\cite{Dalichaouch_et_al_1989, Bauer_2002, Bauer_et_al_2003, Stewart_2001}, see
Fig.\ \ref{fig:21}. Near $x=0.3$ there is a quantum phase transition from the
ferromagnetic phase to the strange-metal phase that has been studied in some
detail by \textcite{Bauer_2002} and \textcite{Bauer_et_al_2003}. These authors
found a value of $\delta=1.5$, in agreement with Eq.\ (\ref{eq:4.35c})
(ignoring the logarithmic corrections to scaling). The specific heat
coefficient for $T\to 0$ is found to diverge logarithmically, or as a very
small power of $T$, which is, within the experimental accuracy, consistent with
Eq.\ (\ref{eq:4.35c}). However, the logarithmic divergence of the specific heat
coefficient is observed to persist away from the quantum critical point in both
the paramagnetic and the ferromagnetic phases. Similar non-Fermi liquid
properties have been found in the paramagnetic phase of other materials that
display a quantum ferromagnetic transition, e.g., in MnSi
\cite{Pfleiderer_Julian_Lonzarich_2001}. We will come back to these
observations in Secs. \ref{subsec:IV.D} and \ref{sec:V}.

\textcite{DiTusa_et_al_2003} have investigated Fe$_{1-x}$Co$_x$S$_2$, which
displays a $T=0$ insulator-metal transition at a very small value of $x$ ($x<
0.001$), followed by a quantum paramagnet-to-ferromagnet transition at
$x\approx 0.032$. The strength of the quenched disorder is substantial in this
system, with values of $k_{\text{F}}\ell$ ranging from 2 to 15 for $x$ between
$0.001$ and $0.17$. The authors have reported scaling of the conductivity near
the ferromagnetic transition that is consistent with Eq.\ (\ref{eq:4.36}).

\subsubsection{Effects of rare regions on the phase transition for Ising system}
\label{subsubsec:smearing}

We now turn to rare regions and nonperturbative disorder effects. In contrast
to the previous subsections, which considered Heisenberg ferromagnets, we will
now discuss a special effect for magnetic order parameters with an Ising
symmetry. In such systems, the rare region effects are much stronger than the
quantum Griffiths phenomena discussed in Sec.\ \ref{subsubsecII.B.6} because of
the coupling between the order parameter and the generic soft modes. To see
this, consider a particular rare region devoid of impurities. Within such a
region, the appropriate paramagnon propagator is given by Eq.\ (\ref{eq:4.7a}),
and the coupling between the magnetization and the conduction electrons is
reflected by the dependence on $\vert\Omega_n\vert$.\footnote{\label{fn:57}
This is sometimes referred to as Landau damping, in analogy to the fate of the
plasmon when it enters the particle-hole continuum. In the absence of such a
coupling, e.g., in the pure spin model given by Eq.\ (\ref{eq:2.75}), one would
have a propagating mode with a frequency dependence on $\Omega_n^2$.} In
imaginary time space, this corresponds to a long-time tail given by $1/\tau^2$,
see Sec.\ \ref{subsubsec:II.B.7}. At $T=0$, every rare region thus maps onto a
clean, classical $(d+1)$-dimensional Ising model that is finite in $d$
dimensions and infinite in one dimension. The interactions between the spins
are short-ranged in the former, and long-ranged, proportional to $1/\tau^2$, in
the latter.

This long-range interaction can have drastic consequences, namely, the sharp
quantum phase transition can be destroyed by smearing. This can be understood
as follows. A one-dimensional Ising model with a $1/r^2$ interaction is known
to possess an ordered phase \cite{Thouless_1969, Cardy_1981}. A rare region can
therefore develop a static order parameter independently from the rest of the
system. This result is consistent with the one obtained by
\textcite{Millis_Morr_Schmalian_2002}, who directly calculated the tunneling
rate of a Griffiths island in an itinerant system and found it to vanish for a
sufficiently large island. Suppose a nonzero average order parameter first
appears on a rare region, rather than in the bulk of the system. Since the
order is truly static, the ordered region will effectively act like a small
permanent magnet imbedded in the system, and it will be energetically favorable
for subsequent rare regions to align their order parameter with the first one
(assuming that the effective interaction between the regions is also
ferromagnetic). As a result, a finite magnetization appears as soon as a finite
volume of rare regions start to order. This magnetization is exponentially
small, and the correlation length does not diverge at this point. However, the
ordered rare regions provide an effective magnetic field seen by the bulk of
the system; therefore, the latter cannot undergo a sharp quantum phase
transition \cite{Vojta_2003a}. Notice that the number of ordered regions
changes continuously with the control parameter, so the smeared transition is
not simply given by the behavior of electrons in a fixed magnetic field.
Rather, order develops in different parts of the system at different values of
the control parameter, and the order parameter close to the smeared transition
is very inhomogeneous in space. The same mechanism also destroys classical
phase transitions in systems with planar defects \cite{Vojta_2003b,
Sknepnek_Vojta_2003}.

The above arguments are valid at $T=0$, where the system is infinite in
imaginary time direction. The behavior at finite temperatures is less well
established. Since the rare regions are far apart, their interaction is very
small. Therefore, the relative alignment between the order parameter on
different rare regions vanishes already at a temperature that is exponentially
small in the density of the rare regions, i.e., double-exponentially small in
the disorder strength. Above this temperature, the rare regions act as
independent classical moments. Whether or not there is a second crossover at
even higher temperatures to quantum Griffiths behavior similar to that in
undamped systems (as discussed in Sec.\ \ref{subsubsecII.B.6}) is not fully
understood; it appears to be a question of the microscopic parameter values
\cite{Millis_Morr_Schmalian_2002, CastroNeto_Jones_2000}.

In systems with continuous order parameter symmetry the rare region effects are
weaker. Specifically, the quantum phase transition will remain sharp because
the rare regions cannot develop static order. This can be seen by mapping each
rare region onto a classical one-dimensional XY or Heisenberg model with
$1/r^2$ interaction. In contrast to the corresponding Ising model, these models
do not have an ordered phase \cite{Kosterlitz_1976, Bruno_2001}. The quantum
Griffiths behavior in the vicinity of the transition has not yet been worked
out in detail, but it is likely to be of a more conventional type.

\subsection{Metal-Superconductor Transition}
\label{subsec:IV.C}

\begin{figure}[t]
\includegraphics[width=6cm]{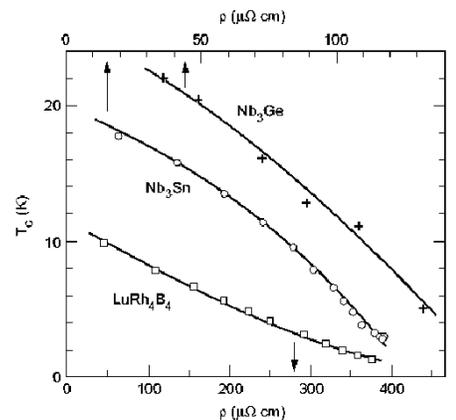}
\caption{\label{fig:22} Superconducting $T_{\text{c}}$ versus normal-state
    resistivity for three superconducting materials. The data are taken from
    \textcite{Rowell_Dynes_1980}; the lines are fits to a
    theory for the $T_{\text{c}}$ degradation that is not important for the
    present discussion. From \protect{\textcite{Belitz_Kirkpatrick_1994}}.
   }
\end{figure}

Sufficiently strong quenched nonmagnetic disorder\footnote{\label{fn:58}{\em
Weak} disorder can actually enhance $T_{\text{c}}$, sometimes substantially so.
Aluminum, and many other low-$T_{\text{c}}$ superconductors, are examples for
this effect.} decreases the critical temperature $T_{\text{c}}$ for the
superconductor-metal transition in bulk conventional
superconductors,\footnote{\label{fn:59} Also very interesting are the analogous
effects in thin superconducting films, where nonmagnetic disorder leads to a
transition from a superconducting phase to an insulating one, with or without
an intermediate metallic phase. This topic is outside the scope of this
review.} see Fig. \ref{fig:22}. At a critical value of the disorder,
$T_{\text{c}}$ vanishes, and there is a quantum phase transition from a
normal-metal phase to a superconducting phase where the concepts of additional
soft modes and GSI effects play a crucial role. Since the generic soft modes in
question, particle-hole excitations again, are massive at nonzero temperature,
this quantum phase transition requires a theoretical description that is
qualitatively different from BCS theory and the theories that describe
fluctuation corrections to it. Such a theory, whose main predictions should not
be hard to check experimentally, is presented below. An alternative, and
physically different, theory has been proposed by
\textcite{Vishveshwara_Senthil_Fisher_2000}. These authors have proposed that
the normal-metal--superconductor transition triggered by disorder leads
generically to a gapless superconducting state, and they assert that the
resulting gapless quasiparticle excitation spectrum will have consequences for
the critical behavior. It remains to be worked out to what extent this
assertion is correct, or to what extent the gaplessness is a property of the
stable fixed point that describes the superconducting phase, rather than the
critical fixed point. We will further comment on this in Sec.\
\ref{par:V.B.2.b} below.

One can construct a theory for the metal-superconductor transition that is
structurally very similar to the one for the disordered ferromagnetic
transition discussed in the previous section \cite{Zhou_Kirkpatrick_2004}, so
we will keep the discussion brief. There are, however, a few crucial
differences. The most important one is that the superconducting order parameter
field $\Psi({\bm x})$, whose expectation value is closely related to the
anomalous Green function and the superconducting gap function, is related to
the bilinear product of fermion fields at equal and opposite frequencies,
\be
\Psi({\bm x}) \sim \psi_{n,\sigma}(\bm{x})\,\psi_{-n,-\sigma}(\bm{x}),
\label{eq:4.37}
\ee
which in turn can be expressed in terms of the matrix elements $q$ of the
matrix $Q$, see Eqs.\ (\ref{eq:2.44}, \ref{eq:2.46}). Since the $q$ are soft
modes, see Eq.\ (\ref{eqs:2.48}), this means that the coupling of the order
parameter to the generic soft modes that is even stronger than in the
ferromagnetic case, and therefore the effects of GSI are even more dramatic.

We write the action as,
\be
{\cal A}[\Psi,Q] = {\cal A}_{\Psi} + {\cal A}_q + {\cal A}_{\Psi,q}.
\label{eq:4.38}
\ee
${\cal A}_{\Psi}$ is analogous to ${\cal A}_{M}$, Eq.\ (\ref{eq:4.2}). It is a
static, local, LGW functional for a three-dimensional XY-model. The order
parameter field one can take to be complex valued, or as having two components
$\Psi_n^{\,r}(\bm{x})$, with $r=1,2$. Structurally it is identical to Eq.\
(\ref{eq:4.2}) with $M_n^i(x)$ replaced by $\Psi_n^{\,r}(x)$. Since we are
dealing with a disordered system, $\Psi$ also carries a replica label. The
fermionic part of the action, ${\cal A}_q$, is identical to the one in the
previous section, except now the particle-particle or Cooperon degrees of
freedom, $r=1,2$, need to be taken into account. In particular, the Gaussian
part of ${\cal A}_q$ is given by Eq.\ (\ref{eq:4.3a}) with $\Gamma$ given by
Eqs.\ (\ref{eqs:4.21}) for $r=0,3$ and by
\bea
{_{r=1,2}^{\ \ \ \ \ i}\Gamma}_{12,34}^{(2)}({\bm k}) &=&
     \delta_{13}\,\delta_{24}\,\Gamma_{12}^{(2,0)}({\bm k})
\nonumber\\
&& \hskip -50pt +\ \delta_{i0}\,
\delta_{1+2,3+4}\,\delta_{\alpha_2\alpha_4}\,\delta_{\alpha_1\alpha_3}\,
            2\pi T G\,{\tilde K}_{\text{c}}
\label{eq:4.39}
\eea
for $r=1,2$. Here $\Gamma^{(2,0)}$ is given by Eq.\ (\ref{eq:4.21b}), and
${\tilde K}_{\text{c}}$ is a repulsive Cooper channel interaction that is
generated by the particle-hole channel electron-electron interactions
$K_{\text{s}}$ and $K_{\text{t}}$ in analogy to the mechanism that generated
${\tilde K}_{\text{t}}$ in Eq.\ (\ref{eq:4.3b}). It is also disorder dependent.
As we will see below, ${\tilde K}_{\text{c}}$ is responsible for driving
$T_{\text{c}}$ to zero. The bare attractive Cooper channel interaction that is
due to, e.g., phonon exchange, is denoted by $K_{\text{c}} = -\vert
K_{\text{c}}\vert < 0$. The coupling between $\Psi$ and $q$ originates from a
term ${\cal A}_{\Psi-Q}$ that can be written, in analogy to Eq.\
(\ref{eq:4.4d}),
\bse
\label{eqs:4.40}
\be
{\cal A}_{\Psi-Q} = c_1\,i\sqrt{T}\int d{\bm x}\ \tr \bigl(\beta({\bm
x})\,Q({\bm x})\bigr).
\label{eq:4.40a}
\ee
The functional form of this term becomes plausible if one realizes that the
order parameter field $\Psi$ acts like an external field that couples to the
particle-particle number density. To bilinear order this yields a contribution
to ${\cal A}_{\Psi,q}$ given by
\be
{\cal A}_{\Psi-q} = -8\,c_{1}\sqrt{T}\sum_{12}\int d{\bm x}
\sum_{r=1,2}\;_{r}{\beta}_{12}({\bm x})\,{_r^0q}_{12}({\bm x}).
\label{eq:4.40b}
\ee
Here $c_1\propto \vert K_{\text{c}}\vert^{1/2}$, and
\be
{_r\beta}_{12}({\bm x}) = \delta_{\alpha_1\alpha_2}\sum_{n}\delta
_{n,n_1+n_2}\Psi_n^{r,\alpha_1}({\bm x}).
\label{eq:4.40c}
\ee
\ese
This structure is in exact analogy to Eqs.\ (\ref{eqs:4.4}, \ref{eqs:4.5}) for
the magnetic case, only the factor of $i$ in Eq.\ (\ref{eq:4.40a}), the sign in
Eq.\ (\ref{eq:4.40b}), and the frequency structure in Eq.\ (\ref{eq:4.40c})
reflect the particle-particle channel. Again in analogy with the magnetic case,
one can further expand ${\cal A}_{\Psi,q}$ in powers of $q$. The next term in
this expansion has an overall structure
\be
{\cal A}_{\Psi-q^{2}}\propto c_2\sqrt{T}\int d{\bm x}\ \tr\bigl(\beta({\bm
x})\,
   q({\bm x})\,q^{\dagger}({\bm x})\bigr).
\label{eq:4.41}
\ee

It is now straightforward to determine the Gaussian propagators. In particular,
the order parameter correlation function reads
\bse
\label{eqs:4.42}
\be
\langle\Psi_n^{r,\alpha}({\bm x})\,\Psi_m^{s,\beta}({\bm x})\rangle
   = \delta_{{\bm k},-{\bm p}}\,\delta_{rs}\,\delta_{\alpha\beta}\,
        \delta_{nm}\,{\cal N}_n({\bm x}),
\label{eq:4.42a}
\ee
with
\be
{\cal N}_n({\bm k}) = \frac{1}{r_0 - C({\bm k},\Omega_n)}\,.
\label{eq:4.42b}
\ee
Here
\be
C({\bm k},\Omega_n) = \frac{c_1^2\,\ln\bigl(\Omega_0/(D{\bm k}^2 +
\vert\Omega_n\vert)\bigr)}{1 + ({\tilde K}_{\text{c}}/H)
\ln\bigl(\Omega_0/(D{\bm k}^2 + \vert\Omega_n\vert)\bigr)}\,,
\label{eq:4.42c}
\ee
where $\Omega_0$ is an ultraviolet frequency cutoff and $r_0$ is the
coefficient of the $\Psi^2$ in ${\cal A}_{\Psi}$. For asymptotically small wave
numbers and frequencies the critical propagator is given by
\be
{\cal N}_{n}(\mathbf{k})= \left[r + \frac{\text{const.}}{\ln\bigl(\Omega_0/
   (D{\bm k}^2 + \vert\Omega_n\vert)\bigr)}\right]^{-1},
\label{eq:4.42d}
\ee
\ese
with ${\text{const.}}>0$ and $r = r_0 - c_1^2\,H/{\tilde K}_{\text{c}}$ the
Gaussian distance from the quantum critical point.

The Gaussian theory represented by Eqs.\ (\ref{eqs:4.42}) has several
interesting properties. First, for given $r_0>0$, $H$, and ${\tilde
K}_{\text{c}}$ there is a quantum phase transition at a critical strength of
$c_1^2\propto \vert K_{\text{c}}\vert$ which yields $r=0$. Second, for fixed
other parameters in the superconducting phase, an increase of ${\tilde
K}_{\text{c}}$, which is an increasing function of disorder, will drive the
system into the normal-metal phase. This is consistent with other theories that
have identified the Coulomb pseudopotential as an important source of
$T_{\text{c}}$-degradation in disordered superconductors
(\onlinecite{Finkelstein_1987, Kirkpatrick_Belitz_1992c}; see
\onlinecite{Belitz_Kirkpatrick_1994} for a discussion of earlier theories of
this effect). Third, as in the ferromagnetic case there are long-ranged order
parameter correlations in the disordered phase away from criticality. Equation
(\ref{eq:4.42d}) implies for the order parameter susceptibility
\be
\chi_{\Psi}(r>0,\vert{\bm x}\vert\to\infty) \propto 1/\vert{\bm x}\vert^d\,
   \ln\vert{\bm x}\vert.
\label{eq:4.43}
\ee
A comparison with Eq.\ (\ref{eq:4.25}) shows that the correlations are of even
longer range than in the ferromagnetic case for all $d>2$. This reflects the
strong coupling of the superconducting order parameter to the generic soft
modes mentioned in connection with Eq.\ (\ref{eq:4.37}).

Most of the critical behavior predicted by the Gaussian theory can simply be
read off Eqs.\ (\ref{eqs:4.42}) \cite{Kirkpatrick_Belitz_1997}. The correlation
length depends exponentially on $r$, rather than as a power law,
\bse
\label{eqs:4.44}
\be
\xi \sim e^{1/2\vert r\vert}.
\label{eq:4.44a}
\ee
For the correlation length exponent this implies $\nu = \infty$. The exponent
$\gamma$ has its mean-field value,
\be
\gamma = 1,
\label{eq:4.44b}
\ee
and the exponents $\eta$ and $z$ for the order parameter ($\Psi$) and the
fermionic ($q$) degrees of freedom, respectively, are
\be
\eta_{\Psi} = 2\quad,\quad \eta_q = 0\quad,\quad z_{\Psi} = z_q = 2.
\label{eq:4.44c}
\ee
\ese
The order parameter susceptibility at criticality decays in real space as
\be
\chi_{\Psi}(r=0,\vert{\bm x}\vert\to\infty) \propto \ln\vert{\bm
x}\vert/\vert{\bm x}\vert^d.
\label{eq:4.45}
\ee
As expected, the critical order-parameter fluctuations are of even longer range
than those reflecting GSI in the disordered phase, Eq.\ (\ref{eq:4.43}).

These results also follow from a tree-level RG analysis of the field theory.
The exponents $\eta$ are related to the scale dimensions of the fields via
\bse
\label{eqs:4.46}
\bea
\left[q({\bm x}\right] &=& -(d - 2 + \eta_q)/2,
\label{eq:4.46a}\\
\left[\Psi({\bm x})\right] &=& -(d - 2 + \eta_{\Psi})/2,
\label{eq:4.46b}
\eea
\ese
As in the magnetic case, Sec.\ \ref{par:IV.B.4.c}, there is a critical fixed
point where $c_1$ is marginal, and the fermions are diffusive, with exponents
given by Eq.\ (\ref{eqs:4.44}). However, in contrast to the magnetic case, the
coupling constant $c_2$ of the term ${\cal A}_{\Psi-q^2}$ is RG irrelevant, and
so are all higher terms in the expansion in powers of $q$. We therefore
conclude that the Gaussian critical behavior is exact. The most obvious
technical reason for this surprising result is the fact that the time scales
for the order-parameter fluctuations and the fermions, respectively, are the
same, which renders inoperative the mechanism that led the possibility of $c_2$
being marginal in Sec.\ \ref{subsubsec:IV.B.4}. Physically, the very long range
of the order-parameter fluctuations reflecting the GSI stabilizes the Gaussian
critical behavior. This is in agreement with the fact that long-ranged order
parameter correlations in classical systems stabilize mean-field critical
behavior \cite{Fisher_Ma_Nickel_1972}.

The renormalized mean-field theory, the equation of state, and the critical
exponent $\beta$ can be discussed in analogy to the magnetic case
\cite{Kirkpatrick_Belitz_1997, Zhou_Kirkpatrick_2004}.

The theory of the metal-superconductor transition presented above is assuming
conventional s-wave spin-singlet superconductors. It is interesting to ask how
the phase transition scenario is modified by exotic (i.e., nonzero angular
momentum, $\ell>0$) pairing. First of all, in contrast to conventional
superconductivity, exotic superconductivity is rapidly destroyed by nonmagnetic
disorder because Anderson's theorem \cite{Anderson_1959} does not hold. This
has been observed, e.g., in ZrZn$_2$ \cite{Pfleiderer_et_al_2001} which is
believed to be a p-wave spin-triplet superconductor, $\ell=1$. Recently, the
resulting quantum phase transition between a dirty metal and an exotic
superconductor has been studied within a Landau-Ginzburg-Wilson approach
\cite{Sknepnek_Vojta_Narayanan_2002}. It turns out that the nonzero order
parameter angular momentum suppresses the effects of GSI on this transition.
The nonanalytic wave number dependence in the Gaussian order parameter
propagator ${\cal N}_n({\bm k})$ takes the form ${\bm k}^{2\ell} \ln {\bm k}$.
Thus, compared to the s-wave case, Eq. (\ref{eq:4.42d}), the nonanalytic term
is suppressed by a factor of ${\bm k}^{2\ell}$. This can be understood as
follows: In the presence of nonmagnetic quenched disorder, the dominant
electronic soft modes are those that involve fluctuations of the number
density, spin density, or anomalous density in the zero angular momentum
channel, while the corresponding densities in higher angular momentum channels
are not soft. Since the different angular momentum modes are orthogonal at zero
wave number, the coupling between a finite angular momentum order parameter and
the zero angular momentum soft modes must involve powers of the wave number
$|{\bm k}|$.

In conclusion, for $\ell>1$, the GSI induced nonanalytic term is subleading
compared to the conventional ${\bm k}^2$ term, and it will not influence the
critical behavior. For $\ell=1$, the nonanalytic term is marginally relevant.
The ultimate fate of the transition has not yet been worked out, mainly because
of (unrelated) complications stemming from the disorder fluctuations similar to
that in disordered itinerant antiferromagnets.

\subsection{Quantum Antiferromagnetic Transition}
\label{subsec:IV.D}

Let us now consider a quantum antiferromagnetic transition, in analogy to the
ferromagnetic one discussed in Sec.\ \ref{subsubsec:III.B.1}. A crucial
difference between these two cases is that in the latter, both the order
parameter field ${\bm M}$ (whose average is the magnetization) and the generic
soft modes are soft at zero wave number, while in the former, the order
parameter field ${\bm N}$ (whose average is the staggered magnetization) is
soft at a nonzero wave number $\vert{\bm p}\vert$. As a result, the
hydrodynamic wave number ${\bm k}$ in the dynamical piece of the ferromagnetic
Gaussian action, Eq.\ (\ref{eq:4.6}), gets replaced by ${\bm p}$, and one has
\cite{Hertz_1976}
\be
{\cal A}_{\text{H}}[{\bm N}] = \sum_k {\bm N}(k)\,(r + c\,{\bm k}^2
     + d\,\vert\Omega_n\vert)\,{\bm N}(k) + O({\bm N}^4).
\label{eq:4.47}
\ee
Here ${\bm k}$ is the wave vector measured from the reference wave vector ${\bm
p}$. The missing inverse wave number in the frequency term reflects the much
weaker coupling, compared to the ferromagnetic case, of the particle-hole
excitations to the order parameter field.\footnote{\label{fn:60} This holds for
a generic shape of the Fermi surface. Special geometric features of the Fermi
surface can cause particle-hole excitations to be soft at the same wave vector
as the order parameter even in an antiferromagnet, see,
\textcite{Abanov_Chubukov_2000} and the discussion below.} Because of this
weaker coupling, one expects the Gaussian approximation to be much better here
than in the ferromagnetic case. Indeed, it is easy to show that the effects
that needed to be taken into account for the ferromagnet are irrelevant with
respect to the mean-field transition described by Eq.\ (\ref{eq:4.47}). One
thus expects a continuous transition with mean-field critical behavior in all
dimensions $d>2$. The finite-temperature properties of this theory have been
worked out in detail by \textcite{Millis_1993}. It is important to remember
that Eq. (\ref{eq:4.47}) assumes that the only relevant soft modes at the
quantum antiferromagnetic transition are the order-parameter fluctuations.

Experimental observations are not in agreement with this expectation, probably
because most metallic materials that display easily accessible quantum
antiferromagnetic transitions are far from being simple
metals.\footnote{\label{fn:61} An interesting counterexample is the case of
Cr$_{1-x}$V$_x$, which is a common transition metal, yet displays properties,
both at its antiferromagnetic quantum critical point and away from it, that is
very similar to the exotic behavior shown by the heavy-fermion materials and
high-$T_{\text{c}}$ superconductors \cite{Yeh_et_al_2002}. We will come back to
this in Sec.\ \ref{sec:V}.} They fall into the class of heavy-fermion
materials; an overview has been given by \textcite{Coleman_et_al_2001}. The
best-studied system is CeCu$_{\text{6-x}}$Au$_{\text{x}}$, which shows a
quantum phase transition to an antiferromagnetic state at a critical gold
concentration $x_{\text{c}}\approx 0.1$ \cite{von_Loehneysen_et_al_1994}. There
are experimental indications for the quantum critical fluctuations being
two-dimensional or quasi-two-dimensional in nature \cite{Stockert_et_al_1998}.
A detailed phenomenological analysis of neutron scattering experiments
\cite{Schroeder_et_al_2000} has shown that the magnetic susceptibility is well
described by the form
\be
\chi({\bm k},\Omega) = a\,\left[(-i\Omega + b\,T)^{\alpha}
                       + \theta({\bm k})^{\alpha}\right]^{-1}.
\label{eq:4.48}
\ee
Here $a$ and $b$ are constants, $\alpha = 0.75\pm 0.05$, and $\theta({\bm k})$
is the wave vector dependent Weiss temperature. This type of behavior is often
referred to as ``local quantum criticality'', as it is believed (although not
universally so, see below) to point to atomic-scale physics as the cause of the
antiferromagnetism, as opposed to the Fermi-liquid or spin-density-wave
description that underlies Eq.\ (\ref{eq:4.47}). The possibility of such a
local quantum critical point was first pointed out by
\textcite{Si_Smith_Ingersent_1999}.

\begin{figure}[t]
\includegraphics[width=7cm]{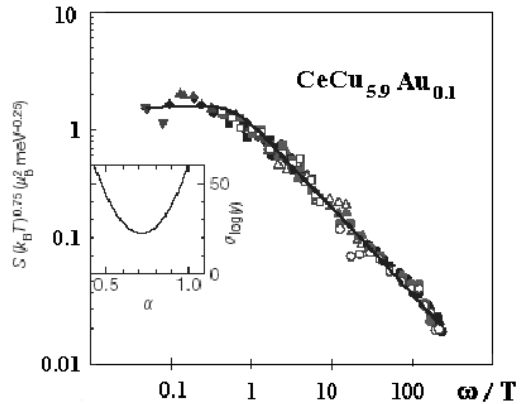}
\caption{\label{fig:23} The magnetic structure factor in CeCu$_{5.9}$Au$_{0.1}$
as measured by neutron scattering at various frequencies and temperatures, as a
function of $\Omega/T$. The inset shows a measure of the scatter of the scaling
plot for various values of the exponent $\alpha$. After
\protect\textcite{Schroeder_et_al_2000}. }
\end{figure}

\begin{figure}[t]
\includegraphics[width=8.6cm]{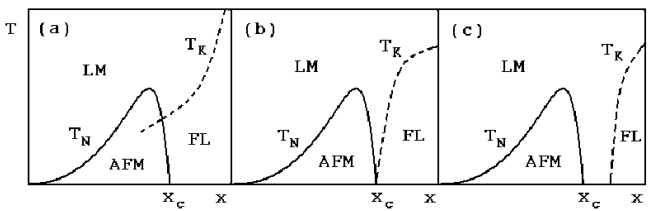}
\caption{\label{fig:24} Schematic phase diagram for CeCu$_{6-x}$Au$_x$.
  Shown are the antiferromagnetic (AFM) phase, the local moments (LM) region, and
  the Fermi liquid (FL) region. The N{\'e}el temperature $T_{\text{N}}$ (solid
  lines) represents a true phase transition, while the Kondo temperature
  $T_{\text{K}}$ (dashed lines) denotes a crossover. The three scenarios shown
  correspond to the three different physical situations discussed in the text.
  After \protect\textcite{Schroeder_et_al_2000}.
}
\end{figure}

Perhaps the most obvious violation of the expected mean-field behavior is the
occurrence of the exponent $\alpha\neq 1$. Other discrepancies between the
observed behavior and the one expected from Hertz theory have been discussed by
\textcite{Coleman_et_al_2001} and \textcite{Si_et_al_2001}. A prominent one is
that frequency and temperature are observed to scale in the same way, see Fig.\
\ref{fig:23}.
This observation is usually referred to as ``$\Omega/T$ scaling'', and it is
expected to hold at a quantum critical point that exhibits hyperscaling
\cite{Sachdev_Ye_1992}. By contrast, in Hertz theory $\Omega$ scales as
$T^{3/2}$ \cite{Millis_1993}.\footnote{\label{fn:62} Naively, one might expect
$\Omega/T$ scaling to simply follow from the fact that the scale dimension of
$T$ is given by the dynamical critical exponent $z$, and hence the same as that
of $\Omega$. In general, this is not true due to the presence of, (1) multiple
temperature scales, and (2) dangerous irrelevant variables. See,
\textcite{Millis_1993, Sachdev_1997}.} Another one is the behavior of the
specific heat coefficient, which is observed to diverge logarithmically, while
in Hertz theory it remains finite and shows a square-root cusp singularity
\cite{Millis_1993}. \textcite{Schroeder_et_al_2000} have proposed the following
physical picture to explain this violation. In CeCuAu, or any heavy-fermion
material, the highly localized f-orbitals of the lanthanide (in this case, Ce)
or actinide provide local magnetic moments, while the more extended s-, p-, and
d-orbitals provide a Fermi surface. An increase in Au doping increases the
hybridization between the localized and extended orbitals, which increases the
coupling between the local moments via the conduction electrons, and hence the
N{\'e}el temperature $T_{\text{N}}$. If the hybridization becomes too strong,
however, the f-electrons are incorporated into the Fermi surface, and the
antiferromagnetism disappears. As a function of doping, $T_{\text{N}}$ thus
goes through a maximum and disappears at a critical doping concentration, see
Fig.\ \ref{fig:24}.
If this were all that happens, Hertz theory should apply. However, there is a
second temperature scale besides $T_{\text{N}}$, namely, the Kondo temperature
$T_{\text{K}}$. Only for $T<T_{\text{K}}$ do the local moments of the
f-orbitals become screened by the conduction electrons, and the hybridization
becomes effective. \textcite{Schroeder_et_al_2000} have proposed that
$T_{\text{K}}$ vanishes at the same critical concentration as $T_{\text{N}}$,
see Fig.\ \ref{fig:24}(b). In this picture, there is no heavy-electron Fermi
surface at the quantum critical point, and Hertz theory is inapplicable. Notice
that, according to this picture, there is nothing technically wrong with Hertz
theory; the reason for its failure is that the model does not contain crucial
slow degrees of freedom. Alternatively, one can imagine a scenario where
$T_{\text{K}}$ remains nonzero into the antiferromagnetic phase, see Fig.\
\ref{fig:24}(a), but the experiments show that this is not the case in CeCuAu.
Finally, it is conceivable, at least in principle, that the local-moment region
might extend to zero temperature in an entire region of the phase diagram, as
shown in Fig.\ \ref{fig:24}(c), although there is currently no explicit theory
that can explain why the unscreened local moments in heavy-fermion systems
would not order at sufficiently low temperatures. Still, it is interesting to
note that there are cases where a non-Fermi liquid phase is observed adjacent
to an antiferromagnetic one, e.g., in URu$_{2-x}$Re$_x$Si$_2$, see Fig.\
\ref{fig:21}.

Theoretically, the quantum antiferromagnetic transition is an open problem that
is under very active consideration. A theoretical description of the above
scenario involves the so-called Kondo lattice problem, i.e., the interaction of
many localized spins with each other and with a band of conduction electrons.
Si and collaborators have studied such a model within a dynamical mean-field
approach \cite{Si_et_al_2001, Si_2003}. In this approximation, these authors
find what they call a local quantum critical point, which has many properties
that are consistent with the observations. The exponent $\alpha$ is
nonuniversal. More recently, \textcite{Senthil_Vojta_Sachdev_2004} have
proposed that the magnetic state is an unconventional spin-density wave where
spin-charge separation has taken place. \textcite{Coleman_Pepin_Tsvelik_2000}
have developed a supersymmetric representation of spin operators that allows
for a treatment of both magnetism and the Kondo effect in the context of a
large-$N$ expansion. These authors have speculated that, if applied to the
Kondo lattice problem, this formalism can give rise to two different types of
fixed points: A weak-coupling fixed point of Hertz type, and a non-Fermi-liquid
fixed point that displays spin-charge separation. There is, however, no
consensus that the existence of a Fermi surface precludes an explanation of the
observations. \textcite{Rosch_et_al_1997} have proposed an explanation in terms
of three-dimensional conduction electrons coupling to two-dimensional
ferromagnetic fluctuations. In a two-dimensional spin-fermion model,
\textcite{Abanov_Chubukov_2000} have found that special nesting properties of
the Fermi surface lead to an exponent $\alpha\approx 0.8$, but their results do
not display $\Omega/T$ scaling. A breakdown of the LGW expansion due to nesting
has also been discussed by \textcite{Lercher_Wheatley_2000}, for a review, see
\textcite{Abanov_Chubukov_Schmalian_2003}. \textcite{Sachdev_Morinari_2002}
have found that similar ``nonlocal'' forms of the dynamic susceptibility are
obtained in two-dimensional models where an order parameter couples to
long-wavelength deformations of a Fermi surface. Notice that all of these
theories involve some coupling of soft modes, generic or otherwise, to the
order parameter, although there currently is no consensus about their nature
and origin. In the order of theoretical ideas listed above, they are, (1) Slow
local moment fluctuations that serve as ``Fermi surface shredders'', or (2)
other fluctuations that destroy the Fermi surface,\footnote{\label{fn:63} These
soft modes are `generic' in the sense of our definition only in the scenario
depicted in panel (c) of Fig.\ \ref{fig:24}.} (3) ferromagnetic fluctuations,
(4) particle-hole fluctuations across a Fermi surface with special geometric
features, and (5) volume and shape deformation of the Fermi surface. We finally
mention that antiferromagnetic quantum criticality, in particular in
two-dimensions, has received much attention in connection with
high-T$_{\text{c}}$ superconductivity, a topic that is beyond the scope of this
article.

\section{Discussion and Conclusion}
\label{sec:V}

In this section we summarize the main theoretical ideas we have presented, as
well as the experimental situation. We also discuss a number of open problems
in the field of quantum phase transitions, and suggest a number of new
experiments to address some of these.

\subsection{Summary of Review}
\label{subsec:V.A}

Early work on quantum phase transitions had suggested that most of them were
conceptually quite simple, since they are related to corresponding classical
phase transitions in a higher dimension. This led to the conclusion that
quantum critical behavior generically would be mean-field like. For a number of
reasons, these conclusions have turned out to be not valid in general. In this
review we have discussed one of the mechanisms that invalidate the mapping of a
quantum phase transition onto a simple classical one in higher dimensions. The
central idea behind this mechanism is as follows. The soft-mode spectrum of
many-body systems is in general different at $T=0$ from the one at $T>0$, since
there are soft modes at $T=0$ that develop a mass at nonzero temperature. These
soft modes lead to long-ranged correlations in entire regions of the phase
diagram, a phenomenon known as generic scale invariance. Physically, both the
generic soft modes and the critical order-parameter fluctuations are equally
important in the long-wavelength limit, and if the coupling between them is
sufficiently strong, the former will influence the leading critical behavior.
This can happen for classical phase transitions as well, but it is less common
since at $T>0$ there are fewer soft modes than at $T=0$. These observation led
to the general paradigm that effects related to generic scale invariance are of
fundamental importance for the theory of {\em generic} quantum critical points.
As a result, quantum phase transitions are typically related to classical phase
transitions in the presence of generic scale invariance, rather than to simple
ones.

In Sec.\ \ref{sec:II} we have reviewed the concept of generic scale invariance
in both classical and quantum systems. Although in both cases examples are
plentiful, in the quantum case this is especially so because of additional
Goldstone modes that exist at zero temperature. We have distinguished between
direct and indirect generic scale invariance effects; the former being
immediate consequences of Goldstone's theorem, conservation laws, or gauge
symmetries, while the latter arise from the former via mode-mode coupling
effects. Classical examples we have discussed include: Goldstone modes in
Heisenberg ferromagnets or analogous systems, which lead to long-ranged
susceptibilities everywhere in the magnetically ordered phase; local gauge
invariance in superconductors or liquid crystals, in which context we have
stressed interesting analogies between statistical mechanics and particle
physics; long-time tails in time correlation functions which determine the
transport coefficients in a classical fluid in equilibrium; and long-ranged
spatial correlations in a classical fluid in a nonequilibrium steady state.
Examples of generic scale invariance in the quantum case all involved
interacting electron systems, namely, weak-localization effects in disordered
materials and the analogous effects in clean ones; and nonequilibrium effects
analogous to those in classical fluids. Throughout this discussion we have
stressed the coupling between the statics and the dynamics in quantum systems,
which leads to long-ranged spatial correlations in quantum systems even in
equilibrium.

In Sec.\ \ref{sec:III} we discussed some classical phase transitions where
generic scale invariance plays a central role. Our first example was the
nematic--smectic-A transition in liquid crystals, which maps onto the classical
superconductor-normal metal transition. In this case the relevant generic soft
modes are the director fluctuations, which are Goldstone modes due to a broken
rotational symmetry. They can drive the transition first order, even though in
their absence one expects a continuous transition. The other classical phase
transition discussed was the critical point in a classical fluid. Here the
generic soft modes are due to conservation laws. In equilibrium, they influence
the critical dynamics only. However, in a fluid subject to shear, they also
couple to the static critical behavior. In this case, long-ranged static
correlations due to the generic soft modes stabilize mean-field critical
behavior below the equilibrium upper critical dimension.

Section\ \ref{sec:IV} was devoted to four examples of quantum phase
transitions. For the first three, namely, the quantum ferromagnetic transition
in clean and disordered itinerant electrons systems, respectively, and the
normal metal-superconductor transition at $T=0$, the generic soft modes are the
particle-hole excitations that cause the weak-localization effects and their
clean counterparts. The clean ferromagnetic case turned out to be analogous to
the classical nematic--smectic-A transition in that the generic soft modes can
lead to a fluctuation-induced first-order transition. The disordered quantum
ferromagnetic transition is analogous to the classical fluid under shear, in
the sense that static long-ranged correlations due to the generic soft modes
stabilize a simple critical fixed point, and the critical behavior can be
determined exactly even in $d=3$. In the case of the superconducting transition
this effect is even stronger, and the critical fixed point is Gaussian. Our
fourth example dealt with the quantum antiferromagnetic transition, where the
critical behavior is currently not understood. There are many indications that,
in the materials studied so far, generic soft modes related to local magnetic
moments exist and couple to the critical order-parameter fluctuations, but a
detailed theory of this effect remains to be worked out.

\subsection{Open Problems, and Suggested Experiments}
\label{subsec:V.B}

We conclude by listing a number of open questions, both theoretical and
experimental, the answers to which would help shed light on some of the
problems we have discussed. Our remarks are necessarily incomplete and
speculative.

\subsubsection{Generic scale invariance, and soft modes}
\label{subsubsec:V.B.1}

There are strong indications that the list of mechanisms for generic scale
invariance in Sec.\ \ref{sec:II} is incomplete, even for relatively simple
systems. Here we briefly mention two examples.

\paragraph{Generic non-Fermi-liquid behavior}
\label{par:V.B.1.a}

As mentioned in Sec.\ \ref{subsubsec:IV.B.4}, behavior not consistent with
Landau's Fermi liquid theory has been observed in a variety of materials far
from any quantum critical point. While such exotic behavior may not come as too
much of a surprise in strongly correlated systems involving rare earths or
actinides, the well documented case of MnSi
\cite{Pfleiderer_Julian_Lonzarich_2001} makes it likely that some more basic
understanding is lacking. The observations clearly show that long-ranged
correlations exist in a large region of parameter space that are not caused by
any quantum phase transition. For instance, the resistivity has a $T^{3/2}$
asymptotic temperature dependence up to a factor of two in parameter space away
from the quantum ferromagnetic transition. This generic long-time tail behavior
cannot be explained in an obvious way by any existing theory. Apart from being
intrinsically interesting, this also raises the question whether there are
unknown generic soft modes that will be important for a complete understanding
of the quantum phase transition. One obvious candidate are slow fluctuations of
local magnetic moments, which have been invoked for many mysterious effects,
from hard-to-understand aspects of metal-insulator transitions (for a review
see, e.g., Sec. IX.B of \onlinecite{Belitz_Kirkpatrick_1994}) to the critical
behavior at the quantum antiferromagnetic transition (see, Sec.\
\ref{subsec:IV.D}). So far, however, there is no detailed theory that is
capable of describing these soft modes.

\paragraph{Soft modes due to nested Fermi surfaces}
\label{par:V.B.1.b}

\textcite{Abanov_Chubukov_2000} have discussed nesting properties of Fermi
surfaces that are important for quantum antiferromagnetic transitions. In the
language of this review, the effect considered by these authors is likely a
change of the properties of the generically soft particle-hole excitations. It
would be useful to explicitly confirm this conjecture, and to develop a general
classification of the effects of Fermi surface geometries on generic soft
modes.

\subsubsection{Aspects of the quantum ferromagnetic transitions}
\label{subsubsec:V.B.2}

\paragraph{Disordered ferromagnets}
\label{par:V.B.2.a}

A detailed experimental study of the critical behavior of both the
thermodynamic and transport properties near the quantum ferromagnetic
transition in a disordered itinerant electron system would be very interesting.
Some recent results on Fe$_{1-x}$Co$_x$S$_2$ \cite{DiTusa_et_al_2003} appear to
be consistent with the log-log-normal corrections to power-law scaling
discussed in Sec.\ \ref{subsec:IV.B}. A more precise determination of the
critical behavior in this or other systems would be of great help in confirming
or refuting the theoretical ideas.

Also of interest would be a systematic study of the destruction of the first
order ferromagnetic transition by nonmagnetic disorder that is predicted by the
theory discussed in Sec.\ \ref{subsec:IV.B}. For instance, the theory predicts
that the tricritical point observed in MnSi and UGe$_2$ will move to lower and
lower temperature with increasing strength of quenched disorder, turn into a
tricritical end point, then a critical end point, and finally the first-order
transition should disappear, see Fig.\ \ref{fig:19}. An experimental check of
this prediction would be very valuable. For instance, it would give an
indication of whether or not an understanding of the non-Fermi-liquid nature of
the paramagnetic phase, mentioned in Sec.\ \ref{subsubsec:V.B.1}, is important
for describing the quantum phase transition.

\paragraph{Itinerant ferromagnets with magnetic impurities}
\label{par:V.B.2.b}

An interesting quantum phase transition to study experimentally would be the
ferromagnetic transition in an itinerant electron system with dilute magnetic
impurities, which would fundamentally change the soft-mode structure in systems
with or without additional nonmagnetic disorder. Such a study would provide an
important check of the soft-mode paradigm, especially if it were possible to
start with no magnetic impurities and study the crossover induced by their
gradual introduction. In the absence of complications due to Kondoesque
effects, one would expect a local LGW theory for the order-parameter
fluctuations to be valid, since the magnetic impurities cut off the soft modes
that strongly couple to the order-parameter fluctuations. Without nonmagnetic
disorder, this would be Hertz's original theory, while in the nonmagnetically
disordered case, rare-region effects would likely play an important role.

It is likely, however, that Kondo screening effects, and interactions between
the impurity sites, will lead to complications not unlike those believed to be
responsible for the observations in quantum antiferromagnets, Sec.\
\ref{subsec:IV.D}, which are not understood. Generally, the physics of local
moments, and their influence on the properties of conduction electrons, is one
of the most important unsolved problems in condensed matter physics.

\paragraph{Clean ferromagnets}
\label{par:V.B.2.c}

The finite-temperature behavior of the observables near the continuous quantum
ferromagnetic transition in clean itinerant electron systems has not been
worked out. Given the existing RG description that was briefly discussed in
Sec.\ \ref{subsubsec:IV.A.4}, and the existing theory of finite-temperature
effects near Hertz's fixed point \cite{Millis_1993}, this should be a
relatively straightforward problem.

\subsubsection{Metal-superconductor transition}
\label{subsubsec:V.B.3}

\paragraph{Experimental study of critical behavior}
\label{par:V.B.3.a}

It would be interesting to experimentally study the quantum phase transition
between a disordered metal and a disordered conventional bulk superconductor.
The theory reviewed in Sec.\ \ref{subsec:IV.C} predicts that the quantum phase
transition is governed by a Gaussian fixed point. If this is correct, there
will be no crossover to mean-field behavior, and the asymptotic critical
behavior will be observable, in contrast to the situation at the thermal
transition. Since the predicted Gaussian critical behavior is radically
different from mean-field critical behavior, this should be easy to observe.

\paragraph{Quantum phase transition to a gapless superconductor}
\label{par:V.B.3.b}

The metal-superconductor transition described in Sec.\ \ref{subsec:IV.C}
assumed that the superconducting states was a conventional gapped state. It
would be interesting to construct an analogous theory for the case of a gapless
superconductor, especially in the light of the suggestion by
\textcite{Vishveshwara_Senthil_Fisher_2000} that the superconducting ground
state in a disordered system will generically be gapless. On general grounds it
is likely that the critical behavior is be the same in both cases if the
transition is approached from the metallic side. However, the gapless
superconducting state will have additional soft modes, and these modes may
influence some of the critical behavior, such as the exponents governing the
equation of state, when the transition is approached from the superconducting
side.

\subsubsection{Aspects of the quantum antiferromagnetic transitions}
\label{subsubsec:V.B.4}

\paragraph{Phenomenological theory for the quantum antiferromagnetic problem}
\label{par:V.B.4.a}

As discussed in Sec.\ \ref{subsec:IV.D}, the quantum antiferromagnetic
transition in real systems remains rather incompletely understood. In most
materials studied so far, a basic ingredient of the problem seems to be
itinerant electrons coupled to local moments. The latter are not screened
since, at or near the quantum critical point, the Kondo temperature vanishes
due to critical fluctuations effects. In the absence of a more fundamental
approach to this very complicated problem, it would be very interesting to
construct a purely phenomenological theory of these coupled fluctuations,
including the soft modes associated with the slow temporal decay of the local
moments.

\paragraph{Simple quantum antiferromagnets}
\label{par:V.B.4.b}

As we mentioned in Sec.\ \ref{subsec:IV.D}, most of the work on the quantum
antiferromagnetic transition has focused on heavy-fermions compounds and
high-$T_{\text{c}}$ superconductors, partly due to the general interest in
these systems. These materials are far from being simple metals, and their
behavior is heavily influenced by local-moment physics, among other
complications. It would be interesting to find and study an itinerant electron
system without local moments that has a quantum antiferromagnetic transition,
provided such materials do indeed exist. The complicated behavior of
Cr$_{1-x}$V$_x$, see footnote \ref{fn:61}, is discouraging in this respect,
although at least some properties of this material have been explained as due
to nesting properties of the Fermi surface \cite{Norman_et_al_2003,
Pepin_Norman_2004, Bazalyi_Ramazashvili_Norman_2004}. In the absence of
disorder, the quantum phase transition in such a systems should be described by
Hertz's theory. The introduction of nonmagnetic disorder would likely lead to a
quantum phase transition where statistically rare events are important.

\subsubsection{Nonequilibrium quantum phase transitions}
\label{subsubsec:V.B.5}

In Sec.\ \ref{subsubsec:III.B.2} we discussed a classical nonequilibrium phase
transition where generic scale invariance plays a role. The general topic of
classical phase transitions in driven systems has received a substantial amount
of attention \cite{Schmittmann_Zia_1995}. In Sec.\ \ref{subsubsec:II.B.5} we
discussed how nonequilibrium situations at zero temperature can lead to
correlations that are of even longer range than in equilibrium. Experimentally,
however, the problem of nonequilibrium quantum phase transitions has received
little attention so far. In quantum Hall systems, which we have not discussed
in this review, electric-field effects have been studied (see,
\onlinecite{Sondhi_et_al_1997}), but in this case the electric field acts as a
relevant operator with respect to the transition. Experiments that show an
actual nonequilibrium quantum phase transition would be of interest.

\subsubsection{Other quantum phase transitions where GSI might play a role}
\label{subsubsec:V.B.6}

In addition to the examples covered in this review, one can think of quantum
phase transitions that have not been investigated so far either theoretically
or experimentally, for which GSI effects are likely to play a central role. For
example, the order parameter for the spin-triplet analog of the
isotropic-to-nematic phase transition that has been proposed to occur in
quantum Hall systems by \textcite{Oganesyan_Kivelson_Fradkin_2001} is expected
to couple to generic soft modes. For this transition, in contrast to its
spin-singlet analog, a simple LGW theory is therefore unlikely to be valid.

\acknowledgments

We have benefited from discussions and correspondence with M. Aronson, E.D.
Bauer, A. Chubukov, P. Coleman, N.G. Deshpande, W. G{\"o}tze, A. Millis, C.
Pepin, P. Phillips, S. Sachdev, Q. Si, and J. Toner. We would also like to
thank our collaborators on some of the topics discussed in this review: J.R.
Dorfman, F. Evers, M.T. Mercaldo, A. Millis, R. Narayanan, J. Rollb{\"u}hler,
J.V. Sengers, S.L. Sessions, R. Sknepnek, and Lubo Zhou. We are grateful to
E.D. Bauer, J.R. Dorfman, M. Dzero, T.C. Lubensky, B.M. Maple, M. Norman, Q.
Si, and J. Toner for comments on a draft version of the manuscript. Part of
this work was performed at the Aspen Center for Physics, and at the Max-Planck
Institute for the Physics of Complex Systems in Dresden, Germany. We would like
to thank both institutions for their hospitality. This work was supported by
the NSF under grant Nos. DMR-99-75259, DMR-01-32555, and DMR-01-32726, and by
the University of Missouri Research Board.

\section*{List of Acronyms}

\noindent GSI \hskip 20pt generic scale invariance\hfill
\break\noindent 
LGW \hskip 12pt Landau-Ginzburg-Wilson \hfill
\break\noindent 
LTT\hskip 21pt long-time tail \hfill
\break\noindent 
RG\hskip 26pt renormalization group


\end{document}